\documentclass{elsarticle}       

\usepackage
[
        a4paper,
        left=2.5cm,
        right=2.5cm,
        top=2cm,
        bottom=2cm,
]
{geometry}

\usepackage{graphicx}
\usepackage{color}
\journal{Archive of Applied Mechanics}
\usepackage{graphicx,epstopdf}
\usepackage{bm}
\usepackage{amsfonts}
\usepackage{amssymb}
\usepackage[version=4]{mhchem}
\usepackage{siunitx}
\usepackage{epsfig}
\usepackage{epstopdf}

\usepackage[normalem]{ulem}
\date{}

\begin{document}
\begin{frontmatter}

\title{Micromechanical Analysis of Hyperelastic Composites with Localized Damage Using a New Low-Memory Broyden-Step-Based Algorithm
}


\author{Nathan Perchikov}
\author{Jacob Aboudi}


\address{School of Mechanical Engineering,
  Faculty of Engineering,
              Tel Aviv University,
  Ramat Aviv, Tel Aviv 69978, Israel
             
}



\begin{abstract}
A multiscale (micro-to-macro) analysis is proposed for the prediction of the finite strain behavior of composites with hyperelastic constituents and embedded localized damage. The composites are assumed to possess periodic microstructure and be subjected to a remote field. At the microscale, finite-strain micromechanical analysis based on the homogenization technique for the (intact) composite is employed for the prediction of the effective deformation. At the macroscale, a procedure, based on the representative cell method and the associated higher-order theory, is developed for the determination of the elastic field in the damaged composite. The periodic composite is discretized into identical cells and then reduced to the problem of a single cell by application of the discrete Fourier transform. 
The resulting governing equations, interfacial and boundary conditions in the Fourier transform domain, are solved by employing the higher-order theory in conjunction with an iterative procedure to treat the effects of damage and material nonlinearity. The initial conditions for the iterative solution are obtained using the weakly-nonlinear material limit and a natural fixed-point iteration. A locally-convergent low-memory Quasi-Newton solver is then employed. A new algorithm for the implementation of the solver is proposed, which allows storing in the memory directly the vector-function history-sequence, which may be advantageous for convergence-control based on specific components of the objective vector-function. The strong-form Fourier transform-based approach employed here, in conjunction with the new solver, enables to extend the application of the method to nonlinear materials and may have computational efficiency comparable or possibly advantageous to that of standard approaches.

\end{abstract}

\begin{keyword}

Finite strain \sep Micromechanical analysis \sep Periodic composites \sep Homogenization \sep Broyden's method \sep Low memory

\end{keyword}
\end{frontmatter}

\section{Introduction}
\label{intro}
It is well known that the effective behavior of composite materials can be established by employing a homogenization technique, which is
based on the assumption that the composite's microstructure is periodic, see \cite{SZ}, for example.
The periodic nature of this type of composites allows the identification of a repeating unit cell which has to be appropriately analyzed in order to
establish its effective behavior which is identical to that of the composite.
However, in the case where it is desired to analyze the elastic field of a composite with embedded local damage (e.g. a crack),
the homogenization technique is no longer applicable.
This is because analyzing a repeating unit cell that includes the localized damage implies that the damage is embedded within the composite in
a periodic manner, which is, of course, hardly a realistic scenario. 

In a series of investigations, \cite{AR12}, \cite{RA12}, \cite{AR13}, \cite{Ab17a} and \cite{BRA}, 
localized damage, in the form of stiff and soft inclusions, cavities, finite, semi-infinite and interfacial cracks in composite materials, has been studied.
In these investigations, a multiscale analysis approach has been developed, aimed to establish the behavior of composites with localized damage.

The multiscale formulation involves micro-to-macro analysis.
In the framework of the microscale analysis, the effective behavior of the undamaged (intact) composite is established.
The micromechanically established solution is employed as the applied `far-field' of the damaged composite.
In these investigations, the high-fidelity generalized method of cells (HFGMC), \cite{AAB},   
micromechanics model has been employed for the analysis on the microscale.
The HFGMC micromechanical theory is based on the homogenization technique for composites that are characterized by periodic microstructure,
such that a repeating unit cell can be identified. This repeating unit cell is discretized into several subcells.
The displacement vector in each subcell is represented by a second-order expansion (which includes the applied far-field)
in terms of local coordinates. The unknown terms (microvariables) in the expansion are determined by the fulfillment of the equilibrium equations,
interfacial displacement and traction continuity, as well as global boundary conditions.
The latter conditions ensure that the displacements and tractions at opposite surfaces of the repeating unit cell represent far-field-induced global displacements -- which translates to identical strains and stresses on opposite surfaces (which essentially owes to symmetry -- a weaker property of periodic undamaged microstructure).
All the equations are imposed in the average (surface-integral sense).
Thus, instead of satisfying the equilibrium equations, for example, at every point of the subcell,
the volume integrals over the subcell of these equations are satisfied (i.e., set equal to zero).
Similarly, the continuity conditions between neighboring subcells are imposed by equating the surface integrals of the relevant field values.
A great advantage of HFGMC theory stems from its ability to predict, in addition to the effective moduli of the composite, also the fields distributions that result from
the application of specific loading to the composite. 

Having established the effective behavior of the undamaged composite, from which the applied far-field on the damaged composite can be determined,
it is possible to proceed to the macroscale analysis, by which the \emph{damaged} composite can be studied.
This composite is subjected to the micromechanically established far-field, which is assumed to hold far away from the damage locus, which is assumed to be limited to one or a few cells, situated far enough from the boundary cells.  
The macroscale derivation involves two distinct analysis steps.
In the first one, the representative cell method, \cite{RN97}, is employed. In the framework of this method, the periodic composite with localized
damage is discretized into several identical cells and then reduced to the problem of a single cell by applying the discrete Fourier transform. 
In the second one, the resulting governing equations, interfacial and boundary conditions in the Fourier transform domain, are solved by employing the higher-order 
theory (HOT), \cite{AAB}. Among the features of the approach is that the sub-problem in the Fourier space obtained for each Fourier transform coefficient (harmonica), can be solved on a separate computer. In other words, the method allows parallel computation (or parallelization) of a major part of the computational  effort required for the solution of the full problem, not unlike the parallelization that can be employed in the construction of the stiffness matrix in Finite Element Method algorithms.

The HOT has been originally developed for the analysis of functionally graded materials. The basic analysis of the governing equations is similar
to that of the HFGMC, but differs in the application of the relevant boundary conditions, which replace the periodic boundary conditions of the HFGMC formulation.
Here too, the boundary conditions that are applied at the composite surfaces, are imposed in the integral sense rather than in a point-wise manner.
Chapters 6 and 11 of \cite{AAB} discuss in detail both the HFGMC and the HOT, and supply various verifications and applications.

Having established by the HOT the elastic field in the transform domain, it is possible to proceed by applying inverse Fourier transform and to obtain the real-space
field at any point of the composite impaired by the localized damage.

The macroanalysis involves calculation of initially unknown eigenstresses.
Therefore, an iterative procedure, based on Banach's contracting mapping method (e.g., \cite{Rall}), had been employed. This iterative procedure converged rapidly,
yielding the field distributions at any location of the damaged composite.  

The aforementioned investigations involved damaged composites with thermoelastic quasistatic fields. Extensions of the multiscale analysis for composites with
localized damage have been performed by \cite{AR14} and \cite{RA16}. 
The effects of various types of localized damage in `smart' composites (piezoelectric, electro-magneto-elastic and thermo-electro-magneto-elastic) have been investigated
by \cite{Ab12}, \cite{Ab17b}, \cite{Ab17c}, \cite{Ab17d} and \cite{Ab18}. 
 
Thus far, the described multiscale analysis has been applied to composites with linear constituents, to predict the behavior of linear composites with various types of localized damage.
Unfortunately, the implementation of this method of analysis for the prediction of damaged composites with hyperelastic constituents undergoing large deformations was unsuccessful.
This owes to the fact that the iterative mapping becomes non-contracting as nonlinearity becomes more pronounced. 
To this end, instead of relying on a naturally emerging fixed-point iterative mapping that becomes non-contracting, a locally convergent generic solver is employed in the present work.  

The most notable general efficient numerical solver for smooth square \emph{systems} of general smoothly-nonlinear equations is Newton's method (see \cite{Newton} for a review). The disadvantage of Newton's method for problems in which the (vector) function the root of which is sought is calculated by a complicated algorithm, is that the associated Jacobian is not available analytically. In such cases one has to use numerical differentiation. The main problem with this approach in terms of algorithmic complexity is that the appropriate step for the calculation of a finite difference is not known in advance and has to be searched for. Moreover, for large systems, the number of calculations is proportional to the number of variables. In terms of memory requirements, the generally non-sparse computed Jacobian matrix has to be stored, which tends to be quite expensive. Finally, the method may diverge if the roots coincide with local extrema of the aforementioned functions, which may be true, especially for oscillatory functions, such as those obtained from approximate inverse Fourier transforms, as in the algorithm described above.

A standard remedy for all the noted disadvantages is using the so-called Quasi-Newton methods. There, approximations for the aforementioned Jacobian is constructed using information from the solution history. Those methods require less computational effort as there is no explicit differentiation stage. The price, however, is a lower convergence rate. One method, proven especially successful in practice is the so-called Broyden method, \cite{Broyden1965}, which suggests an especially clever approach to approximating the Jacobian, such that the convergence rate is optimized. An obvious shortcoming of the direct application of the Broyden method is that it still has high memory demands as the approximate Jacobian needs to be stored in the memory. The solution to this shortcoming has been introduced in the literature in what is known as Limited-Memory, or Low Memory Broyden methods. The economy in memory is possible due to the fact that only an inner product of the inverse approximate Jacobian with the vector function the root of which is sought needs to be stored. Moreover, the approximate Jacobian according to the Broyden method is updated in each solution iteration by a low-rank matrix. This allows one storing only the histories of one or two special vectors, for the execution of the algorithm. One common approach, \cite {Rotten2014} uses two such vectors, requiring the storage of $2Np$ double precision numbers, $N$ being the number of equations to be solved and $p$ being the maximum number of iterations up to reasonable level of convergence. Although this approach is rather popular, as it happens, a more economic algorithm was proposed already in 1970 by Rheinboldt (see \cite{Rheinboldt1970}, \cite{Rheinboldt1998}). In his seminal work, Rheinboldt managed to construct an algorithm requiring the storage of only $Np$ double-precision numbers, which is 50 percents less. The vector histories to be stored in his algorithm were the steps themselves, that is the suggested updates for the state variables themselves.

The algorithm of Rheinboldt is undoubtedly as efficient and economic as one can hope for. However, one step forward can still be made. The solution steps, although being the required output of a solver, are not the natural candidates for stored information. Those can only point on the \emph{steps} of the algorithm, but do not allow following the \emph{success} of convergence  for different vector function components. If one could devise an algorithm which stores directly the function values histories, then it would have been possible to observe the success of convergence for specific equations. For example, for the problem addressed in this paper, one may expect convergence challenge close to the location of the damage, or, alternatively, near specific inclusions in a composite. For a solver that stores directly the function histories, which stand for Cauchy stress components' corrections at each computational spatial cell, one could follow the quality of convergence of the solution for the stress (represented through the eigenstress) at specific `interesting' spatial locations, and perhaps devise certain control or stopping criteria based on such observations.

Having this idea in mind, this work indeed offers a novel low-memory algorithm based on the Broyden step, requiring storage memory for $Np$ double-precision numbers, such that the consecutive vector function histories are stored in the memory directly. The construction of the algorithm assumed the use of the so-called \emph{good} Broyden step, the optimality of which was phenomenologically discussed (and never actually disputed) in the literature, and which is argued for theoretically in Appendix \ref{AppendixB} of this  paper.

This algorithm complements the idea of the use of the back and forth Fourier transform for the micromechanical analysis of hyperelastic composites with localized damage in the volume (area)-averaged strong form. The structure of the remainder of the paper is as follows: Section 2 presents the governing equations, Section 3 discusses the numerical solution approach, including the introduction of the new algorithm for the implementation of the low-memory nonlinear solver, Section 4 demonstrates the application of the approach to a set of specific composites with different geometries, albeit only relatively `nicely-shaped' ones (which is to be overcome in future work), resorting to various materials and damage patterns, and Section 5 concludes.

\section{Governing Equations}
\label{sec:2}
We consider a composite whose constituents are isotropic hyperelastic materials.
The undamaged composite is assumed to possess periodic microstructure forming a doubly periodic array (see Fig. \ref{Figure1}(a)). The composite's deformation due to external loading is described with respect to initial Lagrangian coordinates ($X_1$, $X_2$, $X_3$).

The strain energy function $W$ of a constituent is given by 
\begin{eqnarray} \label{E1}
 \ \ \ \ \ \ \ W = W (I_1, I_2, I_3)
\end{eqnarray}
where $I_1$, $I_2$, $I_3$ are the three standard invariants of the right Cauchy-Green deformation tensor ${\bm C}$. This tensor is expressed in terms of the deformation gradient ${\bm F}$ by
\begin{eqnarray} \label{E2}
\ \ \ \ \ \ \ {\bm C} = {\bm F}^{\top} {\bm F} 
\end{eqnarray}
where the superscript $\top$ denotes the transpose operation.

Let us denote by ${\bm S}$ the second (symmetric) Piola-Kirchhoff stress tensor. It follows that
\begin{eqnarray} \label{E3}
 \ \ \ \ \ \ \ {\bm S} =2 \frac{\partial W}{\partial {\bm C}}
\end{eqnarray}
The nonsymmetric first Piola-Kirchhoff stress tensor is defined by, \cite{Mal}:
\begin{eqnarray} \label{E4}
 \ \ \ \ \ \ \ {\bm T} = {\bm S} {\bm F}^{\top}
\end{eqnarray}
In the absence of body forces, the equilibrium (linear momentum balance) equation is given by
\begin{eqnarray} \label{E5}
\ \ \ \ \ \ \ \nabla \cdot {\bm T} = {\bm 0}
\end{eqnarray}
where the gradient $\nabla$ is calculated with respect to material coordinates.

In the present article, two types of hyperelastic materials are considered. 
For the first one, the compressible version of the  Mooney-Rivlin strain energy for rubber-like
materials, introduced by \cite{SB}, is considered:
\begin{eqnarray} \label{E6}
 \ \ \ \ W = C_1 (\hat {I_1} - 3) + C_2 (\hat{I_2} - 3) + \frac{\kappa}{2} (J - 1)^2
\end{eqnarray}
where $\hat {I_1}=I_1 I^{-1/3}_3$, $\hat {I_2}=I_2 I^{-2/3}_3$, $J =$ det$({\bm F})$, $\kappa$ is the bulk modulus and $C_1$, $C_2$ are material parameters.
For small deformations the material is characterized by the bulk modulus $\kappa$ and the shear modulus $\mu = 2 (C_1 + C_2)$. In the examples shown in the applications section, the values $C_1=0.3$ MPa, $C_2=0.1$ MPa, $\kappa=3$ MPa were employed.
The resulting first Piola-Kirchhoff stress tensor, derived from Eqs. (\ref{E6}), is denoted by ${\bm T}^{MR}$.

The second type of hyperelastic material is characterized by the Murnaghan strain energy function, \cite{Mur},
which is expressed in terms of the invariants of the Cauchy-Green (or the Lagrangian) strain tensor ${\bm E} = \frac{1}{2}({\bm C} - {\bm I})$ as follows:
\begin{equation} 
\label{E7}
\begin{split}
\ \ \ W = \frac{\lambda + 2 \mu}{2} J^2_1 - 2 \mu J_2 + \frac{l +2 m}{3} J^3_1 - 2 m J_1 J_2 + n J_3 
\end{split}
\end{equation}
where
\begin{eqnarray} \label{E8}
J_1 = \textrm{tr} ({\bm E}), \ \ \ \
J_2 = \frac{J^2_1 - \textrm{tr} ({\bm E}^2)}{2}, \ \ \ \
J_3 = \textrm{det} ({\bm E})
\end{eqnarray}
where `tr()' denotes the trace of a square matrix and $\lambda$, $\mu$, $l$, $m$ and $n$ are material constants (later assumed identical for all the subcells and cells associated with the material in question, and with $\lambda$ and $\mu$ corresponding to initial, `ambient', linear-elastic behavior).  In the examples shown in the applications section, the material parameters for the two composite-constituent materials obeying the Murnaghan model as employed, were taken form \cite{Chen}.
The resultant first Piola-Kirchhoff stress tensor is denoted by ${\bm T}^{MUR}$, and its components can be represented in the form:
\begin{eqnarray} \label{E9}
\ \ \ \ \ \ \ \ \ \ \ \ \ T^{MUR}_{ij} = \lambda E_{pp} \delta_{ij} + 2 \mu E_{ij} +  T^{'}_{ij} 
\end{eqnarray}
where $T^{'}_{ij}$ are the additional terms.
The structure of the components of the Cauchy-Green strain tensor $E_{ij} = \left(u_{i,j} + u_{j,i} + u_{k,i} u_{k,j} \right) / 2$
can be utilized to represent the stress tensor $T^{MUR}_{ij}$ as follows:
\begin{eqnarray} \label{E10}
\ \ \ \ \ \ \  T^{MUR}_{ij} = \lambda  u_{p,p} \delta_{ij} + \mu (u_{i,j} + u_{j,i}) + T^{NL}_{ij} 
\end{eqnarray}
where $T^{NL}_{ij}$ are the components of the (added) nonlinear terms, which can be obtained by subtracting the linear terms from the full expression derivable from the strain energy function.

As will be discussed in the following, in the presence of localized damage it is advantageous to represent the constitutive equation of the hyperelastic material as follows:
\begin{eqnarray} \label{E11}
\ \ \ \ \ \ \ T_{ij} = \lambda u_{p,p} \delta_{ij} + \mu (u_{i,j} + u_{j,i}) - T^{e}_{ij} 
\end{eqnarray}
where $T^{e}_{ij}$ denote the components of the eigenstresses, given for the Mooney-Rivlin material by
\begin{eqnarray} \label{E12}
\ \ \ T^{e}_{ij} = \lambda u_{p,p} \delta_{ij} + \mu (u_{i,j} + u_{j,i})  - (1 -D) T^{MR}_{ij}
\end{eqnarray}
In these equations, $u_i$ denote the components of the displacements, $\delta_{ij}$ denote the components of the Kronecker delta and $\lambda$ and $\mu$ are the Lam\'e constants of the material in the small strain limit. In Eq. (\ref{E12}), $D$ is a damage parameter (a binary field taking the values of either 0 --  undamaged region -- or 1 -- damaged region -- assumed to be localized in space in the following), such that:
\begin{equation} 
\label{E13}
\ \ \ \ \ \ \ \ \ \ \ \ \ T_{ij} =\begin{cases} T^{MR}_{ij} \ \ \textrm{for} \ \ D = 0 \\
 0 \ \ \ \ \ \ \ \textrm{for} \ \  D = 1
\end{cases}
\end{equation}

Thus, the representation of the stress components by Eq. (\ref{E11}) in conjunction with Eq. (\ref{E12}), provides the desired requirement that in the region of a crack or a hole, the stresses are zero. 

The stress tensor of the Murnaghan material that includes damage is also given by Eq. (\ref{E11}), provided that the eigenstress tensor is defined in this case by:
\begin{eqnarray} \label{E14}
\ \ \ \ \ \ \ T^{e}_{ij} = D \left( \lambda u_{p,p} \delta_{ij} + \mu( u_{i,j}+u_{j,i}) \right) - (1 -D) T^{NL}_{ij}
\end{eqnarray}
It can be easily verified that
\begin{equation} 
\label{E13}
\ \ \ \ \ \ \ \ \  T_{ij} =\begin{cases} T^{MUR}_{ij} \ \ \textrm{for} \ \ D = 0 \\
 0 \ \ \ \ \ \ \ \ \ \textrm{for} \ \  D = 1
\end{cases}
\end{equation}

\section{Method of Solution}
\label{sec:3}
Far away from the perturbed region within which a crack or a cavity exists, the composite's behavior is governed by its (undamaged) global response resulting from homogenization.
Presently, the finite strain HFGMC, \cite{AAB}, is employed, which establishes the following macroscopic incremental constitutive equation:
\begin{eqnarray} \label{M1}
\ \ \ \ \ \ \Delta \bar {\bm T} = {\bm R}^{*} : \Delta \bar {\bm F} 
\end{eqnarray}
where $\bar {\bm T}$ and $\bar {\bm F}$ are the global (average) first Piola-Kirchhoff stress and deformation gradient tensors, respectively, and ${\bm R}^{*}$ is the effective fourth-order tangent tensor.
Consequently, it is possible to integrate Eq. (\ref{M1}) and establish the composite's stress-deformation (i.e. $\bar {\bm T} - \bar {\bm F}$) response up to any desired externally applied loading level.
 
\begin{figure}[h]
\begin{center}
\includegraphics[scale=0.95,trim={0cm 0.5cm 0cm 0cm},clip]{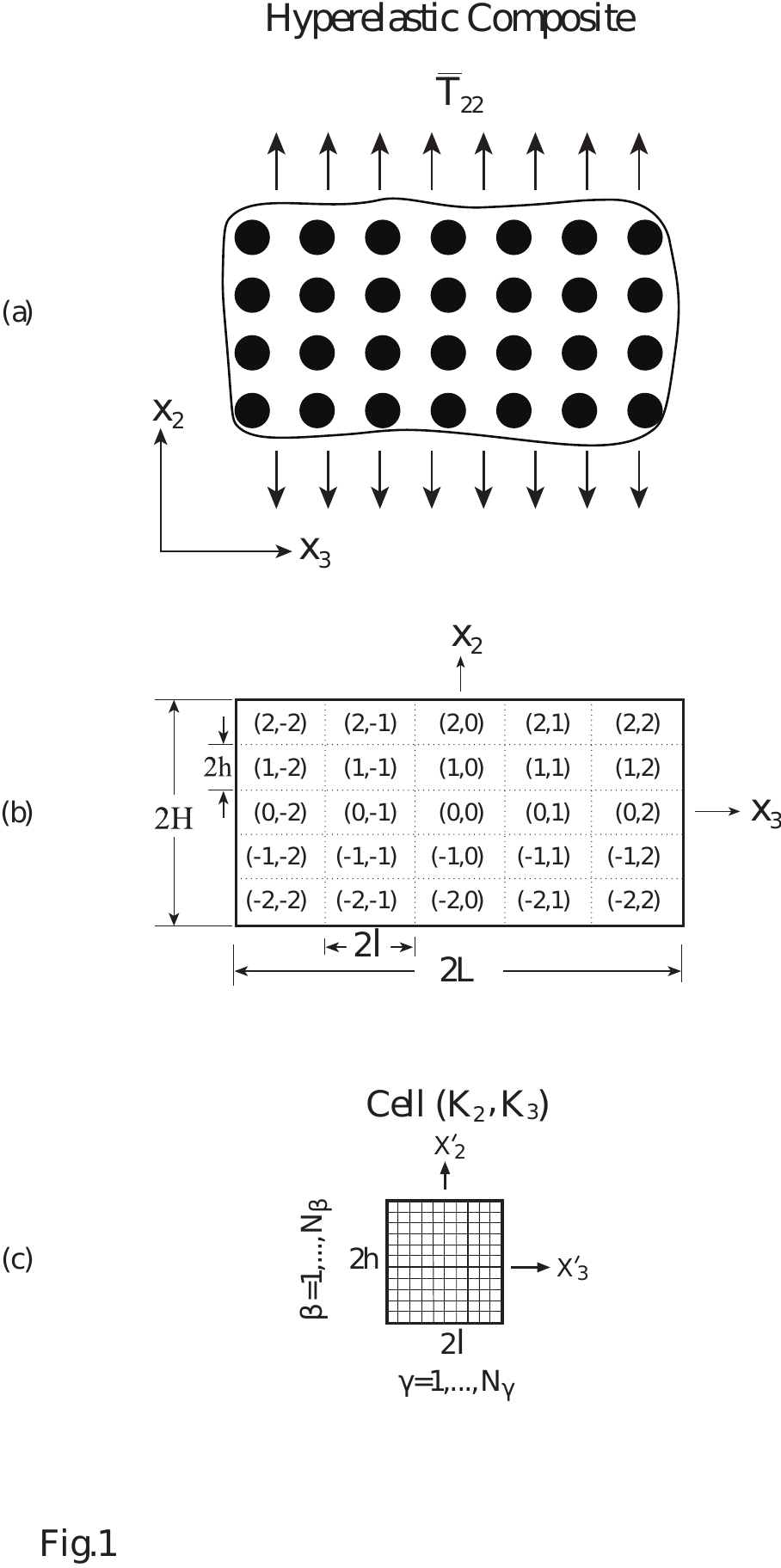}
\end{center}
\caption{(a) A periodically fiber-reinforced composite with a region of localized damage,
                      subjected to constant remote stress $\bar T_{22}$ at infinity.
                  (b) A rectangular domain $2H \times 2L$ of the composite is divided into repeating cells.
                      These cells are labeled by $(K_2, K_3)$ with $-M_2 \le K_2 \le M_2$ and $-M_3 \le K_3 \le M_3$,
                      and the size of every one of them is $2h \times 2l$ (the figure is shown for $M_2=M_3=2$).
                  (c) A characteristic cell $(K_2, K_3)$, in which local coordinates
                      $(X^{'}_2, X^{'}_3)$ with origin at the cell-center are introduced.
                  (d) The characteristic cell, discretized into $N_{\beta}\times N_{\gamma}$ subcells.}
\label{Figure1}
\end{figure}

According to the representative cell method, \cite{RN97}, which is presently generalized for the analysis of hyperelastic composites,    
a rectangular domain $-H \le X_2 \le H$, $-L \le X_3 \le L$ of the composite is considered which includes the perturbed region.
Although this region includes the localized damage, it is assumed that the region is sufficiently extensive relative to the damaged zone, such that the elastic field at its boundaries
is not influenced by the existence of the localized damage. This assumption is standard for the method (see \cite{AR12}). The far-field stress, as obtained for a given far-field strain by the HFGMC homogenization is insensitive to sufficiently localized damage -- in contrast to the exact stress and strain profiles, which are of the main interest in the analysis. In the examples provided in the following, the elastic fields \emph{at the domain boundaries} are verified to be sufficiently close to those obtained for undamaged domains, with relative error of the order of that of the spatial discretization.
Consequently, the boundary conditions that are applied on $X_2=\pm H$ and $X_3=\pm L$ are referred to as the far-field boundary conditions.
This rectangular region is divided into $(2 M_2 +1) \times (2 M_3 +1)$ cells, see Fig. \ref{Figure1}(b) for $M_2=M_3=2$.
Every cell is labeled by $(K_2,K_3)$ with $K_2=-M_2,...,M_2$ and $K_3=-M_3,...,M_3$.
In each cell, local coordinates $(X^{'}_2, X^{'}_3)$ are introduced whose origins are located at the cell center, see Fig. \ref{Figure1}(c). The equilibrium equations in Eq. (\ref{E5}) of the material within the cell $(K_2,K_3)$ take the form
\begin{eqnarray} \label{M2}
\ \ \ \ \ \  T^{(K_2,K_3)}_{kj,k} = {0}, \ \ \ \ j= 1,2,3 \ ; \ \  k=2,3
\end{eqnarray}

The constitutive equation in the cell, Eq. (\ref{E11}), can be written as
\begin{equation} 
\label{M3}
\begin{split}
T^{(K_2,K_3)}_{ij} = \lambda u^{(K_2,K_3)}_{p,p} \delta_{ij} + \mu (u^{(K_2,K_3)}_{i,j} + u^{(K_2,K_3)}_{j,i})  - T^{e(K_2,K_3)}_{ij} 
\end{split}
\end{equation}
where the eigenstress components are given by 
\begin{equation} 
\label{M4}
\begin{split}
T^{e(K_2,K_3)}_{ij} = \lambda u^{(K_2,K_3)}_{p,p} \delta_{ij} +  \mu (u^{(K_2,K_3)}_{i,j} + u^{(K_2,K_3)}_{j,i})  - (1 -D) T^{\text{MR}(K_2,K_3)}_{ij}                      
\end{split}
\end{equation}
for a Mooney-Rivlin material, and
\begin{equation} 
\label{M41}
\begin{split}
T^{e(K_2,K_3)}_{ij} =  D \left[ \lambda u^{(K_2,K_3)}_{p,p} \delta_{ij}  \right.  \left.+ \mu (u^{(K_2,K_3)}_{i,j}+ u^{(K_2,K_3)}_{j,i}) \right]  - (1 -D) T^{NL(K_2,K_3)}_{ij}
\end{split}
\end{equation}
for a Murnaghan material. For a linearly elastic material, on the other hand, the eigenstress components can be verified to take the form
\begin{equation} 
\label{M42}
\begin{split}
 T^{e(K_2,K_3)}_{ij} =  D \left[ \lambda u^{(K_2,K_3)}_{p,p} \delta_{ij} + \mu (u^{(K_2,K_3)}_{i,j}  + u^{(K_2,K_3)}_{j,i} )  \right] 
\end{split}
\end{equation}

The continuity of the tractions acting on the $X_2$ and $X_3$ planes requires that
\begin{equation}
\label{M5}
\begin{split}
\left[ {T}_{2j} ( h, X^{'}_3)  \right]^{(K_2, K_3)} -
   \left [{T}_{2j} (-h, X^{'}_3 ) \right]^{(K_2 +1, K_3)}= 0, 
 K_2=-M_2,...,M_2-1, \ K_3=-M_3,...,M_3
\end{split}
\end{equation}
\begin{equation}
\label{M6}
\begin{split}
\left[ {T}_{3j} (X^{'}_2 ,  l) \right]^{(K_2, K_3)} -
  \left[ {T}_{3j} (X^{'}_2 , -l) \right]^{(K_2, K_3+1)} = 0,    
K_2=-M_2,...,M_2,\ K_3=-M_3,...,M_3-1
\end{split}
\end{equation}

The continuity of displacements between adjacent cells should be imposed, which requires that
\begin{equation} 
\label{M7}
\begin{split}
\left[ {u}_j ( h, X^{'}_3)  \right]^{(K_2, K_3)} -
   \left[ {u}_j (-h, X^{'}_3 ) \right]^{(K_2 +1, K_3)} = 0, 
K_2=-M_2,...,M_2-1,\  K_3=-M_3,...,M_3
\end{split}
\end{equation}
\begin{equation}
\label{M8}
\begin{split}
\left[ {u}_j (X^{'}_2 ,  l) \right]^{(K_2, K_3)} -
  \left[ {u}_j (X^{'}_2 , -l) \right]^{(K_2, K_3+1)} = 0,     
K_2=-M_2,...,M_2,\ \ \ \ \ K_3=-M_3,...,M_3-1
\end{split}
\end{equation}

Next, the far-field boundary conditions must be imposed at the opposite sides $X_2 = \pm H$, $X_3 = \pm L$ of the rectangular domain as shown in Fig. \ref{Figure1}(b).
The far-field tractions at the boundaries are homogenized quantities, insensitive to sufficiently localized damage (limited to a single cell or, at most, several cells), for sufficiently many cells. In addition, \emph{disregarding the damage}, the assumed rectangular array of cells is doubly-periodic, and therefore definitely has at least two symmetry planes. This implies opposite tractions on opposite surfaces, or identical stresses there, and also identical strains, or displacements differing by constants:
\begin{equation} 
\label{M9}
\begin{split}
\left[ {T}_{2j} (h, X^{'}_3) \right]^{(M_2,q)} - \left[ {T}_{2j} (-h, X^{'}_3) \right]^{(-M_2,q)} = 0, \ q=-M_3,...,M_3
\end{split}
\end{equation}
\begin{equation} 
\label{M10}
\begin{split}
\ \ \ \left[ {T}_{3j} (X^{'}_2,l)  \right]^{(p, M_3)} - \left[ {T}_{3j} (X^{'}_2,-l) \right]^{(p, -M_3)} = 0, \ p=-M_2,...,M_2
\end{split}
\end{equation}

The displacements ($j=1,2,3$) at the opposite sides of the rectangular domain, in turn, satisfy 
\begin{equation} 
\label{M11}
\begin{split}
\left[ {u}_j (h, X^{'}_3) \right]^{(M_2,q)} - \left[ {u}_j (-h, X^{'}_3) \right]^{(-M_2,q)} = {\Delta}_{2j}, \ q=-M_3,...,M_3
\end{split}
\end{equation}
\begin{equation} 
\label{M12}
\begin{split}
\left[ {u}_j (X^{'}_2,l) \right]^{(p, M_3)} - \left[ {u}_j (X^{'}_2,-l) \right]^{(p, -M_3)} = {\Delta}_{3j}, \ p=-M_2,...,M_2
\end{split}
\end{equation}
where ${\Delta}_{2j}$ and ${\Delta}_{3j}$ denote the vector of the far-field displacement differences, and are given by
\begin{equation} 
\label{M13}
\begin{split}
 {\Delta}_{21} = 2 H {\bar E}_{21} , \  
{\Delta}_{22} = 2 H \left( \sqrt{1+2 {\bar E}_{22}}-1 \right)  , \    
{\Delta}_{23} = 2 H {\bar E}_{23}                                \\
{\Delta}_{31} = 2 L {\bar E}_{31}                               , \
{\Delta}_{32} = 2 L {\bar E}_{32}                                , \
{\Delta}_{33} = 2 L \left( \sqrt{1+2 {\bar E}_{33}}-1 \right)   
\end{split}
\end{equation}
(calculated at the \emph{material points} referred to in the boundary conditions in Eqs. (\ref{M11}) and  (\ref{M12})),
 $\bar {E}_{2j}$ and $\bar {E}_{3j}$ being the average (Lagrangian) strains of the unperturbed periodic composite, which can be calculated from the (current) average deformation gradient $\bar {\bm F}$ (and assigned back to their corresponding material points). The latter can be obtained from Eq. (\ref{M1}) by applying the finite strain HFGMC for a specified far-field loading. 

A clarification regarding Eqs. (\ref{M11})-(\ref{M13}) is in order. It is assumed that the $X_1'$ direction is `prismatic', there is no discretization and calculation along it, and thus in Eqs. (\ref{M11})-(\ref{M12}), one has only ${\Delta}_{2j}$ and ${\Delta}_{3j}$ (but not $\Delta_{1j}$). As for Eq. (\ref{M13}), the difference of the displacement in the direction $X_1'$ between two points with different $X_2'$ or $X_3'$ values, can be nonzero in the sense that there may be displacement in the $X_1'$ direction in the composite, but, much like the displacements in the other two directions, it is uniform along $X_1'$. Thus the term `plane deformation` only means that one direction ($X_1'$) is `prismatic' (all the elastic field are independent of $X_1'$).

\subsection{Analysis in the Fourier transform space}
\label{sec:3.1}
Thus far, the analysis performed in the real space was described. In the following, the double discrete Fourier transform is applied on the governing equations, constitutive relations, interfacial and boundary conditions.
For the displacement vector, for example, this transform is defined as follows:
\begin{equation} 
\label{TR}
\begin{split}
\hat {u}_j (X^{'}_2, X^{'}_3, \phi_p, \phi_q) = 
\sum^{M_2}_{K_2 = -M_2} \sum^{M_3}_{K_3 = -M_3} {u}^{(K_2, K_3)}_j (X^{'}_2, X^{'}_3) 
 e^{ i (K_2 \phi_p + K_3 \phi_q ) } , \  j=1,2,3
\end{split}
\end{equation}
with $\phi_p = \frac{2 \pi p}{2 M_2 +1},  p=0, \pm 1, \pm 2, ..., \pm M_2, 
\phi_q = \frac{2 \pi q}{2 M_3 +1}, q=0, \pm 1, \pm 2, ..., \pm M_3 $.

The application of this transform on the boundary-value problem in Eqs. (\ref{M2})-(\ref{M12})
for the rectangular domain $-H< X_2< H$, $-L <X_3 <L$, divided into $(2M_2+1) \times (2M_3+1)$ cells,
converts it to the problem for the single representative cell $-h < X'_2 < h$, $-l < X'_3 < l$ with respect to the complex-valued transforms.

For each couple of Fourier harmonics indices, $\lbrace p,q \rbrace$, a separate linear mechanical problem emerges, with a right-hand side known for the current iteration. Each of these separate problems can be solved independently. They can be solved consecutively, by using the same physical memory storage domain. Alternatively, if memory supplies are sufficient and it is computational time which is of the essence, then parallel computing can be employed, and all those Fourier space `single-cell` problems can be solved in parallel, on different processor cores or different computers altogether, assuming the correct parallelization code is written. Continuing with the description of the algorithm, the field equations obtained from the equilibrium and constitutive equations take the form
\begin{eqnarray} \label{T2}
\ \ \ \ \ \ \ \ \ \ \ \ \ \ \ \hat {T}_{kj,k} = {0}, \ \ \ \ j= 1,2,3 \ ; \ \ k=2,3
\end{eqnarray}
\begin{eqnarray} \label{T3}
\ \ \ \ \hat {T}_{jk} = \lambda {\hat u}_{m,m} \delta_{jk} + \mu ({\hat u}_{j,k} + {\hat u}_{k,j}) - {\hat T}^{e}_{jk} 
\end{eqnarray}
where 
\begin{equation} 
\label{T4}
\begin{split}
\ \ \ \hat {T}^{e}_{jk} =  \sum^{M_2}_{K_2=-M_2}  \sum^{M_3}_{K_3=-M_3} T^{e(K_2,K_3)}_{jk} e ^{ i(K_2 \phi_p + K_3 \phi_q) } 
\end{split}
\end{equation}
and the components of the eigenstresses $T^{e(K_2,K_3)}_{jk}$ are given by Eqs. (\ref{M4}), (\ref{M41}) and (\ref{M42}) for Mooney-Rivlin, Murnaghan and linearly elastic materials, respectively.

The continuity of tractions and displacements between adjacent cells, Eqs. (\ref{M5})-(\ref{M8}),
as well as the conditions which relate these variables at the opposite sides of the rectangle, Eqs. (\ref{M9})-(\ref{M12}), take the Bloch form:
\begin{equation} 
\label{T5}
\begin{split}
{\hat T}_{2j} (h, X^{'}_3) = e^{-i \phi_p}  \hat {T}_{2j}(-h, X^{'}_3), \ -l \le X^{'}_3 \le l,  \ j=1,2,3
\end{split}
\end{equation}
\begin{equation} 
\label{T6}
\begin{split}
\hat {u}_j (h, X^{'}_3) = e^{-i \phi_p}  \hat {u}_j (-h, X^{'}_3)  +  \delta_{0,K_3} (2 M_3+1) {\Delta}_{2j} e^{i \phi_p M_2}, -l \le X^{'}_3 \le l
\end{split}
\end{equation}
\begin{equation} 
\label{T7}
\begin{split}
\ \ \ \ \ \ \ \hat {T}_{3j} (X^{'}_2,l) = e^{-i \phi_q}  \hat {T}_{3j}(X^{'}_2,-l), \ -h \le X^{'}_2 \le h, \  j=1,2,3
\end{split}
\end{equation}
\begin{equation} 
\label{T8}
\begin{split}
\hat {u}_j (X^{'}_2,l) = e^{-i \phi_q}  \hat {u}_j(X^{'}_2,-l) + \delta_{0,K_2} (2 M_2+1) {\Delta}_{3j} e^{i \phi_q M_3},   -h \le X^{'}_2 \le h
\end{split}
\end{equation}
where $p = 0,...,\pm M_2$, $q = 0,..., \pm M_3$.

The boundary-value problem in Eqs. (\ref{T2})-(\ref{T8}) is solved using the high-order theory (HOT), \cite{AAB}, by dividing the cell domain $-h\le X^{'}_2 \le h$, $-l \le X^{'}_3 \le l$ 
into a rectangular array of $N_{\beta}\times N_{\gamma}$ subcells (see Fig. \ref{Figure1}(c)). In the framework of the high-order theory, the governing equations, interfacial and boundary conditions are imposed in the integral (averaged) sense. Specifically, traction continuity is imposed in the face-averaged sense, which implies forces continuity for the (Lagrangian) subcells, and the same goes for the displacements (which is equivalent to continuity of center-of-mass displacements of the staggered-mesh subcells -- up to small error related to nonuniform density which is negligible for moderately large strains). Regarding the equilibrium equations -- here integration is performed over the subcell volume, which is equivalent to requiring local balance of surface forces on each subcell. This volume integration cancels out the gradient correction of the stress distribution in the subcell and also the contribution of the eigenstress. The explanation for this is given in the following.

The HOT assumes a parabolic displacement field and a linear strain field in a subcell. As parabolic corrections to the strain are neglected, the subcells have to be small enough for the strain gradient in a subcell to be a small correction to the subcell-average of the strain.
For elastic calculations, the stress should be an algebraic (or transcendental) function of the strain at every point. Since the strain is assumed linear in a subcell, it is necessary that the stress be linear in the coordinates inside the subcell as well. For linear elasticity this is trivial. For nonlinear elasticity, the stress appears formally nonlinear in the coordinates. However, the subcells should be small enough for the stress to be effectively linear in the coordinates, and for parabolic and higher order corrections to be negligible. Regarding the method of solution -- the equilibrium equations and the continuity conditions for the subcells are solved in the Fourier space. The right-hand sides of those equations contain the Fourier transforms of the eigenstresses. Formally, both the equilibrium equations and the continuity conditions for the stresses are written in terms of the Fourier transforms of the total Cauchy (or first Piola-Kirchhoff) stress, which is comprised of the linear stress (proportional to the strain) and the eigenstress' Fourier transform.

The Fourier transform of the eigenstress at the undamaged cells contains only purely nonlinear contribution. For moderate strain (say, no more than 10 to 20 percents, which is a reasonable upper limit of elastic strain, beyond which inelasticity may emerge for material described by the Mooney-Rivlin model -- for material described reasonably by the Murnaghan equation, this estimate, unless purely volumetric strain is assumed, is more than probably a highly-overestimated non-tight upper bound), the purely nonlinear correction (first term is square) gives only a few percents change and thus it is a small correction. In the damaged cells, the eigenstress' Fourier transform is linear in the strain, but it relates to the linear part of the Fourier transform of the stress (the left-hand side of the equations) as the relative area of the damaged cells. For a single damaged cell and even 5 cells in each direction in total, the corresponding correction is no more than a few (say 4) percents. In total, the Fourier transform of the eigenstress is a small correction to the Fourier transform of the linear-elastic part of the stress.

Next, it should be acknowledged that just like the total stress, also the correction introduced by the eigenstress should change linearly with the coordinates inside a subcell (since it is a difference between a coordinate-linear total stress and a coordinate-linear linear-elastic part).

Thus for a proper choice of (small enough) subcells, the (Fourier transform of the) eigenstress would contain a constant part (subcell-average) and a constant-gradient part, linear in the coordinates inside a subcell. For small enough subcells, the coordinate-linear correction is small relative to the subcell average, both for the linear-elastic part of the stress, (for the total stress) and for the eigenstress. 

Therefore, one can identify three orders of magnitude comprising the Fourier transform of the total stress. The zeroth order is the subcell-average of the linear-elastic part (the one proportional to the Fourier transform of the strain -- or the symmetric part of the displacement gradient).

The first-order correction to this part consists of two terms, both proportional to one small parameter. The first term is the gradient correction to the linear-elastic stress, which becomes relatively small for small enough subcells. The second term is the subcell-averaged part of the (Fourier transform of the) eigenstress, which has two parts, one proportional to the (small) square of the effective total strain, and one proportional to the small relative area of the damaged region.
The second-order correction to the Fourier transform of the total stress is the gradient-part of the eigenstress, which is proportional to the product of two small terms, one vanishing with decreasing subcell size (relative, say to some geometric or physical macroscopic length scale), and the other being small for reasonable moderate strain (say 10 percents) and a reasonable number of cells (even 5 in each direction), with the assumption of damage localized on no more than a few cells.

To conclude, the gradient-part of the Fourier transform of the eigenstress is a higher-order (second) correction that can be neglected for enough cells, subcells and not too much damage and strain. Consequently, the \emph{eigenstresses} can be viewed as uniform in a subcell.
As for the subcell-averaged part of the eigenstress -- this contribution cancels out from the equilibrium equation, but remains relevant for the continuity equations. Therefore the equilibrium equations remain linear in the displacements, and can be used for static condensation, much as is the case in \cite{AV15}. On the other hand, the eigenstresses, uniform within each subcell, do make their contribution -- in the stress continuity equations.

\subsection{Inversion of the Fourier transform}
\label{sec:3.2}
Once the solution in the transform domain has been established, the real-space elastic field
can be readily determined at any point in the desired cell
$(K_2,K_3)$ of the considered rectangular region $-H \le X_2 \le H$, $-L \le X_3 \le L$
by the inverse transform formula, which for the displacements, for example, reads:
\begin{equation}
\label{T19}
\begin{split}
{u}^{(K_2, K_3)}_j (X^{'}_2, X^{'}_3) = \frac{1}{(2 M_2+1)(2 M_3+1)}   \sum^{M_2}_{p=-M_2} \sum^{M_3}_{q=-M_3} \hat {u}_j (X^{'}_2, X^{'}_3, \phi_p, \phi_q)
e^{ - i (K_2 \phi_p + K_3 \phi_q ) }
\end{split}
\end{equation}

In the application of the present analysis, the Fourier-transformed eigenstresses $\hat {T}^{e}_{jk}; \  j,k = 1,2,3;$ in Eq. (\ref{T3}) are not known.
Therefore, an iterative procedure similar to the one that has been applied in the linear case, \cite{AR12}, has to be employed as follows: 

1. Start the iterative procedure by assuming that $\hat {T}^{e}_{jk}$ = 0 and solve the boundary value problem with Eqs. (\ref{T2})-(\ref{T8}) in the transform domain. 

2. Apply the inverse transform formula to compute the displacements and stress fields.
The latter can be used to compute the (current) eigenstress components ${T}^{e(K_2,K_3)}_{jk}$ in the real space. 

3. Obtain an improved estimate, $\tilde{T}^{e(K_2,K_3)}_{jk}$, of the real-space eigenstress based on the value computed in stage 2, having in use a certain (convergence) criterion.

4. Compute the Fourier transform of this improved estimate, $\tilde{T}^{e(K_2,K_3)}_{jk}$, as obtained in stage 3, by employing Eq. (\ref{T4}).
 
5. Solve (again) the  boundary-value problem equations in the transform domain by employing this time the just computed values of $\hat{\tilde {T}}^{e}_{jk}; j,k = 1,2,3.$ 

6. Repeat the iterative process until satisfactory convergence in the real-space eigenstress components is obtained. 

For $\tilde{T}^{e(K_2,K_3)}_{jk}=T^{e(K_2,K_3)}_{jk}$, this procedure can be identified with the Banach contracting mapping method (e.g., \cite{Rall}).
In linear problems of composites with localized damage, \cite{AR12}, this procedure has converged for every magnitude of the applied external loading.

In the presently addressed hyperelastic problem however, convergence of this iterative procedure is obtained only for small magnitudes of the far-field, for which the materials are essentially linear.  
Therefore, the aforementioned procedure is utilized only at the first loading increment, and the corresponding convergent solution is used as an initial guess for an iterative procedure which has $\tilde{T}^{e(K_2,K_3)}_{jk}\neq T^{e(K_2,K_3)}_{jk}$, and employs a nonlinear solver providing $\tilde{T}^{e(K_2,K_3)}_{jk}( T^{e(K_2,K_3)}_{jk})$, as elaborated on in subsection 3.4.

\subsection{A brief description of the HOT}
\label{sec:3.3}
In the framework of the HOT, the single representative cell in the Fourier transform domain is divided into $N_{\beta}$ and $N_{\gamma}$ subcells in the $X^{'}_2$ and $X^{'}_3$ directions,
respectively, see Fig. \ref{Figure1}(c).
Each subcell is labeled by the indices $(\beta \gamma)$ with $\beta=1,...,N_{\beta}$ and $\gamma=1,...,N_{\gamma}$,
and may contain a distinct homogeneous material. The initial dimensions of subcell $(\beta \gamma)$ in the $X^{'}_2$ and $X^{'}_3$ directions
are denoted by $h_{\beta}$ and $l_{\gamma}$, respectively. A local initial coordinates system
$(\bar X^{(\beta)}_2, \bar X^{(\gamma)}_3)$ is introduced in each subcell with its origin located at the subcell center. The finite-strain higher-order theory is based on the following quadratic expansions of the displacement vector $\hat {\bm u}^{(\beta \gamma)}$ 
in subcell $(\beta \gamma)$:
\begin{equation} 
\label{H1}
\begin{split}
  \hat {\bm u}^{(\beta \gamma)} = 
 \hat {\bm W}^{(\beta \gamma)}_{(00)} +
\bar{X}^{(\beta )}_{2}  \hat {\bm W}^{(\beta \gamma)}_{(10)}+ 
\bar{X}^{(\gamma)}_{3}  \hat {\bm W}^{(\beta \gamma)}_{(01)}  + \\ 
 +  \frac{1}{2} \left(3 \bar{X}^{(\beta )2}_2 -\frac{h^2_{\beta }}{4} \right)  \hat {\bm W}^{(\beta \gamma)}_{(20)}
 +   \frac{1}{2} \left(3 \bar{X}^{(\gamma)2}_3 -\frac{l^2_{\gamma}}{4} \right)  \hat {\bm W}^{(\beta \gamma)}_{(02)}
\end{split}
\end{equation}

The unknown coefficient $ \hat {\bm W}^{(\beta \gamma)}_{(mn)}$ are determined, as shown in the following,
from the satisfaction of the equilibrium equations, interfacial and boundary conditions.

In the absence of body forces, the equilibrium equations in the subcell,
expressed in terms of the first Piola-Kirchhoff stress tensor, $\hat {\bm T}^{(\beta \gamma)}$, can be represented in the form
\begin{eqnarray} \label{H2}
  \ \ \ \ \ \ \ \ \ \ \ \frac{\partial \hat T^{(\beta \gamma)}_{2j}}{\partial {\bar X}^{(\beta)}_2}
+ \frac{\partial \hat T^{(\beta \gamma)}_{3j}}{\partial {\bar X}^{(\gamma)}_3} = 0, \ \ \ \ \ j=1,2,3
\end{eqnarray}
 
The surface-averages of the tractions are given by
\begin{equation} 
\label{H3}
\begin{split}
 \hat {\bm T}^{\pm(\beta \gamma)}_2 = \frac{1}{l_{\gamma}} \int_{ -l_{\gamma} / 2}^{ l_{\gamma} / 2}
 \hat {\bm T}^{(\beta \gamma)}_2 \left({\bar X}^{(\beta)}_2 = \pm \frac{ h_{\beta}} {2} \right) \ d {\bar X}^{(\gamma)}_3   \\
 \hat {\bm T}^{\pm(\beta \gamma)}_3 = \frac{1}{h_{\beta}} \int_{ -h_{\beta} / 2}^{ h_{\beta} / 2}
 \hat {\bm T}^{(\beta \gamma)}_3 \left({\bar X}^{(\gamma)}_3 = \pm \frac{ l_{\gamma}} {2} \right) d {\bar X}^{(\beta)}_2
\end{split}
\end{equation}
and $ \hat {\bm T}^{(\beta \gamma)}_2$, $ \hat {\bm T}^{(\beta \gamma)}_3$ are column vectors defined by
\begin{equation} 
\label{H4}
\begin{split}
 \hat {\bm T}^{(\beta \gamma)}_2 = \left[ \hat T_{21},  \hat T_{22},  \hat T_{23} \right]^{(\beta \gamma)},
 \ \hat {\bm T}^{(\beta \gamma)}_3 = \left[ \hat T_{31},  \hat T_{32},  \hat T_{33} \right]^{(\beta \gamma)}
\end{split}
\end{equation}

In terms of the surface-averages of the tractions, the equilibrium equation (\ref{H2}) reads
\begin{equation} 
\label{H5}
  \left[  \hat {\bm T}^{+(\beta \gamma)}_2 -  \hat {\bm T}^{-(\beta \gamma)}_2 \right]
+\frac{h_{\beta}}{l_{\gamma}} \left[  \hat {\bm T}^{+(\beta \gamma)}_3 -  \hat {\bm T}^{-(\beta \gamma)}_3 \right] = 0
\end{equation}

This relation expresses the equilibrium equations imposed in the average sense within subcell $(\beta \gamma)$.

The constitutive relations in Eq. (\ref{T3}) can be represented in the form
\begin{eqnarray} \label{CON}
\ \ \ \ \ \ \ \ \ \ \hat T^{(\beta \gamma)}_{jk} = C^{(\beta \gamma)}_{jklm} \hat u^{(\beta \gamma)}_{l,m} - \hat T^{e(\beta \gamma)}_{jk} 
\end{eqnarray}
which provides the following expressions for the components of the
surface-averages of the tractions $ \hat {\bm T}^{\pm(\beta \gamma)}_{2}$ and $ \hat {\bm T}^{\pm(\beta \gamma)}_{3}$ (recalling the argumentation given above for the assumption that eigenstresses are uniform across the subcell to sufficient approximation):
\begin{equation} 
\label{H6}
\begin{split}
\hat {T}^{\pm(\beta \gamma)}_{2j} =  C^{(\beta \gamma)}_{2j12}
\left(  \hat W^{(\beta \gamma)}_{1(10)} \pm \frac{3 h_{\beta}}{2}  \hat W^{(\beta \gamma)}_{1(20)} \right)
+\\+C^{(\beta \gamma)}_{2j22}
\left(  \hat W^{(\beta \gamma)}_{2(10)} \pm \frac{3 h_{\beta}}{2}  \hat W^{(\beta \gamma)}_{2(20)} \right) 
+C^{(\beta \gamma)}_{2j32}
\left(  \hat W^{(\beta \gamma)}_{3(10)} \pm \frac{3 h_{\beta}}{2}  \hat W^{(\beta \gamma)}_{3(20)} \right) 
+\\+C^{(\beta \gamma)}_{2j13}  \hat W^{(\beta \gamma)}_{1(01)}
 + C^{(\beta \gamma)}_{2j23}  \hat W^{(\beta \gamma)}_{2(01)}
 + C^{(\beta \gamma)}_{2j33}  \hat W^{(\beta \gamma)}_{3(01)} -  \hat T^{e}_{2j}, \  j,k,l=1,2,3
\end{split}
\end{equation}
\begin{equation} 
\label{H7}
\begin{split}
\hat {T}^{\pm(\beta \gamma)}_{3j} =  C^{(\beta \gamma)}_{3j12}  \hat W^{(\beta \gamma)}_{1(10)} + C^{(\beta \gamma)}_{3j22}  \hat W^{(\beta \gamma)}_{2(10)}
 + C^{(\beta \gamma)}_{3j32}  \hat W^{(\beta \gamma)}_{3(10)}  +\\ 
+C^{(\beta \gamma)}_{3j13}
\left(  \hat W^{(\beta \gamma)}_{1(01)} \pm \frac{3 l_{\gamma}}{2}  \hat W^{(\beta \gamma)}_{1(02)} \right)+
C^{(\beta \gamma)}_{3j23}
\left(  \hat W^{(\beta \gamma)}_{2(01)} \pm \frac{3 l_{\gamma}}{2}  \hat W^{(\beta \gamma)}_{2(02)} \right) +\\
+C^{(\beta \gamma)}_{3j33}
\left(  \hat W^{(\beta \gamma)}_{3(01)} \pm \frac{3 l_{\gamma}}{2}  \hat W^{(\beta \gamma)}_{3(02)} \right) -  \hat T^{e}_{3j}, \ \ j,k,l=1,2,3
\end{split}
\end{equation}

Substitution of Eq. (\ref{H6})-(\ref{H7}) in (\ref{H5}) provides the three relations:
\begin{equation} 
\label{H8}
\begin{split}
C^{(\beta \gamma)}_{2j12}  \hat W^{(\beta \gamma)}_{1(20)}
  +C^{(\beta \gamma)}_{2j22}  \hat W^{(\beta \gamma)}_{2(20)}
  +C^{(\beta \gamma)}_{2j32}  \hat W^{(\beta \gamma)}_{3(20)}  \\
+C^{(\beta \gamma)}_{3j13}  \hat W^{(\beta \gamma)}_{1(02)}
 + C^{(\beta \gamma)}_{3j23}  \hat W^{(\beta \gamma)}_{2(02)}
 + C^{(\beta \gamma)}_{3j33}  \hat W^{(\beta \gamma)}_{3(02)} = 0, \ j=1,2,3
\end{split}
\end{equation}

Just like the surface-averaged tractions, the surface-averaged displacements can be defined by
\begin{equation} 
\label{H9}
\begin{split}
 \hat {\bm u}^{\pm(\beta \gamma)}_2 = \frac{1}{l_{\gamma}} \int_{ -l_{\gamma} / 2}^{ l_{\gamma} / 2}
       \hat {\bm u}^{(\beta \gamma)} \left( {\bar X}^{(\beta)}_2 = \pm \frac{ h_{\beta}} {2} \right) \ d {\bar X}^{(\gamma)}_3  \\
 \hat {\bm u}^{\pm(\beta \gamma)}_3 = \frac {1}{h_{\beta}} \int_{ -h_{\beta} / 2}^{ h_{\beta} / 2}
       \hat {\bm u}^{(\beta \gamma)} \left( {\bar X}^{(\gamma)}_3 = \pm \frac{ l_{\gamma}} {2} \right) \ d {\bar X}^{(\beta)}_2
\end{split}
\end{equation}

These surface-averages $ \hat {\bm u}^{\pm(\beta \gamma)}_i$, $i=1,2,3$, can be related to the microvariables
$ \hat {\bm W}^{(\beta \gamma)}_{(mn)}$; $(mn)=0,1,2$; in the expansion in Eq. (\ref{H1}):
\begin{equation} 
\label{H10}
\begin{split}
\ \  \hat {\bm u}^{\pm(\beta \gamma)}_2 =  \hat {\bm W}^{(\beta \gamma)}_{(00)}
                     \pm \frac{h_{\beta}}{2}  \hat {\bm W}^{(\beta \gamma)}_{(10)}
               +       \frac{h^2_{\beta}}{4}  \hat {\bm W}^{(\beta \gamma)}_{(20)} ,  \\ 
 \hat {\bm u}^{\pm(\beta \gamma)}_3 =  \hat {\bm W}^{(\beta \gamma)}_{(00)}
                     \pm \frac{l_{\gamma}}{2}  \hat {\bm W}^{(\beta \gamma)}_{(01)}
               +       \frac{l^2_{\gamma}}{4}  \hat {\bm W}^{(\beta \gamma)}_{(02)}
\end{split}
\end{equation}

Manipulations of Eq. (\ref{H10}) by subtractions and additions yield
\begin{equation} 
\label{H11}
\begin{split}
 \hat {\bm W}^{(\beta \gamma)}_{(10)} = \frac{1}{h_{\beta}}
               \left[  \hat {\bm u}^{+}_2 -  \hat {\bm u}^{-}_2 \right]^{(\beta \gamma)},   \ \ \ 
\hat {\bm W}^{(\beta \gamma)}_{(01)} = \frac{1}{l_{\gamma}}
               \left[  \hat {\bm u}^{+}_3 -  \hat {\bm u}^{-}_3 \right]^{(\beta \gamma)}
\end{split}
\end{equation}
\begin{equation} 
\label{H12}
\begin{split}
\ \ \ \ \ \ \  \hat {\bm W}^{(\beta \gamma)}_{(20)} = \frac{2}{h^2_{\beta}}
               \left[  \hat {\bm u}^{+}_2 +  \hat {\bm u}^{-}_2 \right]^{(\beta \gamma)}
             - \frac{4}{h^2_{\beta}}  \hat {\bm W}^{(\beta \gamma)}_{(00)}              ,  \ 
\ \ \ \ \ \ \hat {\bm W}^{(\beta \gamma)}_{(02)} = \frac{2}{l^2_{\gamma}}
               \left[  \hat {\bm u}^{+}_3 +  \hat {\bm u}^{-}_3 \right]^{(\beta \gamma)}
             - \frac{4}{l^2_{\gamma}}  \hat {\bm W}^{(\beta \gamma)}_{(00)}
\end{split}
\end{equation}

The expressions for $ \hat {\bm W}^{(\beta \gamma)}_{(00)}$ in terms of the surface-averaged displacements can be determined from the equilibrium equation, Eq. (\ref{H5}).

The final form which expresses the equilibrium and constitutive equations is given by 
\begin{eqnarray} \label{H13}
\ \ \left\{ \begin{array}{c}
 \hat {\bm T}^{\pm}_2 \\
 \hat {\bm T}^{\pm}_3 \end{array} \right\}^{(\beta \gamma)}  =
\left[ \begin{array}{c}
{\bm K}  \\
\end{array} \right]^{(\beta \gamma)} \
\left\{ \begin{array}{c}
 \hat {\bm u}^{\pm}_2 \\
 \hat {\bm u}^{\pm}_3       \end{array} \right\}^{(\beta \gamma)}
-\left\{ \begin{array}{c}
 \hat {\bm T}^{e}_2 \\
 \hat {\bm T}^{e}_3       \end{array} \right\}^{(\beta \gamma)}
\end{eqnarray}
where $[\bm K]^{(\beta \gamma)}$ is a matrix of the 12th-order whose elements depend on the dimensions of the subcell and the (zero-stress-limit tangent) elastic isotropic stiffness tensor components $C^{(\beta \gamma)}_{jklm}$
of the material filling this subcell. These relations are employed for the enforcing of the continuity of tractions between subcells as well as the tractions' boundary conditions in Eqs. (\ref{T5}) and (\ref{T7}). 

Along with with the interfacial continuity conditions of the displacements between the subcells and the boundary conditions in Eqs. (\ref{T6}), (\ref{T8}),
a system of $12 N_{\beta} N_{\gamma}$ algebraic equations in the surface-averaged displacements $\hat {\bm u}^{\pm (\beta \gamma)}_2$ and $\hat {\bm u}^{\pm (\beta \gamma)}_3$ is obtained. These equations are linear for every `guess' (iteration) of the Fourier-transformed eigenstresses. These are the equations to be solved in stage 1 as described in the previous subsection. The following subsection presents and describes the approach to solving the overall iterative problem.

\subsection*{3.4 Convergent iterative solution of the nonlinear system of equations}

Within the aforementioned approach, there exists a system of equations in the real space that needs to be solved for each loading increment for the mechanics to be correct. The advantage of the application of the Fourier transform method is in that the linear system to be solved is a small one, i.e. it does not require a lot of memory and computational time. The drawback is that there is an unknown `right-hand' side (for each wavenumber). Since the unknown `right-hand` side is the Fourier transform of the eigenstress, the unknowns for the underlying nonlinear system can only be the eigenstresses (or their Fourier transforms, which is just adding a linear operator to a nonlinear one, so it is generally the real-space eigenstresses). The nonlinear equations that have to be satisfied are the eigenstress update equations, Eqs. (\ref{M4}), (\ref{M41}) or (\ref{M42}), depending on the material, in which $T_{ij}^{MR(K_2,K_3)}$ or $T_{ij}^{NL(K_2,K_3)}$ for a given iteration are perceived as functions of the eigenstress $T^{e(K_2,K_3)}_{ij}$ at the previous iteration, obtained after forward Fourier transforms, solution of a set of linear problems in the Fourier space, an inverse Fourier transform, computation of the deformation gradient and then application of the nonlinear hyperelastic functionals. This underlying nonlinear system can be written as follows:
\begin{equation} 
\label{B1}
  ^{(k)}{T}^{e(K_2,K_3)}_{ij} =  g\left(^{(k-1)}{T}^{e(K_2,K_3)}_{ij}\right)
\end{equation}
(here the function $g(x)$ is a nonlinear function for any nonlinear material, such as the Mooney-Rivlin or Murnaghan material).

The mapping in Eq. (\ref{B1}) has a fixed-point corresponding to the solution of the mechanical problem, which for stable hyperelastic materials exists and is unique. This fixed-point may be unstable, in the sense that the mapping may not be contracting in the vicinity of the fixed-point. However, the fixed-point exists and it is a root of the equation
\begin{equation} 
\label{B2}
 f\left(T^{e(K_2,K_3)}_{ij}\right)\triangleq g\left(T^{e(K_2,K_3)}_{ij}\right) - T^{e(K_2,K_3)}_{ij} =0
\end{equation}

Thus, instead of trying to find a fixed-point of the mapping in Eq. (\ref{B1}), one may look for a root of Eq. (\ref{B2}). The advantage is that the aforementioned mapping may be unstable near the fixed-point, whereas for the root-finding problem, general (at least locally) convergent algorithms are known. 

At this point it is important to note that the root-finding algorithm is to be called at the level of the two-scales discretization, that is after the entire domain is divided into cells and each cell is divided into subcells, such that the vector of unknowns is $x_m^{(n)}\triangleq{^{(l,k)}_{(\beta,\gamma)}T_{ij}^{e(K_2,K_3)}}$, it is sought independently for every loading iteration $l$, it changes during the root-finding process counted by the iteration $k$, and it has $d=d_T N_{\beta}N_{\gamma} \times(2M_2+1)(2M_3+1)$ components ($d_T$ components for every subcell in every cell in the domain, where $d_T \le 9$ and the number of independent stress components is determined by the geometry). 
Since the loading is assumed to be applied incrementally and one can choose arbitrarily small increments, the first increment can always be taken from the linear part of the hyperelastic stress strain relation. Then for the first increment one can use a rigorously linear material with moduli equal to the initial tangent moduli of the hyperelastic material. This initial problem can be solved using the mapping in Eq. (\ref{B1}) and convergence is then guaranteed. One can then use the obtained eigenstress vector as an initial guess for the next loading increment for which the full hyperelastic material is recovered and a locally-convergent root-finding algorithm is employed. In other words, there is a way to guarantee sufficiently good initial guesses for locally-convergent root-finding algorithms. For the second loading increment it is the convergence of the Banach method which would supply a good enough starting point. For the following increments, a good-enough starting point for any consecutive increment would be supplied by good-enough convergence of the solver for the previous increment. Thus the only requirement for overall convergence is the use of a locally-convergent algorithm. 

Furthermore, since we do have good initial guesses here and on the other hand there are two scales in the problem, which may render the vector of unknowns extremely large, one would seek the most efficient locally convergent root-finding algorithm with weak dimension-dependent complexity. Although in the examples considered in this work, the number of equations that need to be solved is of the order of several hundreds of thousands, which is not particularly challenging for a modern work-station computer, clearly, increased resolution for adequate description of fibers with non-trivial cross-section, or three-dimensional problems, would require already tens to hundreds of millions of equations to solve, which is still computationally hard.

Due to the high dimensionality of the problem and the fact that the error in the solution of the Fourier-space problems in intermediate iterations may introduce incorrect frequencies which may result in oscillatory intermediate solution in the real space, using the approach of numerical minimization seems inadvisable here, as many local minima may be encountered. Instead, a direct multidimensional root-finding algorithm would be better-suited. The direct multidimensional locally-convergent root-finding algorithm with the highest convergence rate is Newton's method. 

Exact application of Newton's method requires the Jacobian of the vector of equations in each iteration, or its inverse. The Jacobian is usually a full-rank matrix (for a regular multidimensional space). Such a matrix would have $d^2$ entries, which may be extremely demanding in terms of memory for high-resolution calculations for the discussed multiple-scale approach. 

This can be described mathematically as follows. The standard Newton update step for finding the nearest root of the equation $\textbf{f}(\textbf{x})=\textbf{0}$, starting from $\textbf{x}=\textbf{x}_0$ is:
\begin{eqnarray} \label{B3}
\ \ \ \ \ \ \ \ \ \textbf{x}_{k+1}={\mathcal{G}}\left(\textbf{x}_{k}\right)=\textbf{x}_k-\textbf{B}_k\textbf{f}_k
\end{eqnarray}
where $\textbf{B}_k=(\nabla\textbf{f}^{\top}_k)^{-\top}$ corresponds to the exact Newton step case.
The inverse of the Jacobian can be spectrally decomposed, without loss of generality, according to the following formula:
\begin{eqnarray} \label{B4}
\ \ \ \ \ \ \ \ \ \ \textbf{B}_k=\sum_{n=1}^d \lambda_n\bar{\textbf{v}}_n\hat{\textbf{v}}_n^{\top}=\sum_{n=1}^d \textbf{u}_n\textbf{v}_n^{\top}
\end{eqnarray}
where $\bar{\textbf{v}}_n$ and $\hat{\textbf{v}}_n^{\top}$ are the right and left eigenvectors of the inverse Jacobian, respectively, and $\lambda_n$ are the eigenvalues. The second decomposition has a different normalization and is more convenient for the derivations employed in the following. The idea here is that if 
\begin{eqnarray} \label{B5}
\ \ \ \ \ \ \ \ \ \ \ \ \ \ \ \ \ \ \left \Vert\lbrace\nabla[{g(\textbf{x})]^{\top}\rbrace^{\top}}\right\Vert_s^{\textbf{x}=\textbf{x}^*}\ge1
\end{eqnarray}
where $\Vert\cdot\Vert_s$ indicates the spectral norm, then another fixed-point iteration is constructed, for which
\begin{eqnarray} \label{B6}
\ \ \ \ \ \ \ \ \ \ \ \ \ \ \ \ \ \ \left \Vert\lbrace\nabla[{\mathcal{G}(\textbf{x})]^{\top}\rbrace^{\top}}\right\Vert_s^{\textbf{x}=\textbf{x}^*}<1
\end{eqnarray}

Now, the exact Newton step has two problems. First, especially in the case of oscillatory functions -- as ones obtained by an inverse Fourier transform of an approximate expression -- the Jacobian may vanish at local extrema, which renders the step divergent. This difficulty is often circumvented by the introduction of the so-called Quasi-Newton step, where the Jacobian is substituted with a finite difference calculated by use of previous solution steps, which are always finite. This approach is employed also in the present work. The second problem with the Newton step, is that it requires storing $d$ $d$-dimensional vectors in memory, one for each of the orthogonal directions of the $d$-dimensional state space. These vectors are the ones shown in the sum in Eq. (\ref{B4}). Moreover, the calculation of $d$ finite differences in each direction for each of the nonlinear equations in the system, without knowing \emph{a priori} the sufficient step for stable convergent calculation of such finite differences, may be extremely time-consuming. Therefore, the approach of Quasi-Newton methods is to replace the $d$ orthogonal finite difference approximations of the derivatives of the components of $\textbf{f}$ around the current state in each iteration by a smaller number of vectors related to the solution history. These vectors do not form an orthogonal basis and they are not exactly local. Nevertheless, they may be useful in approximating the Jacobian (or its inverse) in some sense. This is the idea behind the Quasi-Newton method. Derivation of the Quasi-Newton method equations, along with the relevant argumentation, in a form suitable for the following derivation of the final algorithm employed here, is given in \ref{AppendixA}

In the following, we present an algorithm of the Quasi-Newton type, which is essentially a version of the so-called good Broyden method, albeit implemented in an low-memory form directly storing function values, as derived within the present work. Argumentation for the optimality of the good Broyden method, suitable for the following derivation of the final form of the algorithm employed here, is given in \ref{AppendixB} The details of the derivation of the new low-memory form of the good Broyden method (the DFVS-LM-GBM algorithm) is provided in \ref{AppendixC}
 
\subsubsection*{3.4.1 The Directly Function-Value Storing Low-Memory Good Broyden Method}

The update scheme of the algorithm is standard, 
\begin{eqnarray} \label{B11a}
\ \ \ \ \ \ \textbf{x}_{k+1}=\textbf{x}_k+\textbf{s}_{k+1}
\end{eqnarray}

The step (where $k$ denotes the `current' iteration number of the Quasi-Newton solver) is given by:
\begin{eqnarray} \label{B33}
\ \ \ \ \ \textbf{s}_{k+1}=\frac{\beta_k}{\beta_k-\alpha_k}\textbf{c}_k 
\end{eqnarray}
where $\textbf{c}_n$ has the following explicit component-wise-defined expression ($N=d$):
\begin{eqnarray} \label{B49}
c_i^{(k)} =  \sum_{m=1}^{k-1}H_{k-1,m}F^{(k)}_{i,m+1}, \forall \ i\le N,   k > 2,   i,k  \in \mathbb{N}
\end{eqnarray}
and the matrix $\textbf{F}^{(k)}$ is given by:
\begin{eqnarray} \label{B31}
\ \ \ \ \ \ \ \ \ \ \textbf{F}^{(k)}=[\textbf{f}_1,\textbf{f}_2,..,\textbf{f}_{k-1},\textbf{f}_k]
\end{eqnarray}
($\textbf{f}$ being the vector function the root of which is sought -- in our case the vector of residues of the real-space eigenstress decomposition-update equations,  Eqs. (\ref{M4}), (\ref{M41}) or (\ref{M42}), with the definition in Eq. (\ref{B2})).

The following initial conditions should be used:
\begin{eqnarray} \label{B30a}
\ \ \ \ \ \ \ \ \ \ \ \ \ \ \ \textbf{s}_{2}=\textbf{f}_1, \ \textbf{c}_2=\textbf{f}_2
\end{eqnarray}

Regarding the calculation of the matrix $\textbf{H}$, the following definition is introduced:
\begin{equation} \label{B46}
\phi_{nm} \triangleq 
\begin{cases} \frac{\gamma_{n+2}^{(m+1)}}{\beta_{m+1}-\alpha_{m+1}}, \  \textrm{if}  \ m < n < k \\ \ \ \ \ \ \ \ 0 \ \ \ \ \ \ , \ \textrm{otherwise}
\end{cases}  ; m,n  \in \mathbb{N}
\end{equation}

Then, the following (lower triangular) matrix is defined in a component-wise fashion:
\begin{eqnarray} \label{B47}
\ \ \ \ \ \ \ M_{nm} \triangleq \delta_{nm}-\phi_{nm}, \forall \ n,m<k \  ; \ n,m \  \in \mathbb{N} 
\end{eqnarray}
where $\delta_{nm}$ is Kronecker's delta and where the $k\times k$ matrix $\textbf{M}$ has to be stored in memory for each iteration, at the same physical address for every consecutive number of the solver step iteration $k$, each time replacing the previous registry, occupying no more than the size of a $p^2$ array (where $p=k_{\text{max}}$). 

Next, exact matrix inversion is performed for the defined $p\times p$ matrix, requiring $\mathcal{O}(p^3)$ complexity (still polynomial), yielding $\textbf{H}$:
\begin{eqnarray} \label{B48}
\ \ \ \ \ \ \ \textbf{H}\triangleq\textbf{M}^{-1}
\end{eqnarray}

Finally, the three scalar quantities in the above formulas should be iterated according to the following equations:
\begin{equation} 
\label{B38}
\begin{split}
\ \ \ \ \ \alpha_{n} = \frac{\beta_{n-1}}{\beta_{n-1}-\alpha_{n-1}}\left[\vphantom{\frac{\lambda_{n-1} }{\beta_{n-1}-\alpha_{n-1}}}\mu_n^{(n-1)}+\frac{\lambda_{n-1} }{\beta_{n-1}-\alpha_{n-1}}\gamma_n^{(n-1)}\right], \ n\ge3
\end{split}
\end{equation}
\begin{eqnarray} \label{B39}
\ \ \ \ \ \ \ \ \ \ \ \ \ \ \ \beta_{n} = \frac{\beta_{n-1}^2\lambda_{n-1}}{(\beta_{n-1}-\alpha_{n-1})^2}, \ n\ge3
\end{eqnarray}
\begin{equation} 
\label{B40}
\begin{split}
 \gamma_{n}^{(m)} = \frac{\beta_{m-1}}{\beta_{m-1}-\alpha_{m-1}}\left[\vphantom{\frac{\lambda_{n-1} }{\beta_{n-1}-\alpha_{n-1}}}\mu_n^{(m-1)} +\frac{\lambda_{m-1} }{\beta_{m-1}-\alpha_{m-1}}\gamma_n^{(m-1)}\right], \ n,m\ge3
\end{split}
\end{equation}
\begin{equation} 
\label{B41}
\begin{split}
\ \ \ \ \ \ \ \ \ \lambda_n = \nu_{nn}^{(n-1)}+2\frac{\gamma_n^{(n-1)}\mu_n^{(n-1)}}{\beta_{n-1}-\alpha_{n-1}}+\frac{\left[\gamma_n^{(n-1)}\right]^2\lambda_{n-1}}{(\beta_{n-1}-\alpha_{n-1})^2}, \ n\ge3
\end{split}
\end{equation}
\begin{equation} 
\label{B42}
\begin{split}
\mu_{n}^{(m)} = \nu_{nm}^{(m-1)}+\frac{\gamma_n^{(m-1)}\mu_{m}^{(m-1)}}{\beta_{m-1}-\alpha_{m-1}}+\frac{\gamma_{m}^{(m-1)}\mu_{n}^{(m-1)}}{\beta_{m-1}-\alpha_{m-1}}+\frac{\gamma_n^{(m-1)}\gamma_{m}^{(m-1)}\lambda_{m-1}}{(\beta_{m-1}-\alpha_{m-1})^2}, \ m,n\ge3
\end{split}
\end{equation}
\begin{equation} 
\label{B43}
\begin{split}
\nu_{nl}^{(m)} = \nu_{nl}^{(m-1)}+\frac{\gamma_n^{(m-1)}\mu_l^{(m-1)}}{\beta_{m-1}-\alpha_{m-1}}+\frac{\gamma_l^{(m-1)}\mu_n^{(m-1)}}{\beta_{m-1}-\alpha_{m-1}}+\frac{\gamma_n^{(m-1)}\gamma_l^{(m-1)}\lambda_{m-1}}{(\beta_{m-1}-\alpha_{m-1})^2}, \ l,m,n\ge3
\end{split}
\end{equation}

These six equations for the scalar variables required for the implementation of the method need six initial conditions, which are obtained from the six permutations of scalar products involving the three initial values of the vector function the root of which is sought, as follows:
\begin{equation} 
\label{B44}
\begin{split}
\alpha_2=\textbf{f}_1^{\top}\textbf{f}_2, \ \beta_2=\textbf{f}_1^{\top}\textbf{f}_1, \ \lambda_2=\textbf{f}_2^{\top}\textbf{f}_2,  \gamma_n^{(2)}=\textbf{f}_1^{\top}\textbf{f}_n, \ \mu_n^{(2)}=\textbf{f}_2^{\top}\textbf{f}_n, \ \nu_{nl}^{(2)}=\textbf{f}_l^{\top}\textbf{f}_n, \ n,l\ge 3
\end{split}
\end{equation}

\subsubsection*{3.4.2 Summary of the description of the nonlinear solver}

To conclude the description of the solver proposed in this work, we note that the Directly Function-Value Storing Low-Memory Good Broyden Method (DFVS-LM-GBM) is implemented by taking the state vector update as prescribed by Eq. (\ref{B11a}), using the step as given by Eq. (\ref{B33}), with the auxiliary direction vector as given in a component-wise fashion in Eq. (\ref{B49}), employing Eq. (\ref{B46})-(\ref{B48}), along with the definition in Eq. (\ref{B31}). In addition, in every iteration, the system of six scalar equations, namely, Eqs. (\ref{B38})-(\ref{B43}), has to be sub-iterated in an internal loop from 2 to $k$ for every successive value of $k$, starting from initial conditions (updated for every $k$), as given by Eqs. (\ref{B44}). Of course,  Eqs. (\ref{B30a}) should be used for the initial step and auxiliary direction.  

This concludes the description of the nonlinear solver. An extra low memory version was proposed here, which uses a total size of $\mathcal{O}(p N)+\mathcal{O}(N)+\mathcal{O}(p^2)$  of double precision entries (one notes that the two $(k-1)\times (k-1)$ matrices $\textbf{M},\textbf{H}$, as well as the matrix which $n,l$-components are denoted by $\nu_{nl}^{(m)}$, are not stored in separate addresses for every state-vector component $i$, but rather only for one, say the first component, and are rewritten to that same address for every consecutive step-iteration $k$). This is asymptotically half of the memory used by the standard low-memory method, for $1\ll p\ll N$. 

On the issue of computational complexity, it can be said that this matter is less problematic for quasi-static problems with large spatial resolution, a memory limit being a more stringent requirement, however for the least, polynomial complexity should be guaranteed with respect to all dimensionless problem parameters much larger than unity, or the problem becomes computationally NP-hard. The algorithm presented in this subsection indeed guarantees polynomial complexity, the toughest problem to solve in a given iteration being the inversion of a lower triangular matrix of about a hundred (or $p$) columns , which is not particularly hard and obviously has polynomial complexity. Beyond that, the solver uses a triple loop of complexity $p^2N$ with standard arithmetic operations. Last, calculation of initial conditions requires $p^2N$ operations for every solver-step iteration,which increases the complexity to the order of $p^3N$. The matrix inversion complexity is $p^3$ for every iteration, which yields a $p^4$ complexity, approximately, and the overall estimate is thus complexity of $\mathcal{O}(p^3N)+\mathcal{O}(p^4)$, which is, of course, polynomial in $p$ (and $N$-linear). 

\begin{figure}[tp]

\includegraphics[width=0.95\textwidth,trim={1cm 1cm 1cm 1cm},clip]{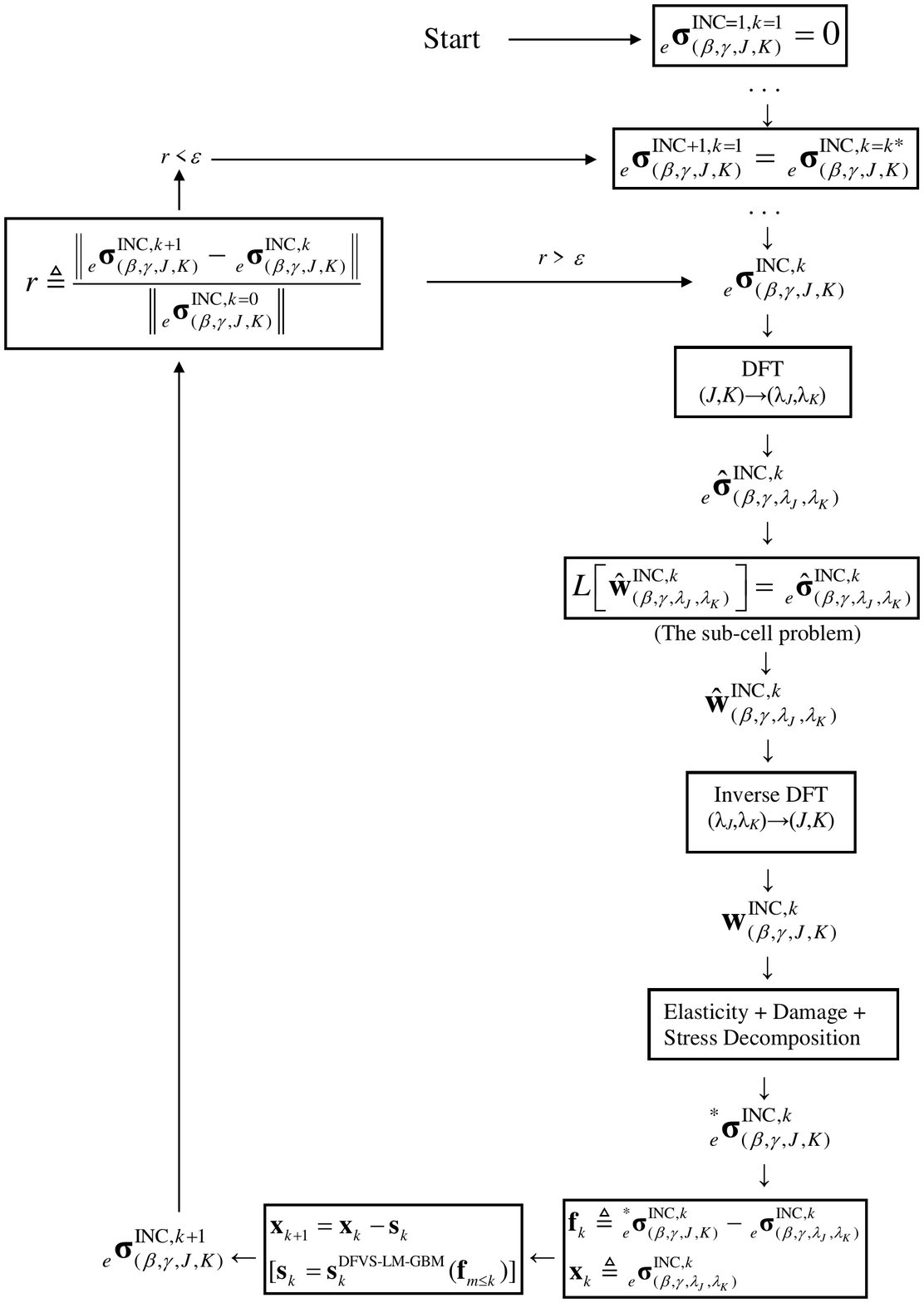}

\caption{Block diagram of the algorithm. Here, for clarity, the cells indices are denoted by $J,K$ instead of $K_2,K_3$, and the loading increment $l$ is denoted by the superscript `INC', unlike in the body of the text of the relevant subsections.}

\label{Figure2}       
\end{figure}

A block diagram of the overall solution algorithm, addressing the multiple-scale mechanical treatment with the mathematical solver as an integral part is depicted in Fig. \ref{Figure2}.

\section{Applications}
\label{sec:4}
The approach described in the previous section was applied to the analysis of stress distribution in typical examples of composites with hyperelastic constituents distributed doubly-periodically, with various examples of localized damage. In all cases a quasi-statically applied `far-field' stress loading of 20kPa--200kPa in the `vertical' direction $X_2$ was assumed (in fact, a far-field displacement boundary condition was applied for which the desired far-field stress as aforementioned was obtained). In one of the directions the composites were assumed to be prismatic (so-called plane-deformation assumption). The number of (structurally-identical) cells taken was chosen as the minimum number required for convergence in terms of spatial attenuation of the effect of local damage on the profiles dictated by the far-field loading. This number was set to five in each of the two directions for the majority of the calculations. Convergence with the number of cells, assuming a square array, was checked for the case of homogeneous Mooney-Rivlin material with a (roughly) octagonal cavity. The result for three cells (in each direction) turned out to be noticeably different, whereas the result for seven cells turned out to be almost identical to the five by five case. The number of subcells for solution by the HOT was set to eleven in each direction, for all cases except for the example of a single ``crack'' (quotation marks here and onward denote the fact that instead of an exact crack the computation considers a line of subcells in which the stress -- but not the density --  is explicitly set to zero by taking $D=1$) in either homogeneous Mooney-Rivlin material, or a composite of two constituents described by the Murnaghan model. For the two latter cases, the number of subcells was eleven in the vertical direction and ten in the horizontal direction (ten columns). For the case of the single ``crack'' in homogeneous Mooney-Rivlin material, convergence with the number of subcells was checked by taking 20 by 21, instead of 10 by 11 subcells. It was found that reasonably correct results are obtained already for 10 by 11 subcells, with reasonable convergence in the spatial profiles (and hence also the stress-strain profiles). This way, when viewed as a structure, the computational domain (at least for the case of homogeneous material with localized damage) was discretized at the maximum examined resolution into a 100 by 100 `effective subcells' grid.

After running several sets of problems with increasing numbers of subcells and cells, it was empirically observed that computational time increases linearly with the total number of subcells in the domain. This is typical for explicit incremental-stepping algorithms. The following can be said about memory and CPU requirements for the algorithm when comparing to the standard (FEM) approach. A common strategy for solving the problem of finding the mechanical response of composites with hyperelastic constituents, is to apply the loading incrementally, with increments sufficiently small such that the equations of elasticity could be linearized in each increment. Then, the tangent modulus is derived, the spatial domain is divided into finite elements and the equilibrium equations are solved in the weak form, which for linear response within an increment produces the correct solution, as minimization converges to the unique minimum. If minimization algorithms are used, then the increments only need to be small enough for the gradient of the objective function to be \emph{monotonic} in each increment, which is a less strict condition. Then, typically, the L-BFGS iterative formula is used, which is the minimization counterpart of the low memory Broyden-step Quasi-Newton algorithm employed for solving nonlinear equations. If the equation form of the FE method is used rather than minimization, which is possible for the tangent-modulus based strategy, then the linear system can either be solved exactly using a sparse-matrix based Gauss elimination for the stiffness matrix, or iterative linear solvers based on variants of the Gauss-Seidel approach, using preconditioning. In any case, sparse exact solvers are usually linear in the number of elements in terms of complexity, for narrow-bandwidth matrices, and iterative solvers, say of the Quasi-Newton type, are sub-linear in complexity (they have super-linear convergence). 
Therefore, in terms of equation solution, both the FEM and the suggested method are linear in the system size, with perhaps different coefficients. As for CPU required for the construction of the equations, again the FEM is linear in system size for stiffness matrix construction, and the proposed algorithm involves back and forth discrete Fourier transforms with solution of linear systems for each harmonic \emph{independently}. This is also linear in system size, with perhaps a different prefactor than for the FEM. In terms of memory usage, the sparse exact solver requires storage size linear in the resolution, and the low memory BFGS or Broyden both also require linear storage. Therefore, asymptotically, the proposed method is comparable with the FEM solution. The implementation here was done in-house, using the Fortran programming language, with no use of black-box commercial software. The potential advantage can come from the fact that in the proposed method, the increments of the loading do not have to be so small as to make the equations \emph{monotonic} in the variables within each increment, since, unlike in the FEM approach, it is not needed for convergence to hold theoretically. Instead, the increments need only to be small enough for the starting point to be sufficiently close to the solution. Then, convergence would hold theoretically within each increments for the nonlinear solver. This \emph{may} become a less stringent requirement than monotonicity conservation, which might allow taking larger increments. The latter, in turn, should not have an effect on the number of iterations needed for convergence. This advantage is inherited by the Broyden step from Newton's method. The number of iterations for convergence is asymptotically system-size independent, but it can add a constant factor to the complexity. This factor can be different than the corresponding factor in the FEM approach. Thus, asymptotically the proposed approach is equivalent to FEM in terms of memory and CPU, for identical loading increments, but \emph{may} allow taking larger increments. One technical advantage of the proposed algorithm is that if for a given increment convergence is achieved, then the solution is at hand, whereas using the FEM, a solution may be obtained by (local) minimization, which is not representative of strong-form equilibrium, and one may have to check for smaller increments, verifying that the solution does not change. Hence the assertion in the abstract that the proposed method can be computationally comparable and perhaps advantageous in certain applications. Regarding the issue of being more problem-specific in terms of treatment of complex geometry and not as robust as FEM in terms of meshing, it should be noted that the parametric HFGMC \cite{Ch14} was recently developed for this purpose. In any event, more meticulous comparison of the proposed and the standard methods would be done in subsequent work.

In the considered examples, only one cell in a typical direction contains the damage, making the assumption of subcell-uniform eigenstress justified. The far-field loading was applied incrementally (without augmenting the hyperelastic model by rate sensitivity) up to about ten percents of uni-directional elongation in each case, which is approximately of the same order in all strain measures. Figures $k$(a) with $k$ ranging from 3 to 10 present a schematic view of the specimen considered in each case, with the damage depicted and a control point shown for which a stress-strain curve is plotted (beneath it). The schematic view is intended to be associated with the engineering problem to be solved, and not necessarily with the specific numerical discretization for which colormaps of the results are shown subsequently. For this reason for the cases when a crack in the material is considered, the schematic view shows a thin line for a crack, instead of a rectangle with an aspect ratio corresponding to the line of subcells representing the crack. It is understood that in the limit of sufficiently high resolution, the width of the ``crack'' will converge to zero. The same holds for the length of the ``crack'', which may extend farther than, say, a cavity or fiber in its vicinity, but only by a subcell in each direction, which will vanish in sufficiently high resolution. Thus one-subcell-large geometric discrepancies between the schematic view and the computed cases are tolerated. The stress-strain curve below the schematic view shows the first Piola-Kirchhoff stress, representing the force per unit unstrained area, and the square `Lagrangian' strain, which is a reasonable dimensionless measure of relative deformation. In all cases, a dashed line shows hypothetical linear behavior with ambient elastic constants, as a reference, implying developed material nonlinearity (or, owing to the large-strain formulation, combined material and geometric nonlinearity) taking place. It can be noted that the curves are slightly concave in most cases. This owes to the fact that the shown strain is quadratic in the deformation gradient and a force-displacement curve would have been a approximately a square of the shown quasi-linear curve, at least for large enough strain, or slightly less convex than a parabola, but still convex, as expected for a hyperelastic material. Figures $k$(b) show maps of stress distribution in a cross-section of the composite, for the normal vertical component of the first Piola-Kirchhoff stress, $T_{22}$ in units of MPa in the form of a `cold-warm' color map. Figures $k$(c) show the same for the normal vertical component of the square `Lagrangian' strain measure, $E_{22}$. In all cases the color maps make the emergence of stress and strain concentration close to the damage loci evident. The points chosen for demonstration of local material stress-strain relations, as depicted in Figs. $k$(a) are always as close as possible to the stress concentration maxima loci. Discussion of specific results is given in the following.

\begin{figure}[tp]
\begin{center}

{\includegraphics[scale=0.6,trim={0cm 1.25cm 0cm 0cm},clip]{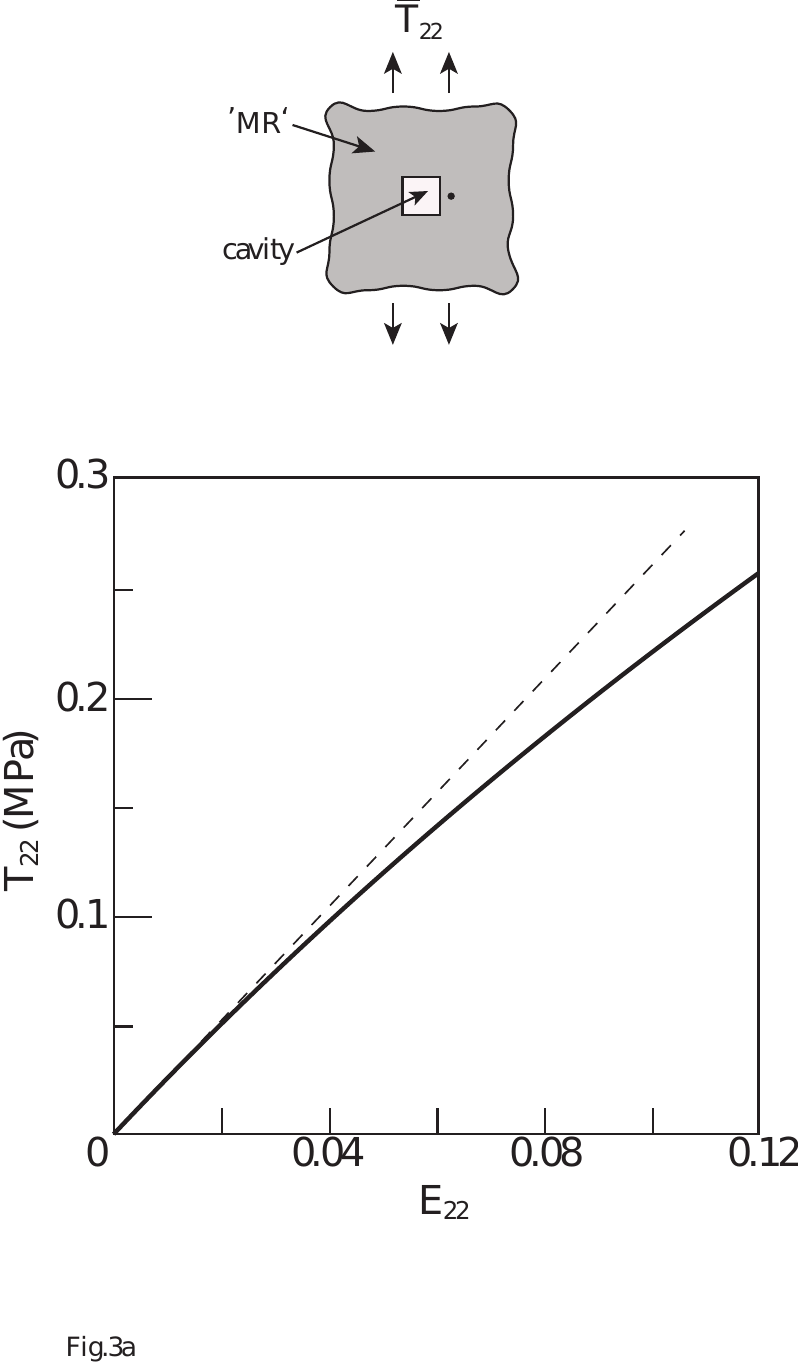} \\
\includegraphics[scale=0.5,trim={0.5cm 1cm 0cm 0.5cm},clip]{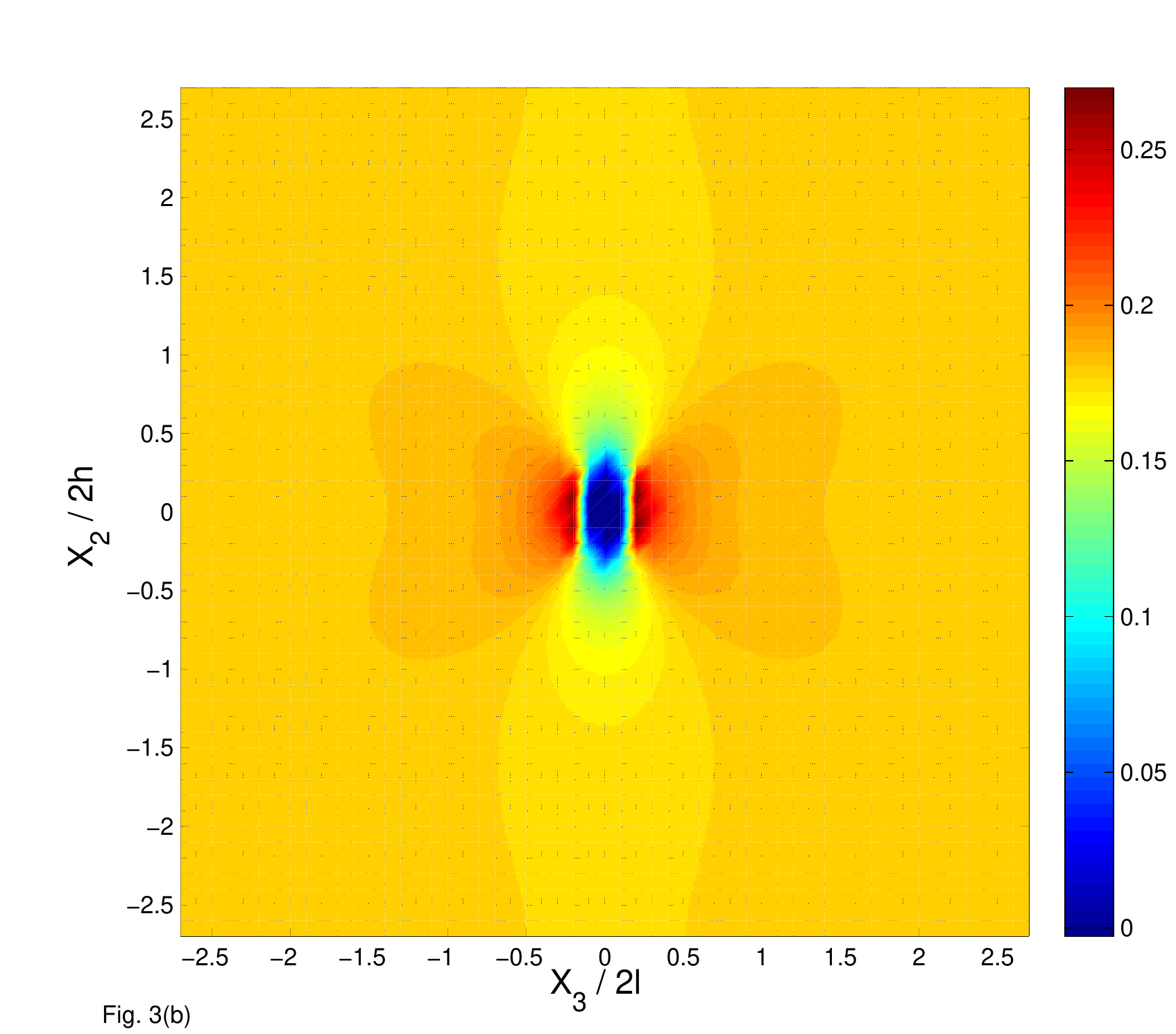} \\
\includegraphics[scale=0.5,trim={0.5cm 1cm 0cm 0cm},clip]{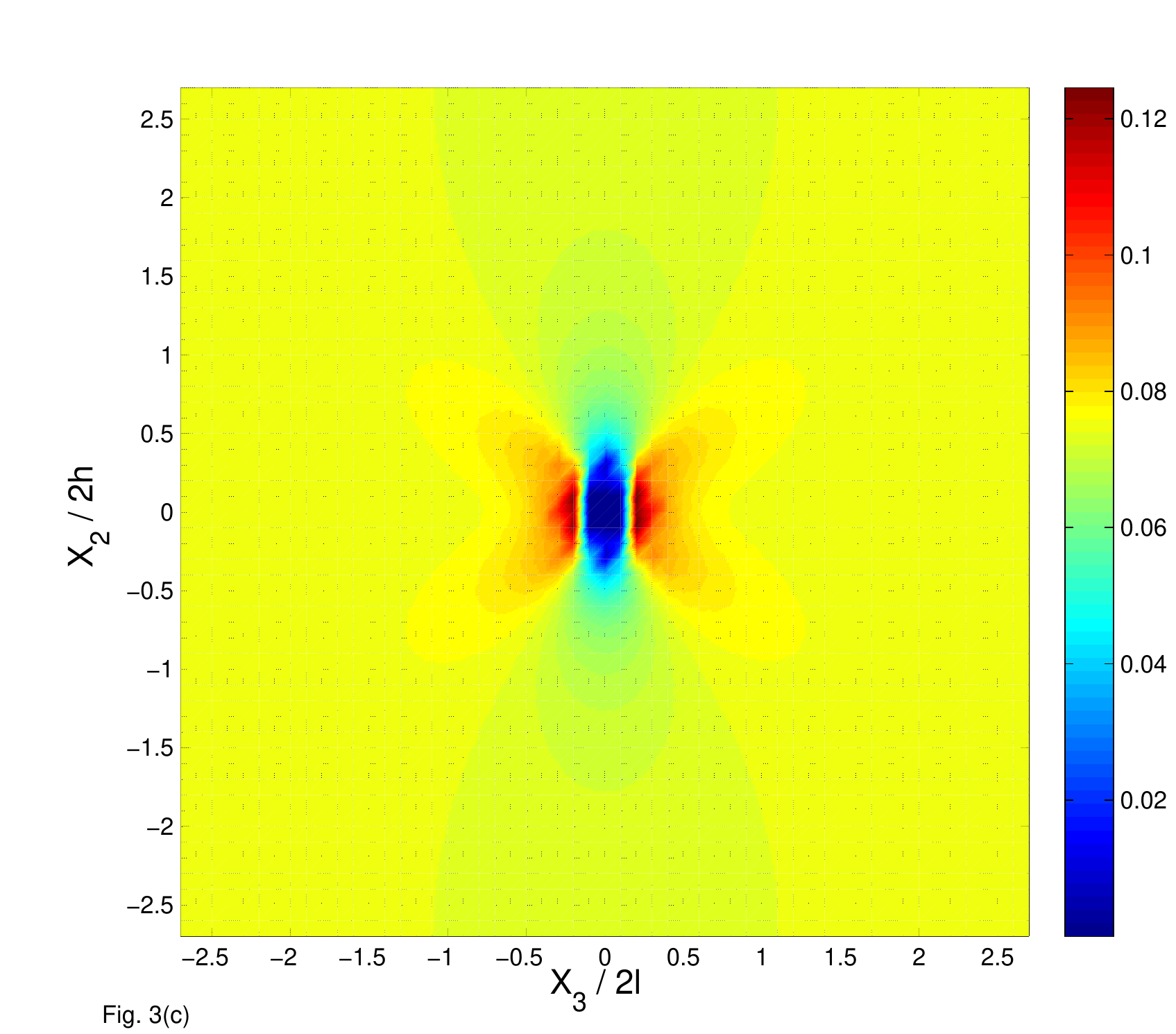}}

\end{center}
  \begin{picture}(0,0)
  \put(203,410){$T_{22}$ (MPa)}
  \put(212,207){$E_{22}$}
  \put(228,212){$X_3/2l$}
  \put(8,212){$(b)$}
  \put(8,420){$(a)$}
  \put(228,10){$X_3/2l$}
  \put(8,10){$(c)$}
  \end{picture}
\caption{(a) An infinite Mooney-Rivlin material with an embedded square cavity, subjected to remote stress $\bar T_{22}$.
                      Normal finite stress-strain response at the indicated point adjacent to the cavity.
                  (b) and (c) The distribution of the normal Piola-Kirchhoff stress $T_{22}$ and large strain $E_{22}$, respectively (here and onward, lack of field-jumps across material interfaces, slight appeared asymmetry and zero strain inside cavities are due to postprocessing). }
\label{Figure3}
  \end{figure}
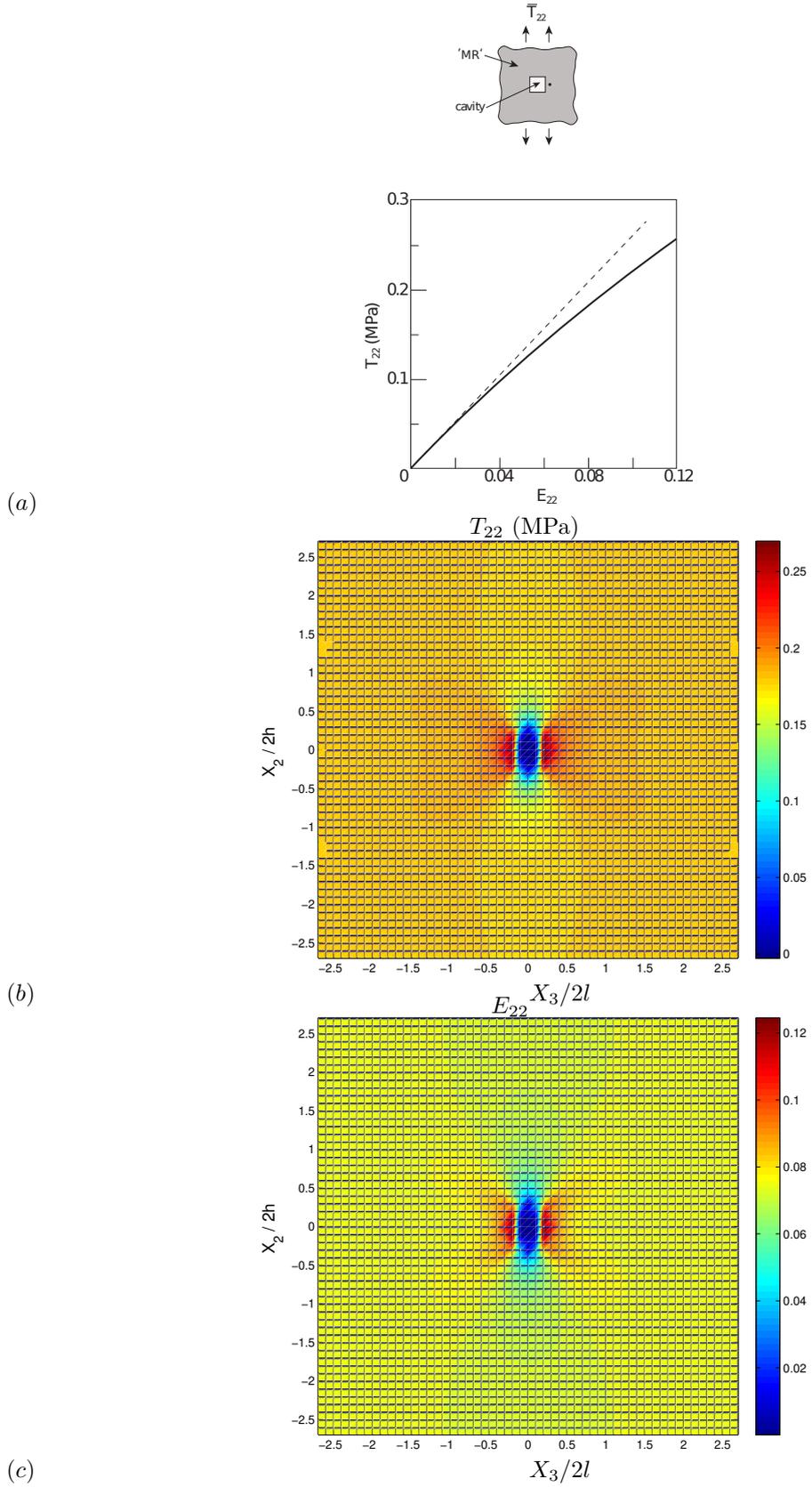

\begin{figure}[tp]
\begin{center}

{\includegraphics[scale=0.55,trim={0cm 1.25cm 0cm 0cm},clip]{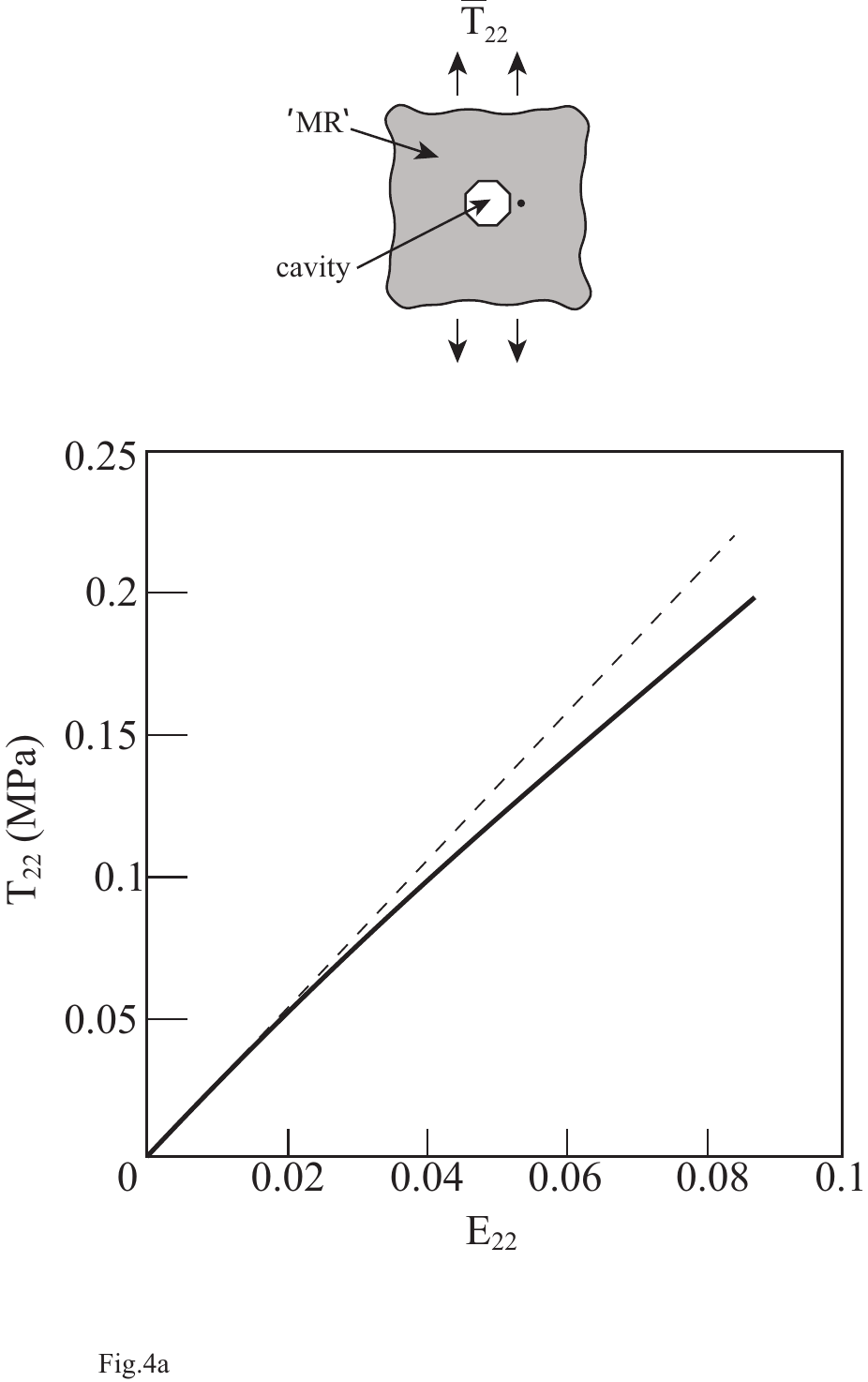} \\
\includegraphics[scale=0.5,trim={0.5cm 1cm 0cm 0.5cm},clip]{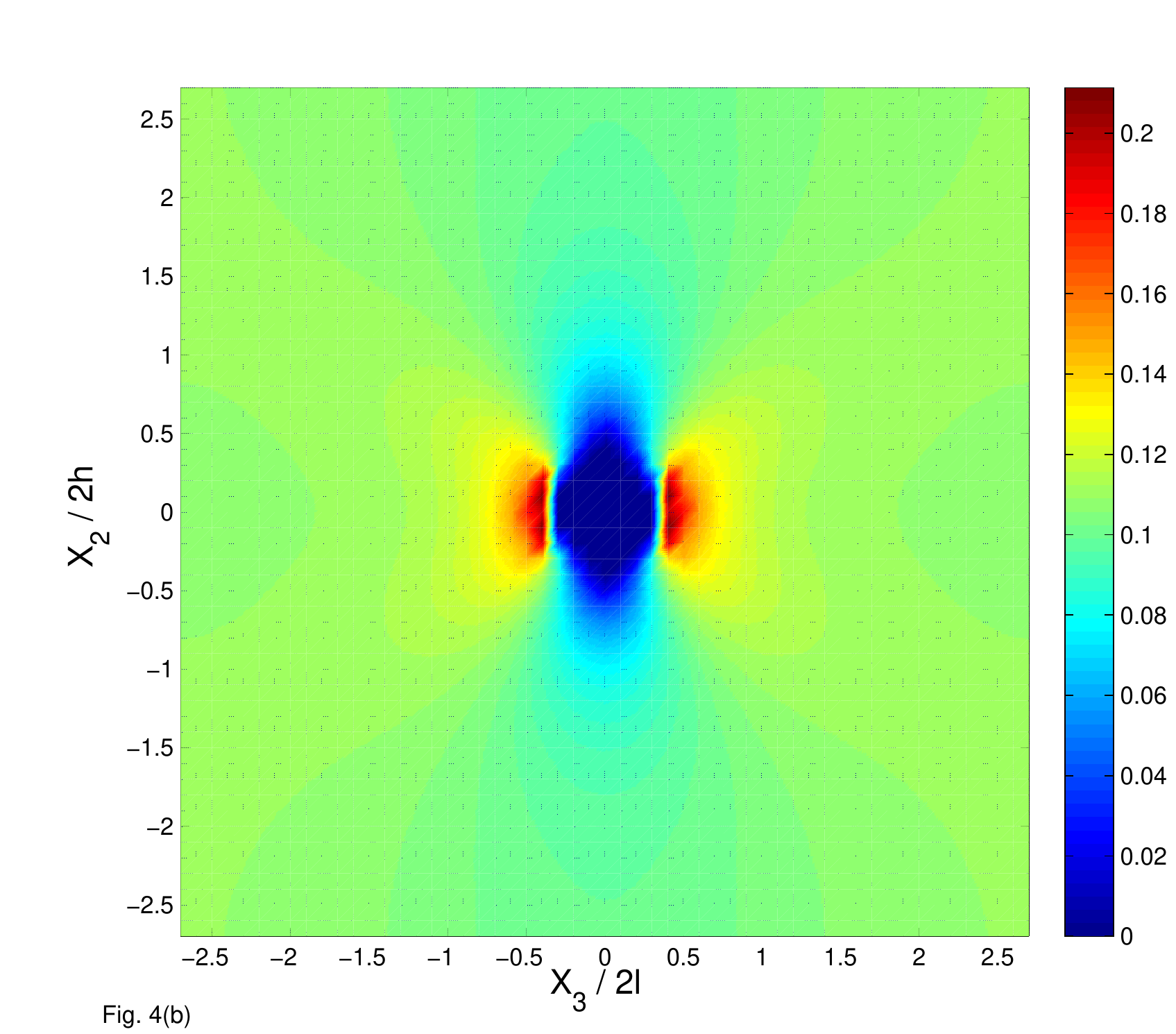} \\
\includegraphics[scale=0.5,trim={0.5cm 1cm 0cm 0cm},clip]{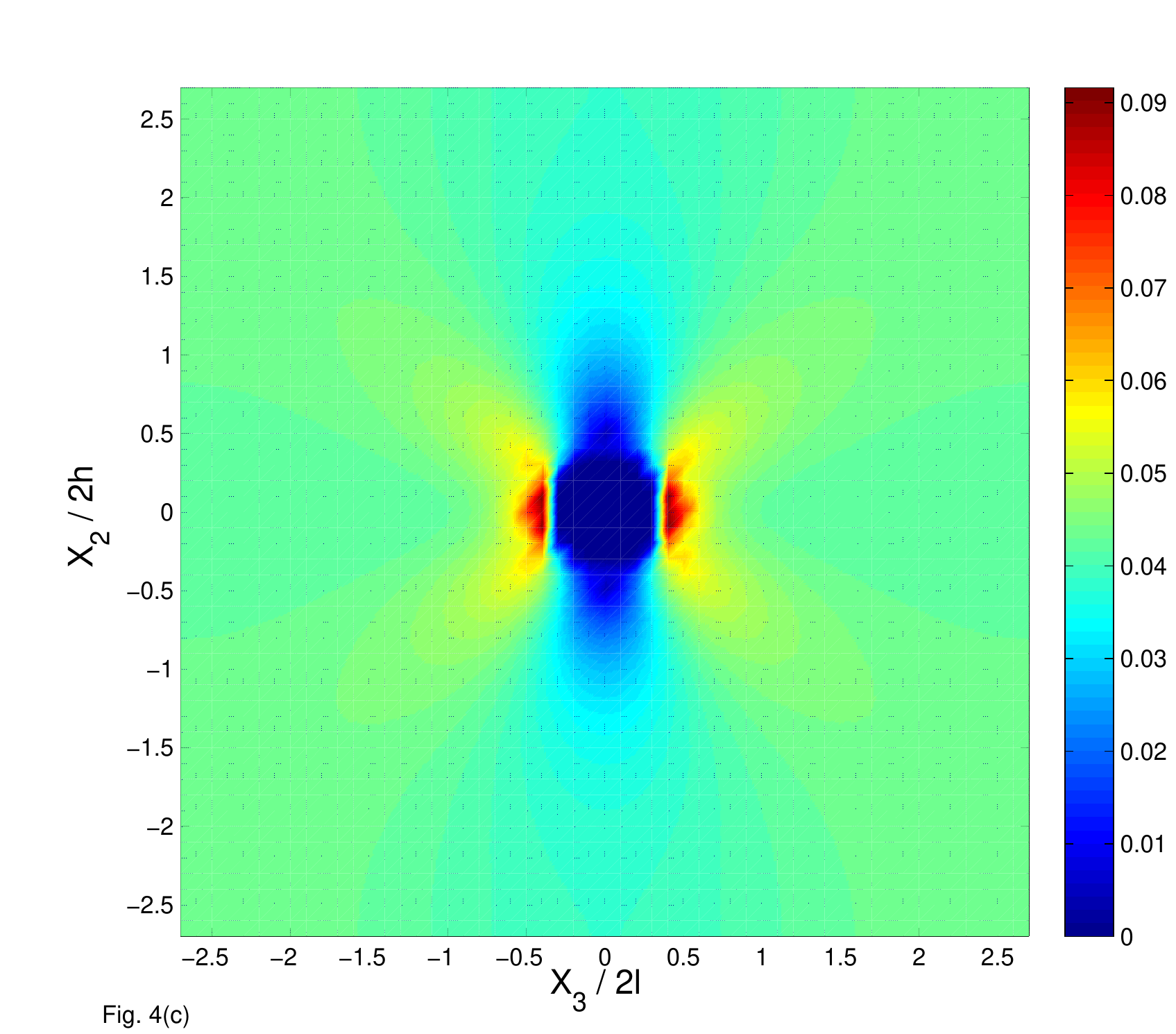}}

\end{center}
  \begin{picture}(0,0)
  \put(203,410){$T_{22}$ (MPa)}
  \put(212,207){$E_{22}$}
  \put(228,212){$X_3/2l$}
  \put(8,212){$(b)$}
  \put(8,420){$(a)$}
  \put(228,10){$X_3/2l$}
  \put(8,10){$(c)$}
  \end{picture}
\caption{(a) An infinite Mooney-Rivlin material with an embedded (roughly) octagonal cavity, subjected to remote stress $\bar T_{22}$.
                      Normal finite stress-strain response at the indicated point adjacent to the cavity.
                  (b) and (c) The distribution of the normal Piola-Kirchhoff stress $T_{22}$ and large strain $E_{22}$, respectively.   }
\label{Figure4}
  \end{figure}

\begin{figure}[tp]
\begin{center}

{\includegraphics[scale=0.61,trim={0cm 1.1cm 0cm 0cm},clip]{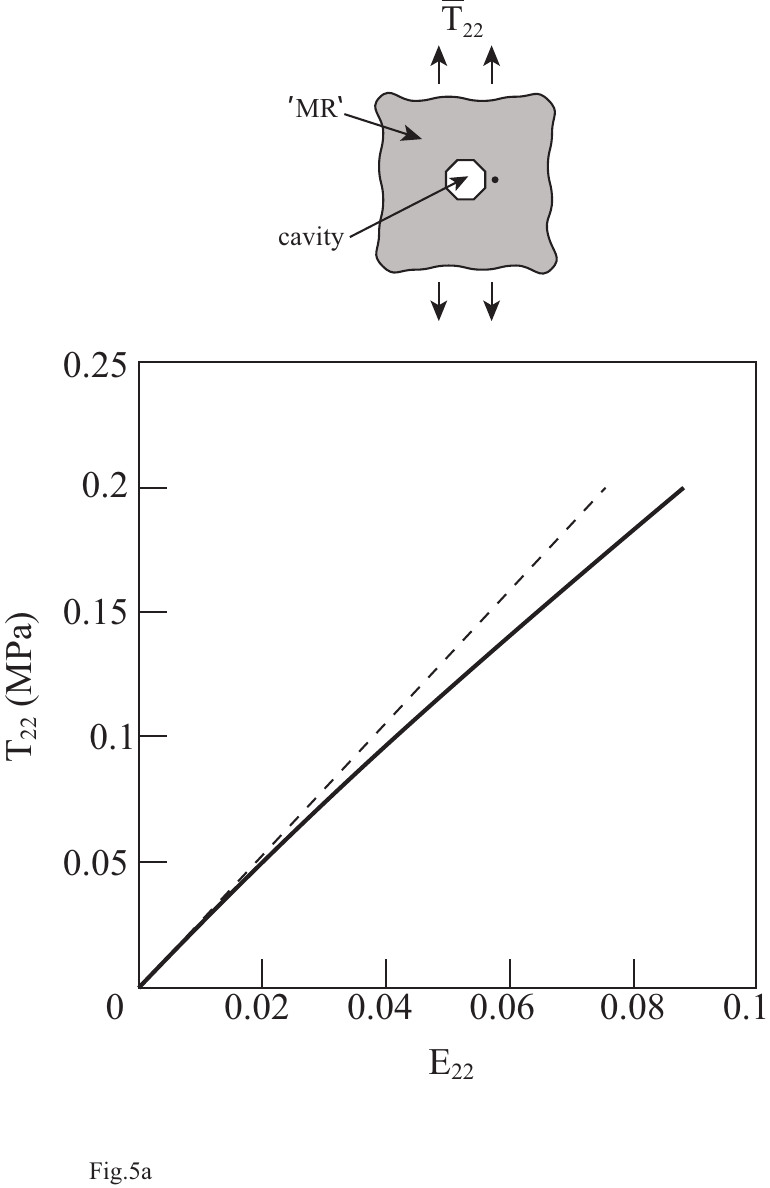} \\
\includegraphics[scale=0.5,trim={0.5cm 1.29cm 0cm 0.5cm},clip]{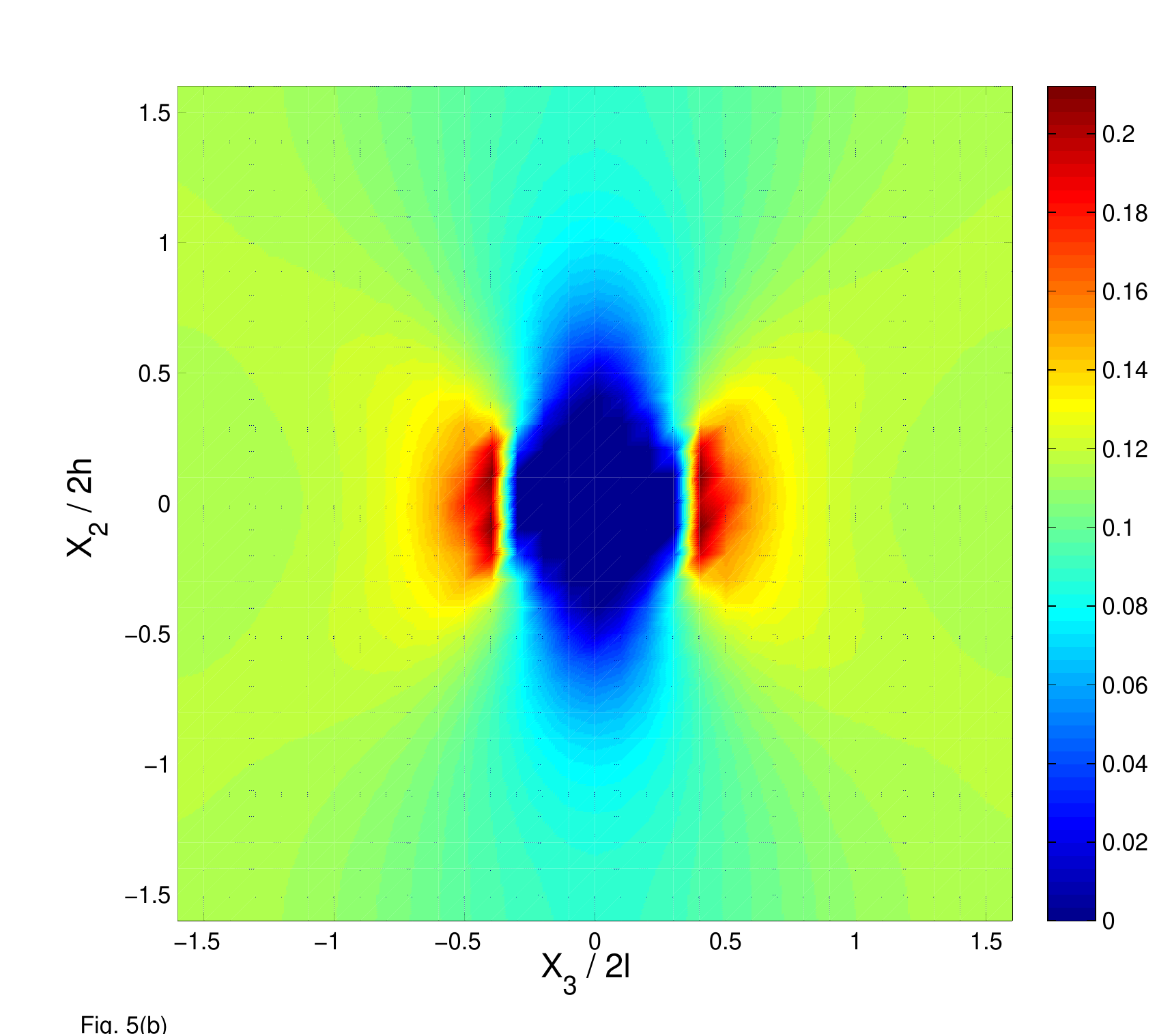} \\
\includegraphics[scale=0.5,trim={0.5cm 1.27cm 0cm 0cm},clip]{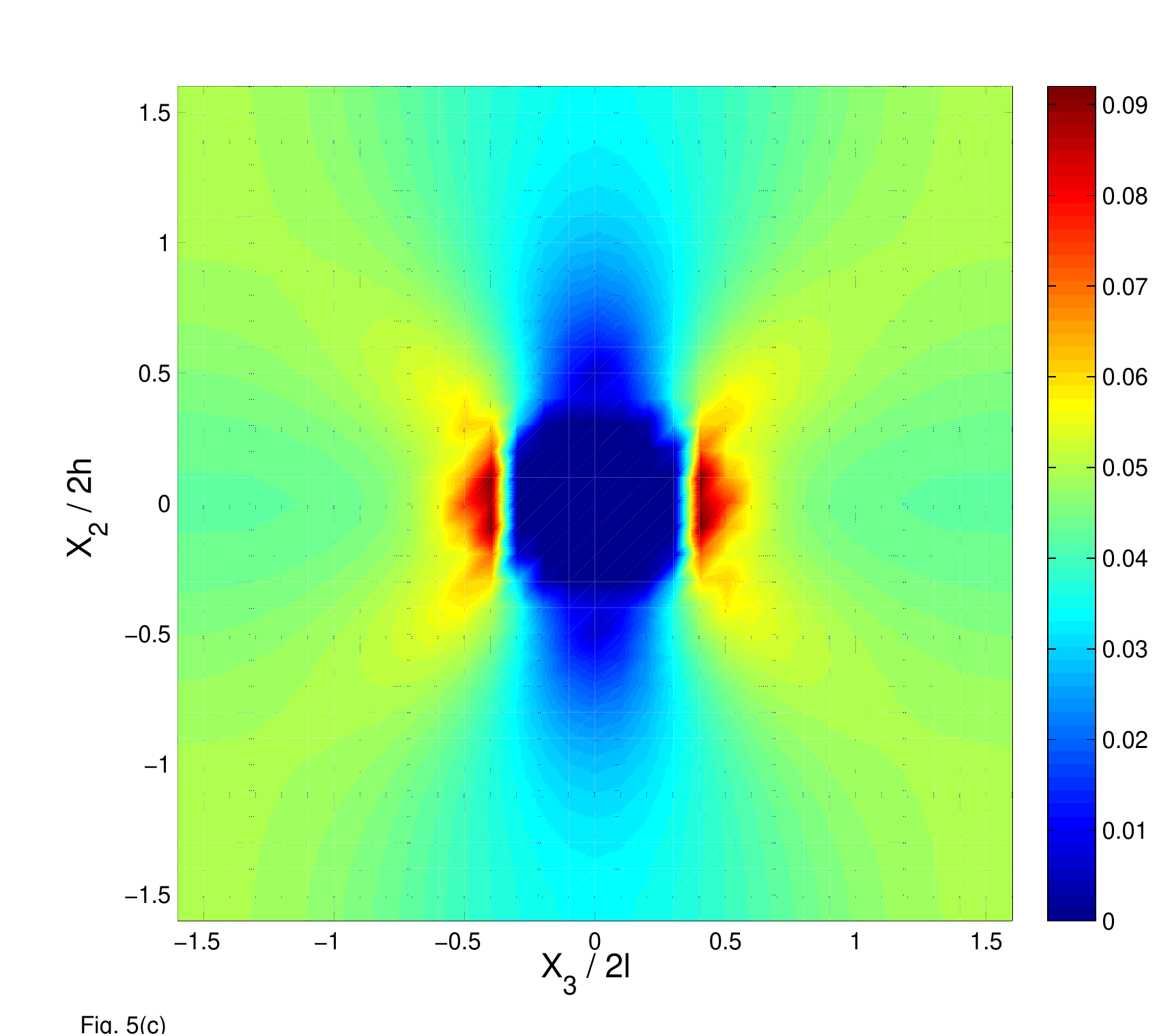}}

\end{center}
  \begin{picture}(0,0)
  \put(203,410){$T_{22}$ (MPa)}
  \put(212,207){$E_{22}$}
  \put(228,214){$X_3/2l$}
  \put(8,212){$(b)$}
  \put(8,420){$(a)$}
  \put(228,10){$X_3/2l$}
  \put(8,10){$(c)$}
  \end{picture}
\caption{(a) An infinite Mooney-Rivlin material with an embedded (roughly) octagonal cavity (void-fraction larger by 5/3 than in Fig. \ref{Figure4}), subjected to remote stress $\bar T_{22}$.
                      Normal finite stress-strain response at the indicated point adjacent to the cavity.
                  (b) and (c) The distribution of the normal Piola-Kirchhoff stress $T_{22}$ and large strain $E_{22}$, respectively.   }
\label{Figure5}
  \end{figure}

Figure \ref{Figure3} shows uniaxial stretch of Mooney-Rivlin material with a square-section cavity represented by 5 subcells in each directions. Nonlinearity for strains increasing from zero is apparent, as well as stress concentration with a maximum value of about 1.5 for the stress and slightly higher for the strain. Figure \ref{Figure4} shows a similar result for a larger and more round cavity. For this example, in the central cell in an array of 5 by 5 cells, a roughly octagonal cavity was created, as follows. A 5 by 5 array of (fully damaged) subcells was placed in the center of the cell. Then four linear arrays of 3 additional (fully damaged) subcells each were added symmetrically below, above, to the right and to the let of the central 5 by 5 array. The resulting domain has four faces three subcell lengths long and four faces with roughly a similar length, hence the term `octagonal'. One should remember that the method of analysis is volume integral, and some roughness in the domain boundary is less influential than in a finite-difference method. Larger stress and strain concentration is observed than for the square cavity case, as expected for a composite with a smaller material volume fracture. A maximum value of about 2 can be observed. The effect of the cavity shape is, again, not very pronounced due to the volume-integrated approach of the method of solution. Figure \ref{Figure5} shows a case similar to that of Fig. \ref{Figure4}, only for fewer cells (which implies a larger damaged area fraction). As expected, a higher stress and stress concentration factor, of about 3 is observed. Moreover, apparently the chosen number of cells, namely 3 in each direction, is insufficient for observing asymptotic behavior far from the damage locus.

\begin{figure}[tp]
\begin{center}

{\includegraphics[scale=0.61,trim={0cm 1.1cm 0cm 0cm},clip]{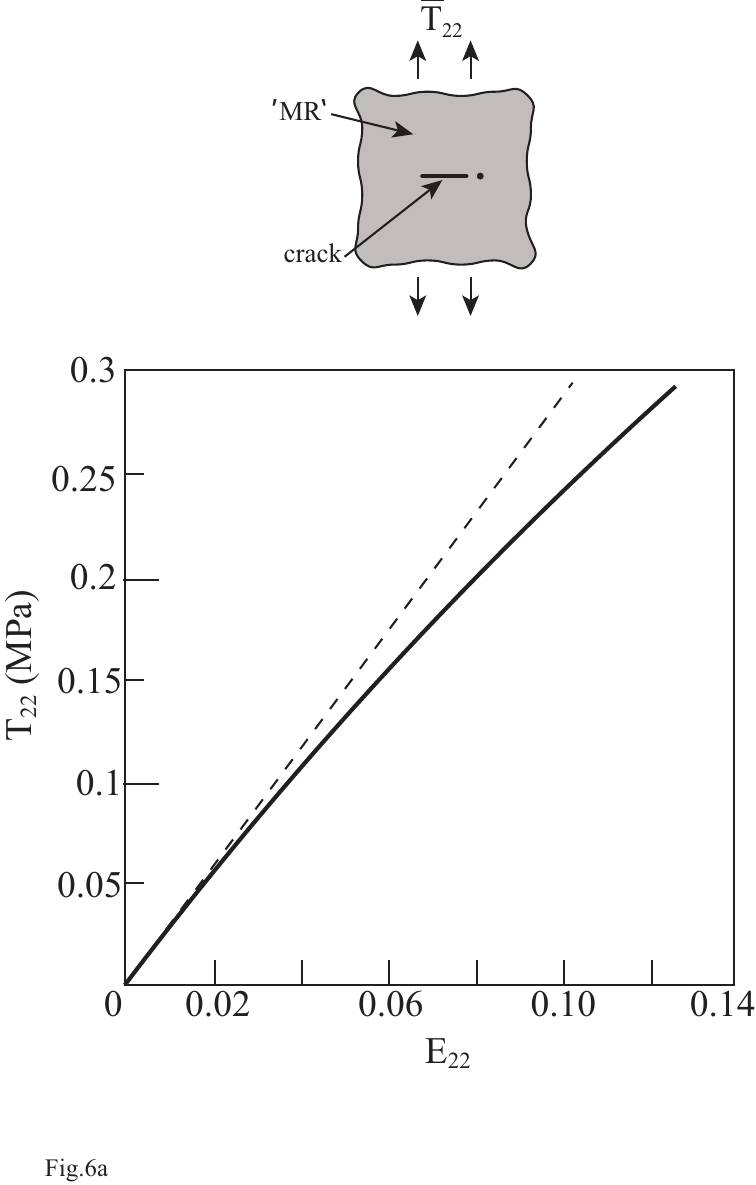} \\
\includegraphics[scale=0.5,trim={0.5cm 1cm 0cm 0.5cm},clip]{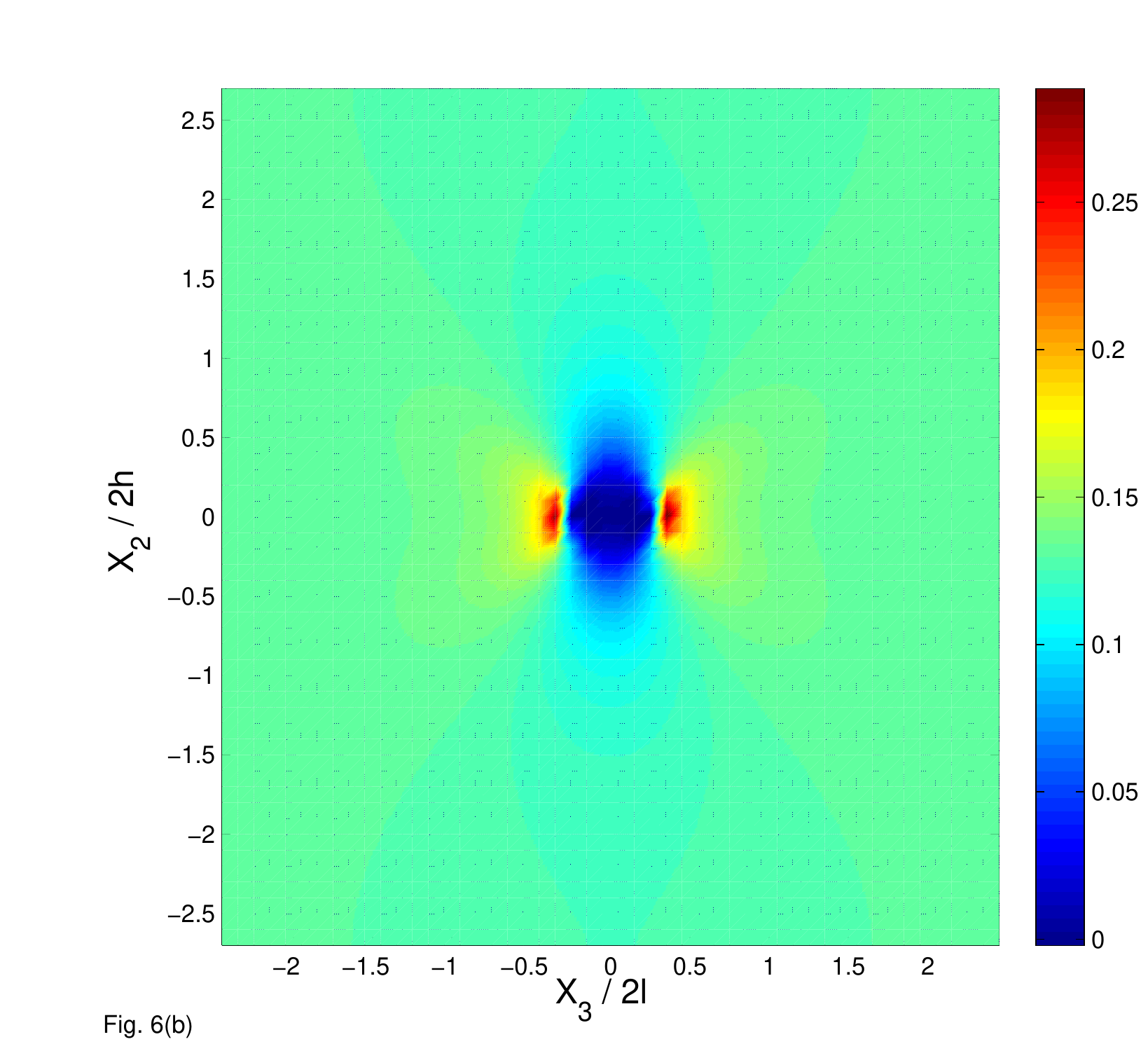} \\
\includegraphics[scale=0.5,trim={0.5cm 1cm 0cm 0cm},clip]{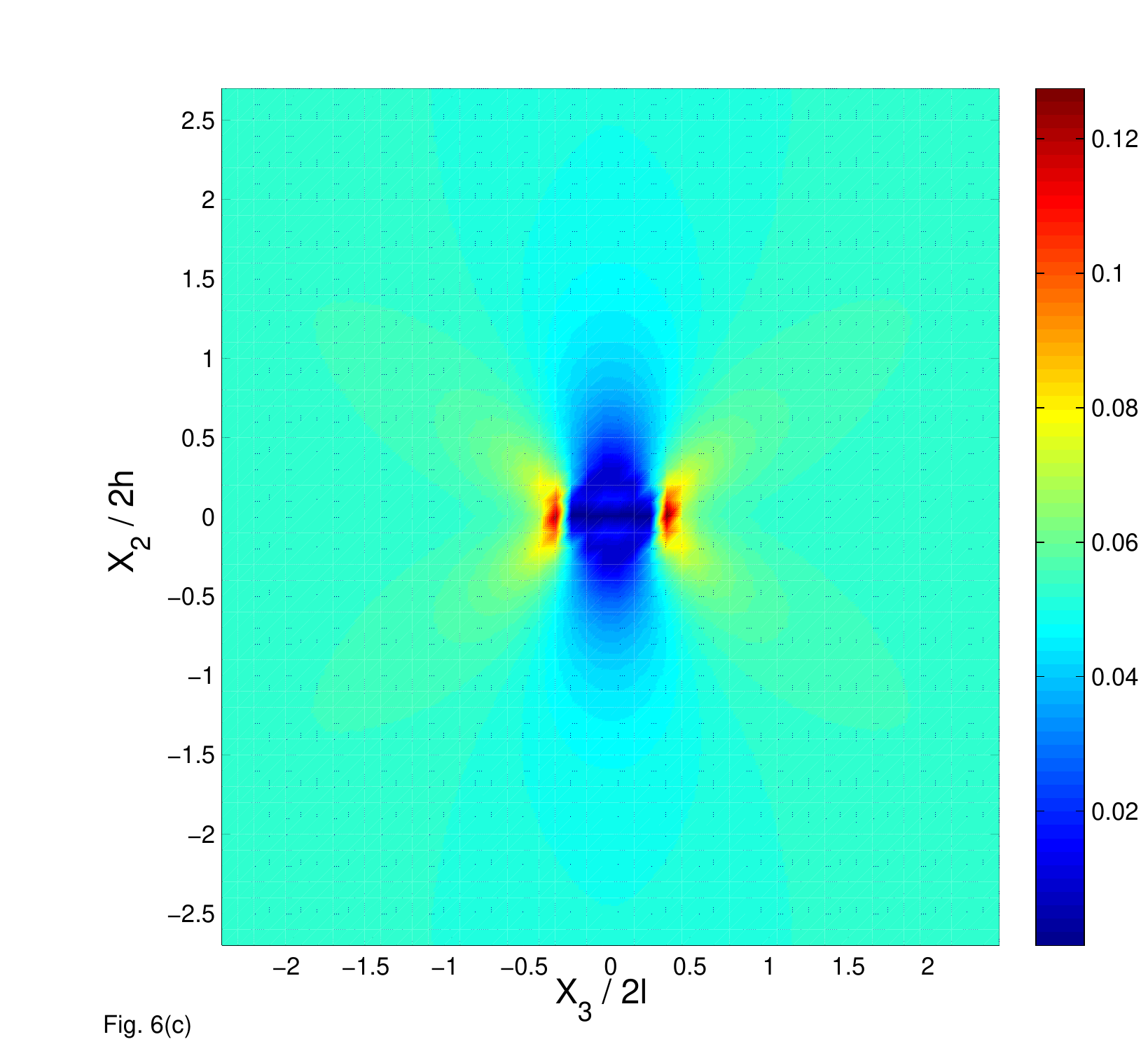}}

\end{center}
  \begin{picture}(0,0)
 \put(203,410){$T_{22}$ (MPa)}
  \put(212,207){$E_{22}$}
  \put(228,212){$X_3/2l$}
  \put(8,212){$(b)$}
  \put(8,420){$(a)$}
  \put(228,10){$X_3/2l$}
  \put(8,10){$(c)$}
  \end{picture}
\caption{ (a) Infinite Mooney-Rivlin material with a ``crack'', subjected to remote stress $\bar T_{22}$.
                      Normal finite stress-strain response at the indicated point adjacent to the ``crack''.
                  (b) and (c) The distribution of the normal Piola-Kirchhoff stress $T_{22}$ and large strain $E_{22}$, respectively.  }
\label{Figure6}
  \end{figure}

Figure \ref{Figure6} shows the case with a finite ``crack'' in a Mooney-Rivlin material, stretched in the first mode. The ``crack'' is represented by a single line of six fully damaged subcells, positioned symmetrically in the central cell. One observes sufficiently developed nonlinearity, attained asymptotic stress and strain distribution and concentration of stress and strain close to the ``crack'' tips, with a maximum value of about 3. Figure \ref{Figure6d} compares a stress profile on the ``crack'' axis for the hyperelastic material to the so-called K-field expected for a linear material. One observes the much higher stress localization for the hyperelastic material, with stronger-than power law decay (apparent on a log scale not shown here). The higher level of (stress) energy localization for the nonlinear case is not surprising. An important remark is to be made here. Within the framework of the chosen approach, cracks are represented by a line of subcells, in which all stress components vanish due to setting $D=1$ there. In terms of stress such entities are as good as `elongated voids', but unlike voids they have the original material density and they obey displacements continuity just as the original material. Clearly, the stress and strain concentration close to the `tips' of such ``cracks'' will markedly underestimate the stress and strain concentrations around the tips of real cracks. However, the common and basically reasonable understanding of this representation is that the stress and strain at the tips of the ``crack'' are not computed anyway. What is computed is the stress and strain averaged over a subcell for the nonlinear elastic part and a linear trend line over the range of a subcell for the linear-elastic part. This linear profile starting from zero at the edge damaged subcell in the ``crack'' and reaching an attenuated value at the neighboring subcell cannot capture a peak anyway. The understanding is that upon increasing resolution, points closer to the tip of the ``crack'' are attained, and along with that also the value of the stress at the tip is gradually attained.

\begin{figure}[tp]
\begin{center}

{\includegraphics[scale=0.6,trim={0cm 1cm 0cm 0cm},clip]{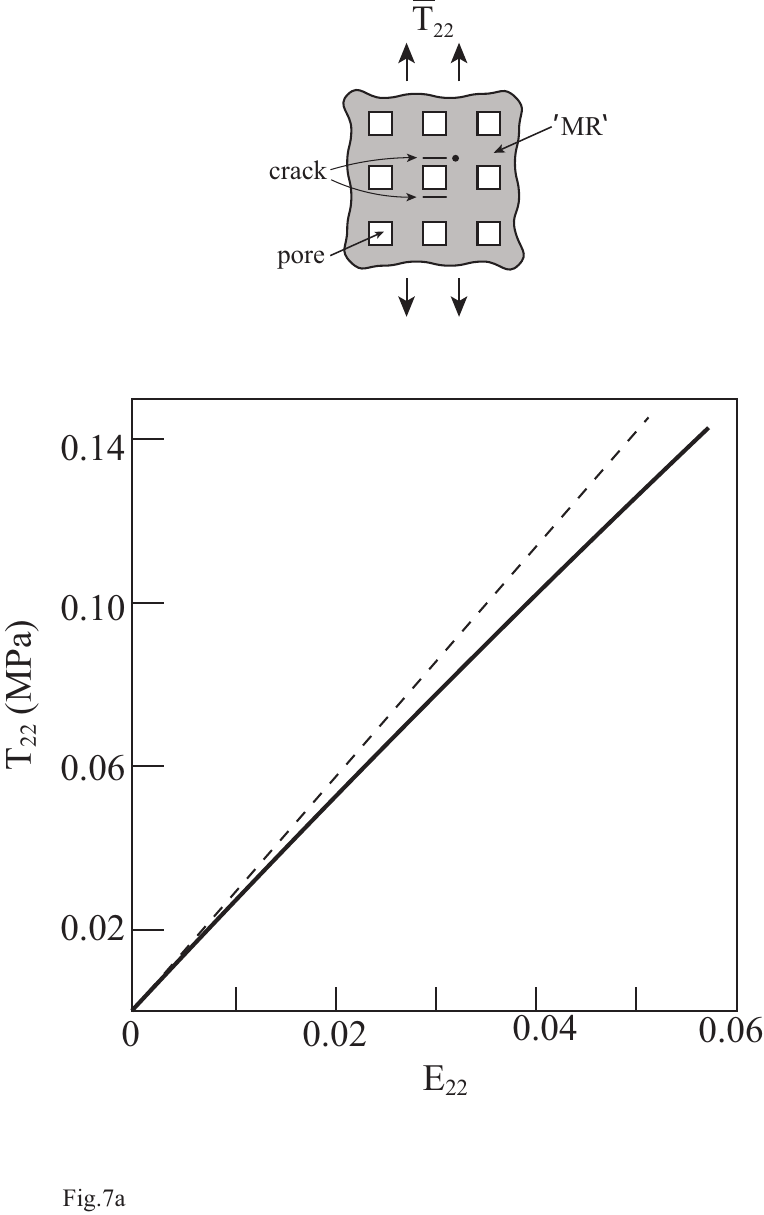} \\
\includegraphics[scale=0.5,trim={0.5cm 1cm 0cm 0.5cm},clip]{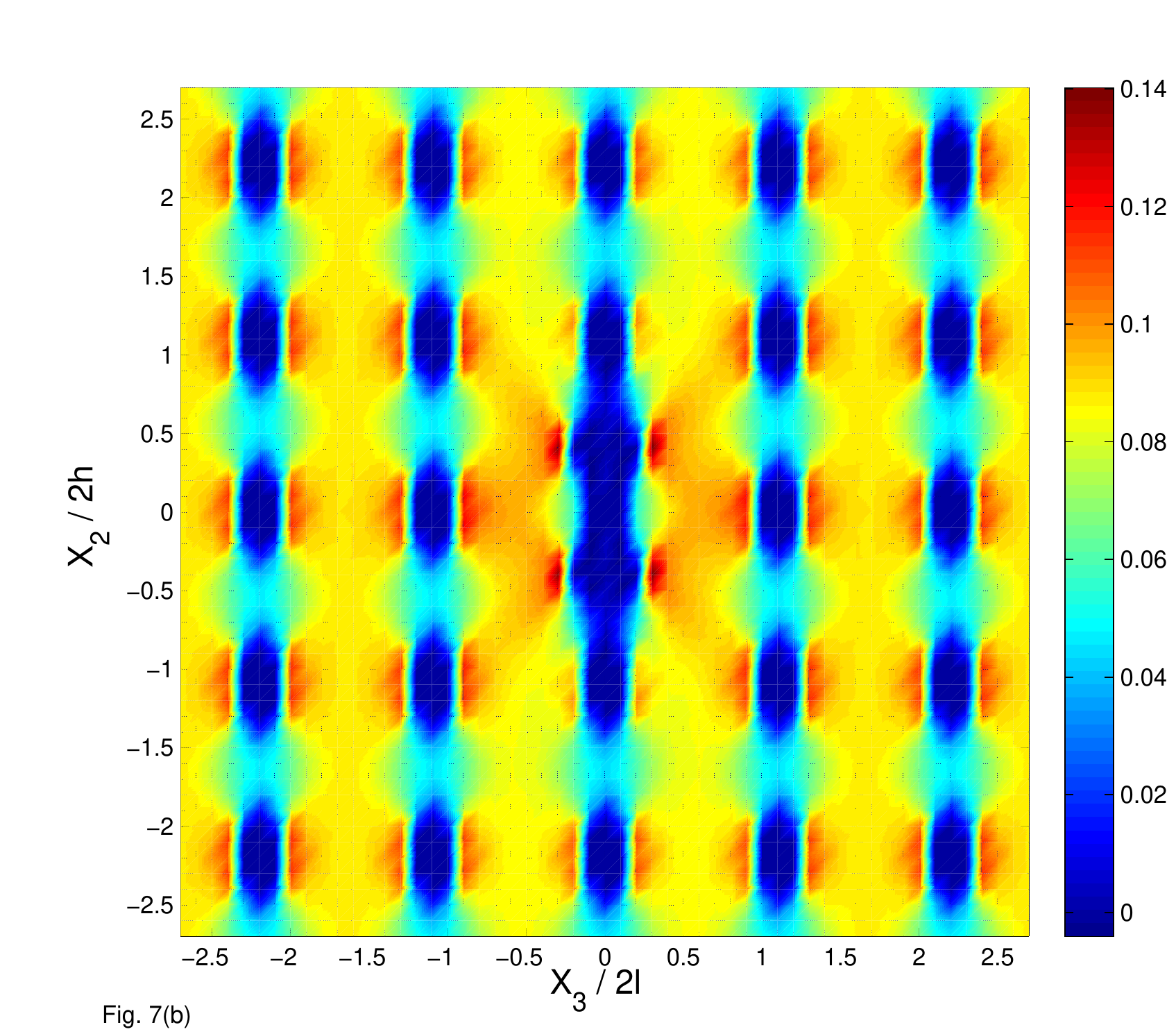} \\
\includegraphics[scale=0.5,trim={0.5cm 1cm 0cm 0cm},clip]{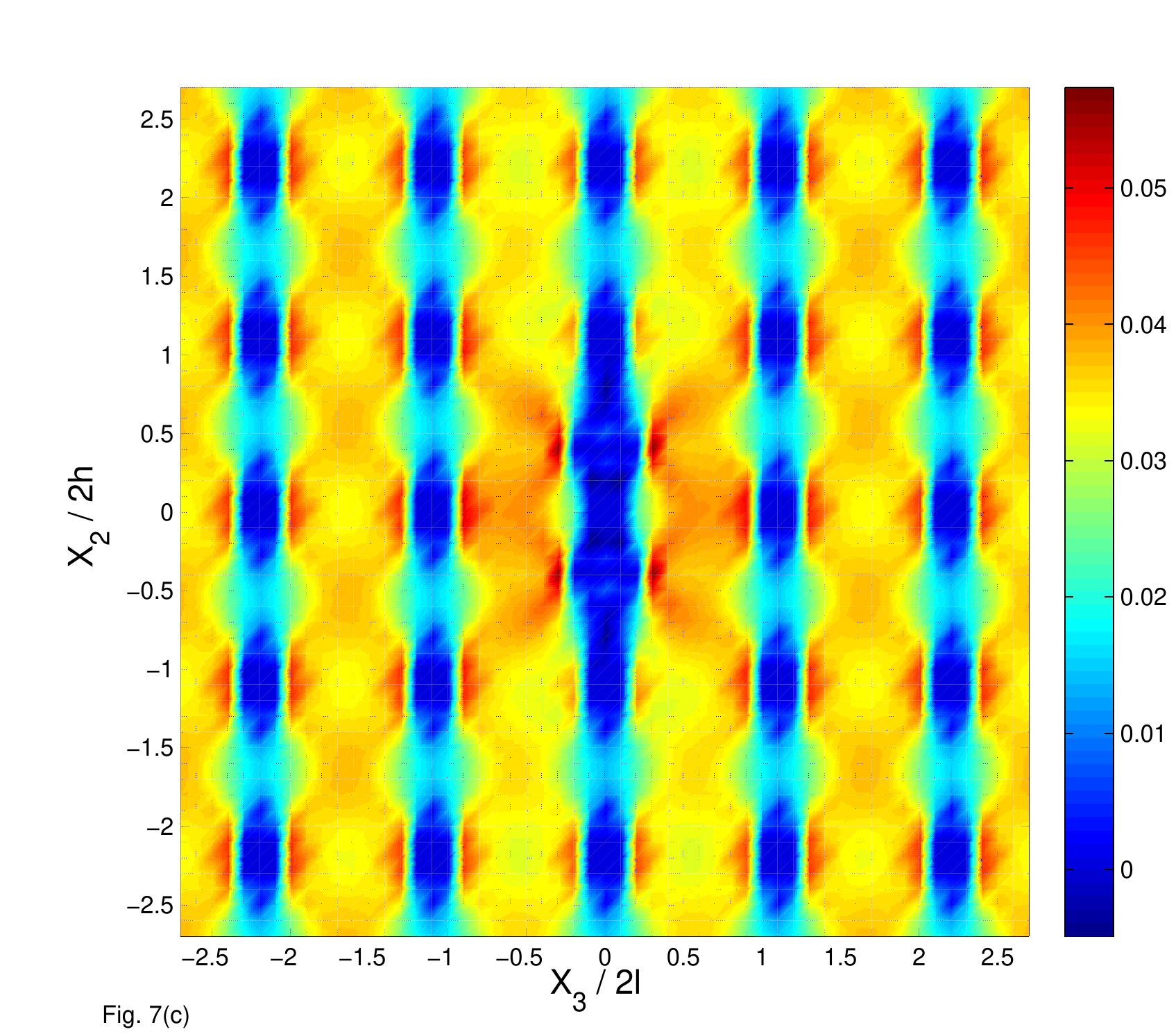}}

\end{center}
  \begin{picture}(0,0)
 \put(203,410){$T_{22}$ (MPa)}
  \put(212,207){$E_{22}$}
  \put(228,212){$X_3/2l$}
  \put(8,212){$(b)$}
  \put(8,420){$(a)$}
  \put(228,10){$X_3/2l$}
  \put(8,10){$(c)$}
  \end{picture}
\caption{ (a) Porous Mooney-Rivlin material with two ``cracks'', subjected to remote stress $\bar T_{22}$.
                      Normal finite stress-strain response at the indicated point adjacent to the upper ``crack''.
                  (b) and (c) The distribution of the normal Piola-Kirchhoff stress $T_{22}$ and large strain $E_{22}$, respectively.   }
\label{Figure7}
  \end{figure}

Figure \ref{Figure7} shows the case of porous Mooney-Rivlin material with a doubly-periodic array of square-section prismatic cavities, each represented by 3 by 3 arrays of subcells of zero-stiffness material, positioned symmetrically in each cell. The localized damage in this case is manifested in the form of two identical finite ``cracks'', each represented by a single line of five fully damaged subcells located symmetrically below and above one of the cavities (the central one) with a vertical distance of two subcells between the cavity and the subcells of a ``crack''. This is the first example of the present work of localized damage in a hyperelastic composite, albeit with the second phase being void. The stress and strain maps show periodic distributions sufficiently far from the damage locus and the apparent stress and strain concentration close to the four tips of the two finite ``cracks''. One observes that the ``cracks'' do ``release'' the stress and strain in the material below and above them, as one might expect. The strain map reveals an interesting ``zig-zag'' pattern of higher strain around the ``cracks'', a sign of ``interaction'' of (void) inclusions. It should be noted that the schematic view in the top of the figure shows the crack and pores to be of equal length, even though in the exhibited computation the crack was represented by a line of five subcells, and the pores are all three-subcells-long. The reason for the appearing discrepancy is that for five by five subcells pores the porosity is already too high to be realistic, and a ``crack'' three-subcells-long would have an aspect ratio too small to represent a line defect, whereas the length mismatch between the pores and the ``crack'' poses no special problem. For twice higher resolution, a six-subcells-long pore would be closer to a five or seven-subcells-long ``crack'' -- hence the schematic view.

\begin{figure}[tp]
\begin{center}

{\includegraphics[scale=0.63,trim={0cm 1.cm 0cm 0cm},clip]{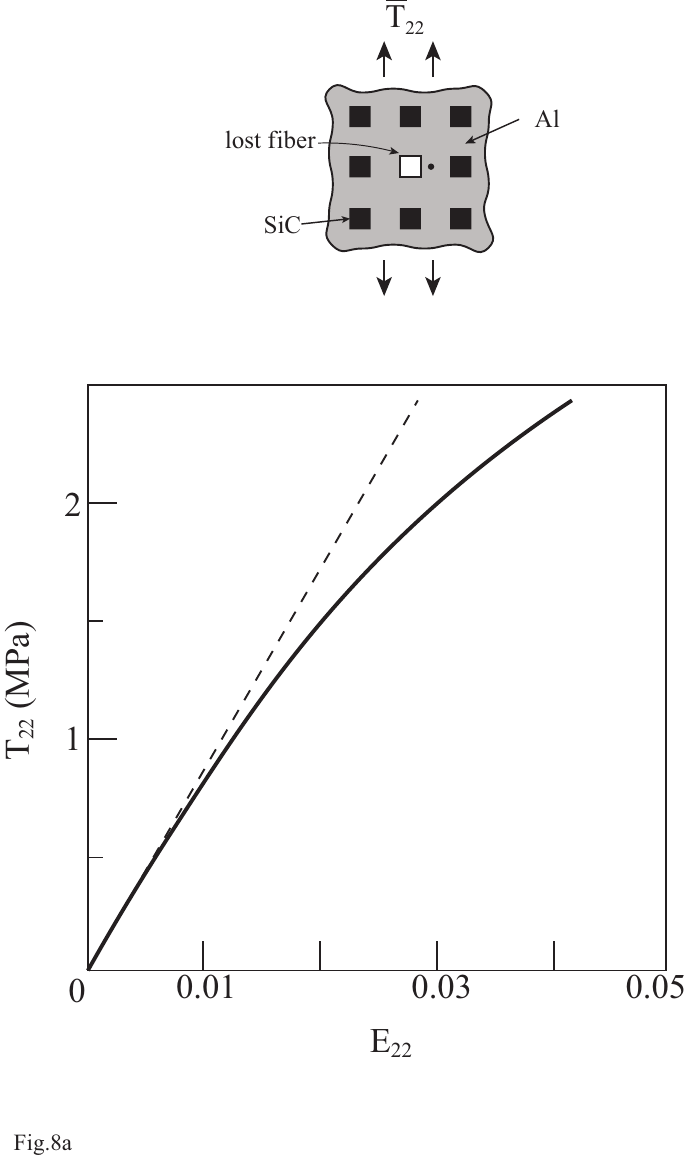} \\
\includegraphics[scale=0.5,trim={0.5cm 1cm 0cm 0.5cm},clip]{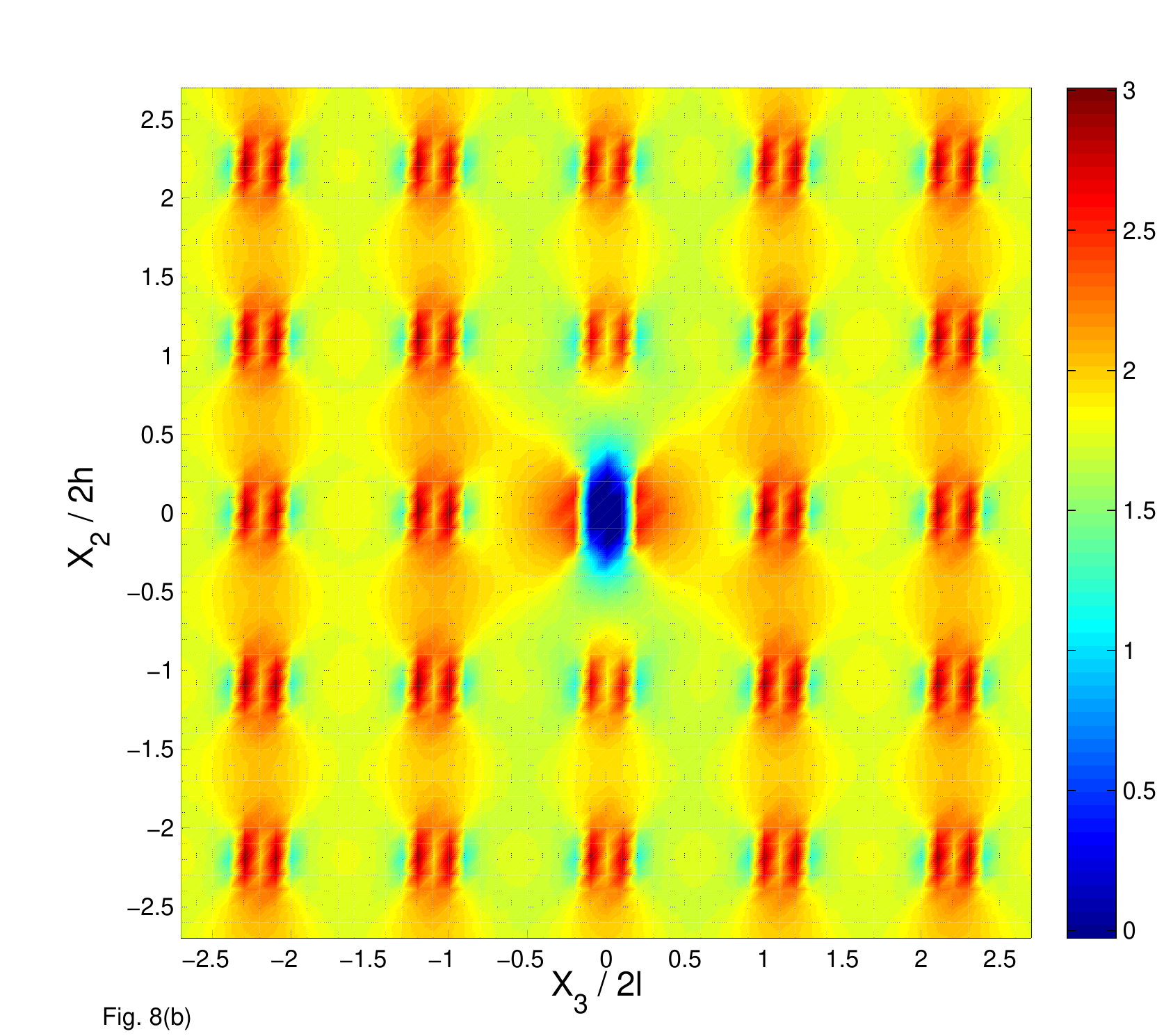} \\
\includegraphics[scale=0.5,trim={0.5cm 1cm 0cm 0cm},clip]{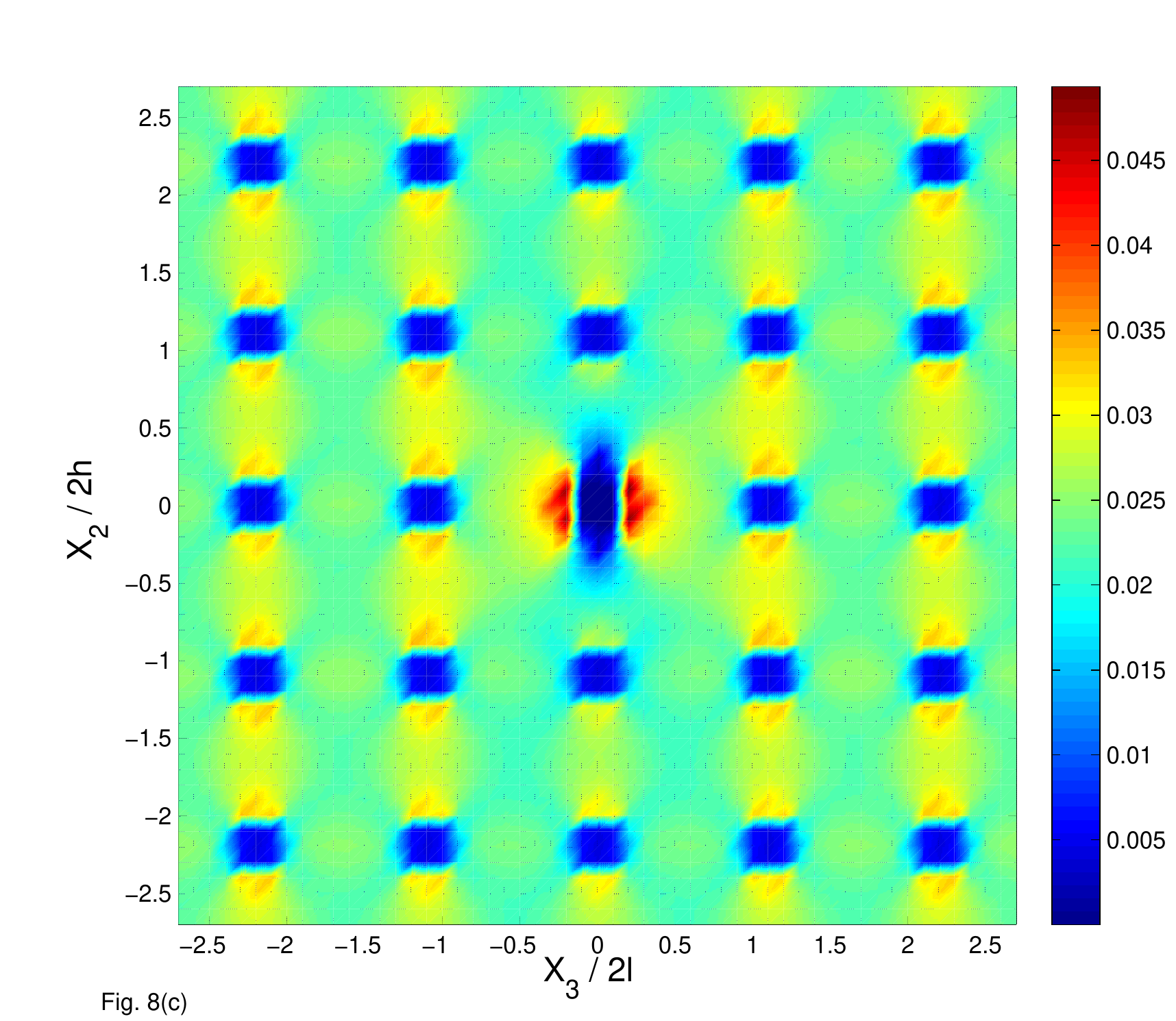}}

\end{center}
  \begin{picture}(0,0)
 \put(203,410){$T_{22}$ (MPa)}
  \put(212,207){$E_{22}$}
  \put(228,212){$X_3/2l$}
  \put(8,212){$(b)$}
  \put(8,420){$(a)$}
  \put(228,10){$X_3/2l$}
  \put(8,10){$(c)$}
  \end{picture}
\caption{ (a) `Silicon carbide'/`aluminum' composite with a lost fiber, subjected to remote stress $\bar T_{22}$.
                      Normal finite stress-strain response at the indicated point adjacent to the lost fiber.
                  (b) and (c) The distribution of the normal Piola-Kirchhoff stress $T_{22}$ and large strain $E_{22}$, respectively.   }
\label{Figure8}
  \end{figure}
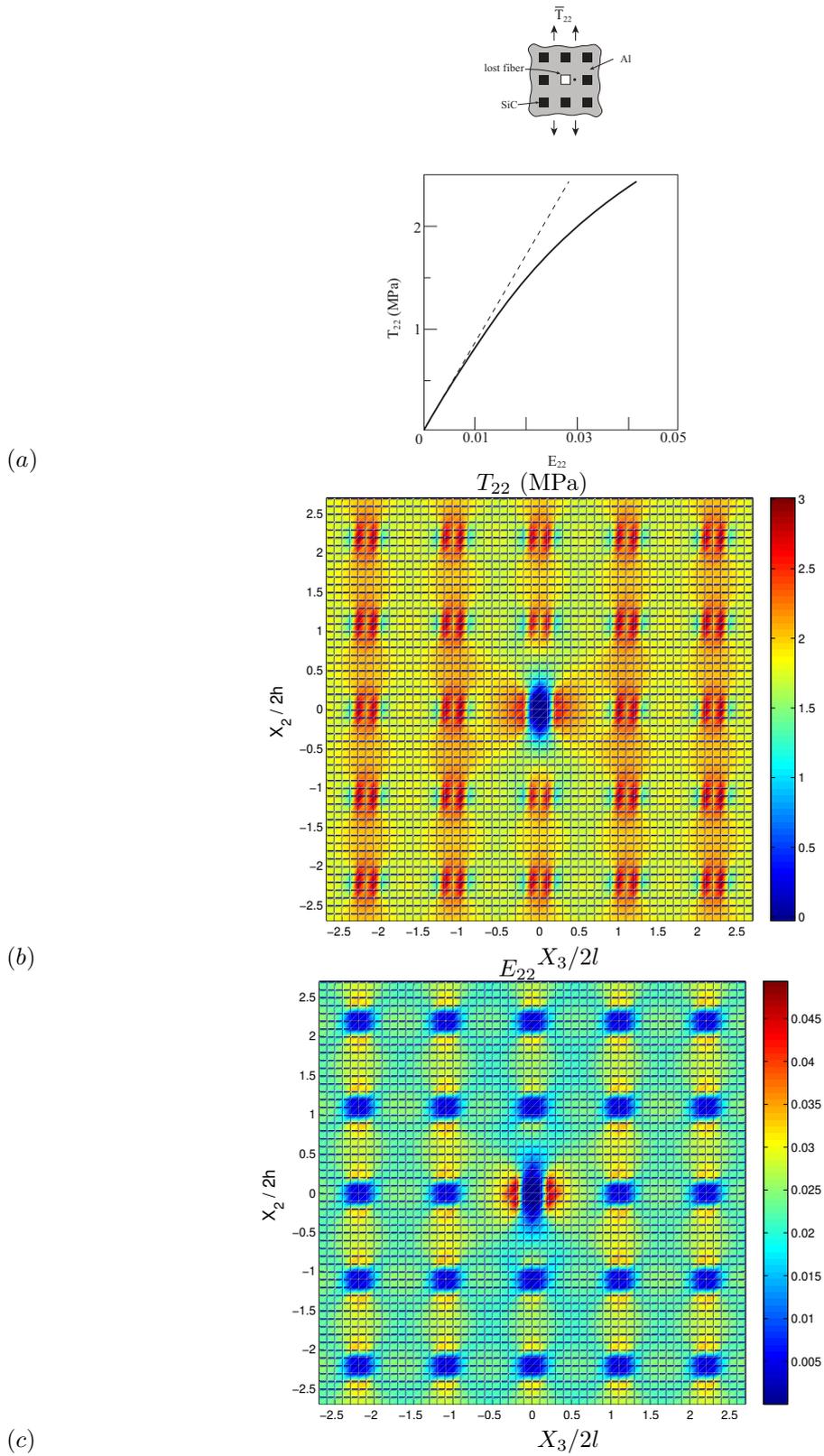

\begin{figure}[tp]
\begin{center}

{\includegraphics[scale=0.61,trim={0cm 0.5cm 0cm 0cm},clip]{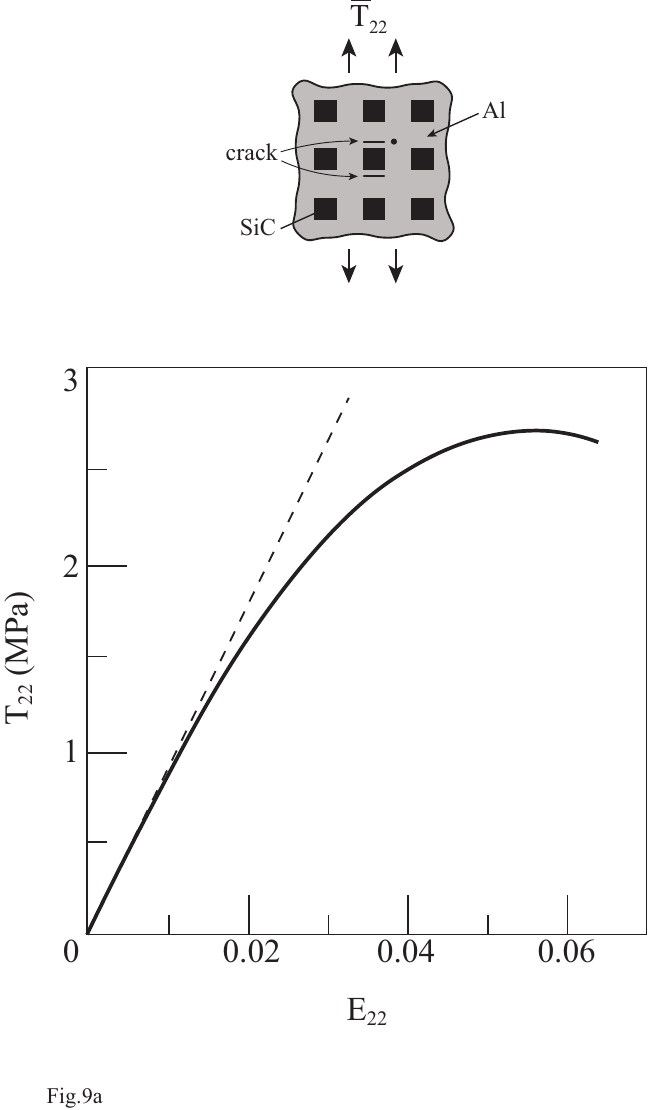} \\
\includegraphics[scale=0.5,trim={0.5cm 1cm 0cm 0.5cm},clip]{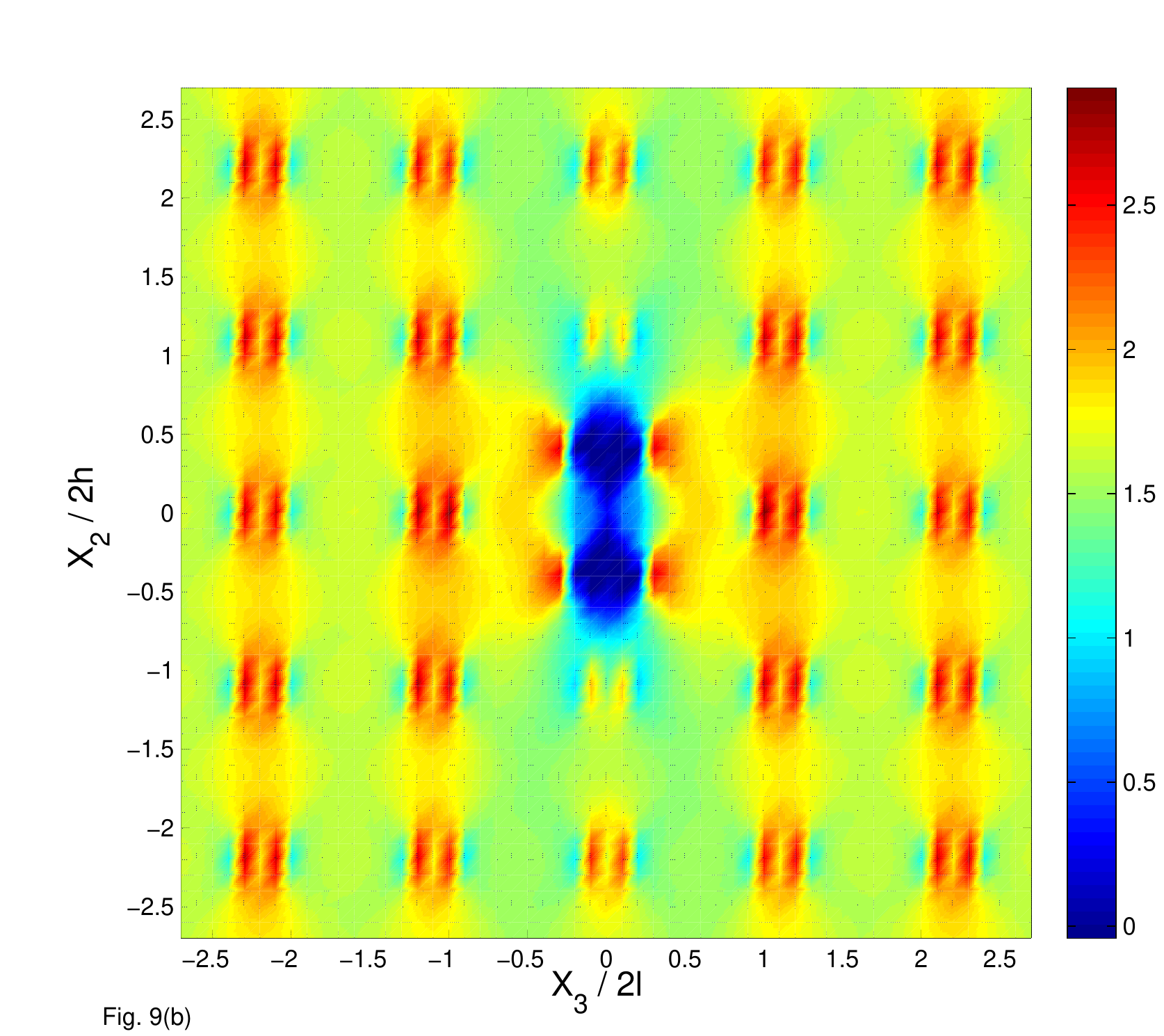} \\
\includegraphics[scale=0.5,trim={0.5cm 1cm 0cm 0cm},clip]{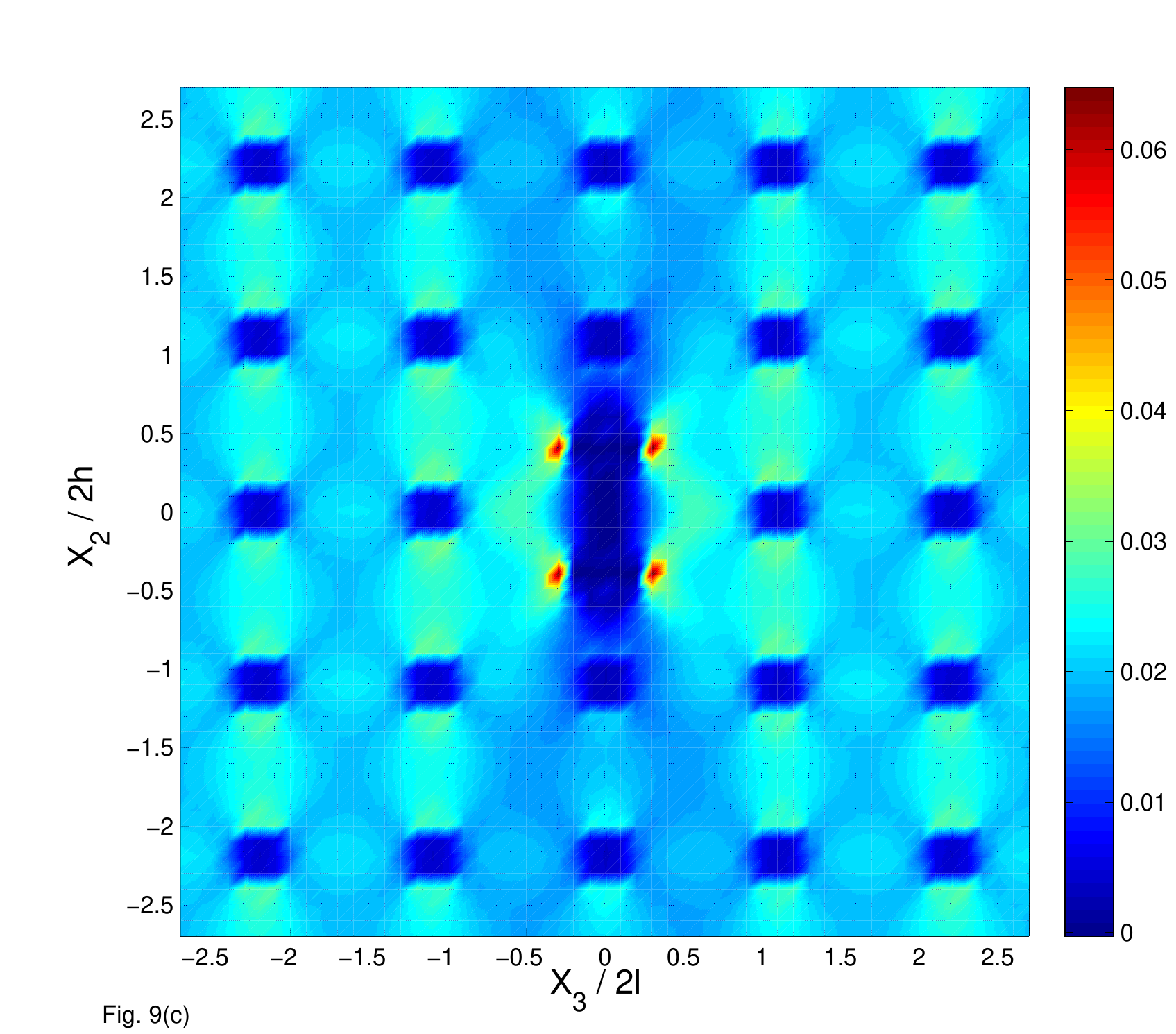}}

\end{center}
  \begin{picture}(0,0)
  \put(203,410){$T_{22}$ (MPa)}
  \put(212,207){$E_{22}$}
  \put(228,212){$X_3/2l$}
  \put(8,212){$(b)$}
  \put(8,420){$(a)$}
  \put(228,10){$X_3/2l$}
  \put(8,10){$(c)$}
  \end{picture}
\caption{  (a) `Silicon carbide'/`aluminum' composite with two ``cracks'', subjected to remote stress $\bar T_{22}$.
                      Normal finite stress-strain response at the indicated point adjacent to the upper ``crack''.
                  (b) and (c) The distribution of the normal Piola-Kirchhoff stress $T_{22}$ and large strain $E_{22}$, respectively.  }
\label{Figure9}
  \end{figure}
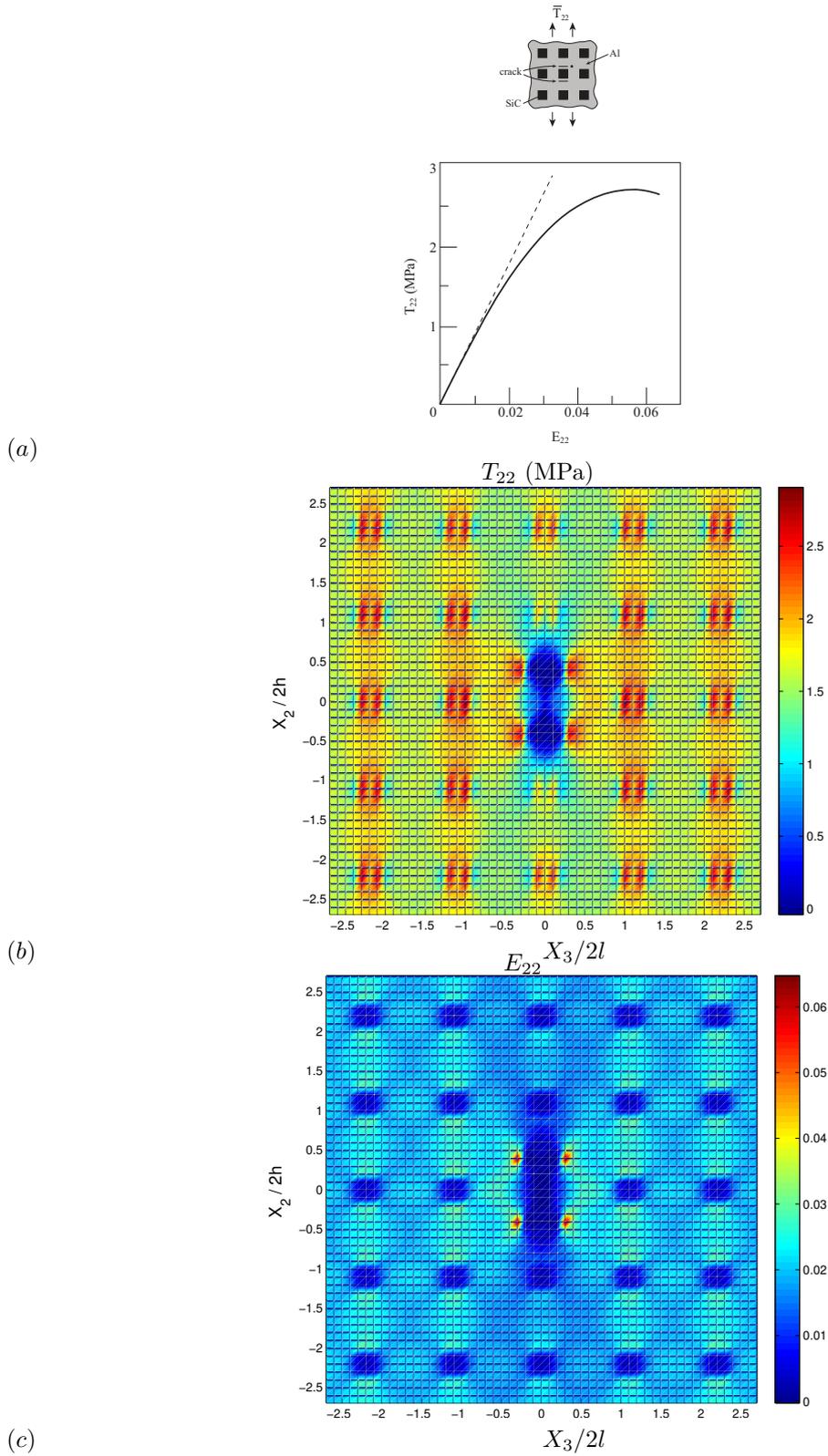

Figures \ref{Figure8}--\ref{Figure9} present the first case of a strictly composite material, consistent of two material phases described by the Murnaghan constitutive equations. It should be noted at this point that the aim of the specific examples is to illustrate how composites with hyperelastic constituents can be analyzed by employing a nonlinear solver. Thus what is needed is calibrated model of hyperelastic energy for a matrix and fiber inclusions. Since hyperelastic energy is expressed through the arbitrary measure of Lagrange quadratic strain, it is impossible to obtain a meaningful expression without either multiple scale modeling or experimental calibration. The latter is chosen here, and hence a specific material couple is taken, using the work of \cite{Chen}. However, it so happens that the materials described by the Murnaghan equation in \cite{Chen}, aluminum and silicon carbide, due not exhibit finite strain hyperelasticity, only stretching elastically to lower strains than those necessary to illustrate pronounced  nonlinearity. One possible alternative is to use the example of rubber, described by the Mooney-Rivlin equation. That was indeed executed. In order to show another example, with another constitutive equation, for the description of two-constituent composite, reference is made to \cite{Chen}. We therefore opt to use the model and numbers cited in \cite{Chen}, where the materials are designated as aluminum and silicon carbide, but imply that we understand the materials as virtual/artificial ones, describing actual natural materials at smaller strains, and behaving asymptotically consistently, but somewhat arbitrary at larger strain. We do so with the sole purpose of exemplifying the convergence of the algorithm in treating hyperelastic materials with developed nonlinearity. For simplicity, we will refer to those half-artificial materials as `aluminum' and `silicon carbide', using the quotation marks to distinguish the names from those referring to real aluminum and silicon carbide (except for in the figures themselves). 

Figure \ref{Figure8} addresses the case of localized damage manifested as a lost fiber, that is a case of a doubly-periodic fiber-reinforced bulk of material, where after production, one fiber was somehow `pulled-out' and there is void in its stead. The fiber is represented by an array of 3 by 3 subcells, positioned symmetrically in a cell. Figure \ref{Figure8}(a) reveals a less convex curve than for the Mooney-Rivlin material, which is reasonable, since metals and ceramic materials are more linear in the reversible range than polymers and rubbers (the shown curve resembles a square root in square strain, which means approximately linear strain--relative displacement dependence, unlike the nearly quadratic one for the MR material). The last remark is clearly somewhat speculative, since the calculated regimes are hardly realistic for actual metals and ceramics. However, if one refers to the constitutive models employed to describe rubber on one hand and metal/ceramic composites on the other, clearly the Mooney-Rivlin model, which is more convex in terms of force displacement dependence than the extrapolation of the Murnaghan equation -- which is more linear for small strains than the MR model, as confirmed by small-strain experiments -- to larger strains, then the aforementioned comparative judgment can be made in reference to the mathematical models inspired by the actual materials and extrapolated with a certain amount of consistency, rather than to the actual materials themselves. Figure \ref{Figure8}(b) shows a nearly periodic stress map, with maximum stresses in the fibers close to their interfaces with the matrix. Moreover, the stress concentration around the cavity is lesser, and the cavity also reduces the stress below and above it, which is reasonable. Figure \ref{Figure8}(c) is very interesting. For the applied uniaxial stress, a material with a nonzero Poisson ratio (for small strains), develops biaxial strain state. Indeed, the strain map looks like superposition of two cases, for each of which the cross-section looks like an array of parallel rods, every other of which looks like a composition of two types of rods interchanged periodically. This picture is perturbed by the cavity in the center, which obviously `releases' the strain below and above it, but shows strain concentration at the `walls' of the cavity, which take about twice the average strain, locally.

Figure \ref{Figure9} shows the classical case of a composite material with damage, namely, a doubly periodic array of `silicon carbide' fibers (the same as in the example with the `lost' fiber') in an `aluminum' matrix (both described by the Murnaghan constitutive equations), with two finite ``cracks'' stretched in the first mode, just below and above a fiber (with the same ``crack'' geometry as in the example with the porous material). A stress-strain curve near a ``crack'' tip shows high stress concentration, as expected due to the combined effect of the presence of a rigid inclusion and a ``crack''. The concavity is indicative of linearity of force--relative displacement dependence up to the maximum point. The emergence of non-monotonicity owes to the fact that the plotted stress and strain values are not invariants and the fiber and the ``crack'' induced large bi-axiality (in other words, the non monotonicity is related emergence of dominant horizontal-plane strains. The maximum reached strain of six percents is sufficient for considering the problem one of finite strain, in light of, for example, the aforementioned stress-strain dependence non-monotonicity. The stress distribution map in Fig. \ref{Figure9}(b) shows the expected periodic pattern far from the damage loci, with a stress concentration factor of about 2 at the `silicon carbide' inclusions. In addition, there is the interesting effect of `stress-screening' in the region between the ``cracks''. This is not surprising -- instead of the loading work to result in straining the material between the ``cracks'', it results in opening the ``cracks'' in the first mode, creating stress concentration at their tips.  The pattern shown in Fig. \ref{Figure9}(c) is similar, apart from the absence of strain concentration at the (relatively more) rigid inclusions. The crack-tip stress singularity regularized in a physical specimen by inelasticity, is regularized in the computational analysis by the fact that there is no crack tip at all, strictly speaking, but rather a square cavity one subcell in size. Thus the error in describing the crack tip singularity is encompassed in the error in the geometrical representation of the crack by a row of damaged subcells (Figs. \ref{Figure6d} and \ref{Figure10d}, discussed in more detail below, compare the obtained ``crack''-tip stress fields to standard K-fields of exact cracks, showing reasonable correspondence). In the same time, the fact that corner effects at, say, square-section cavities, as the one given in the example in Fig. \ref{Figure3}, where the entire cavity is represented by a relatively small number of subcells, are not very pronounced, owes to the geometry-regularizing nature of the volume-integrated PDE-solving approach followed in the present work.

\begin{figure}[tp]
\begin{center}

{\includegraphics[scale=0.62,trim={0cm 0.5cm 0cm 0cm},clip]{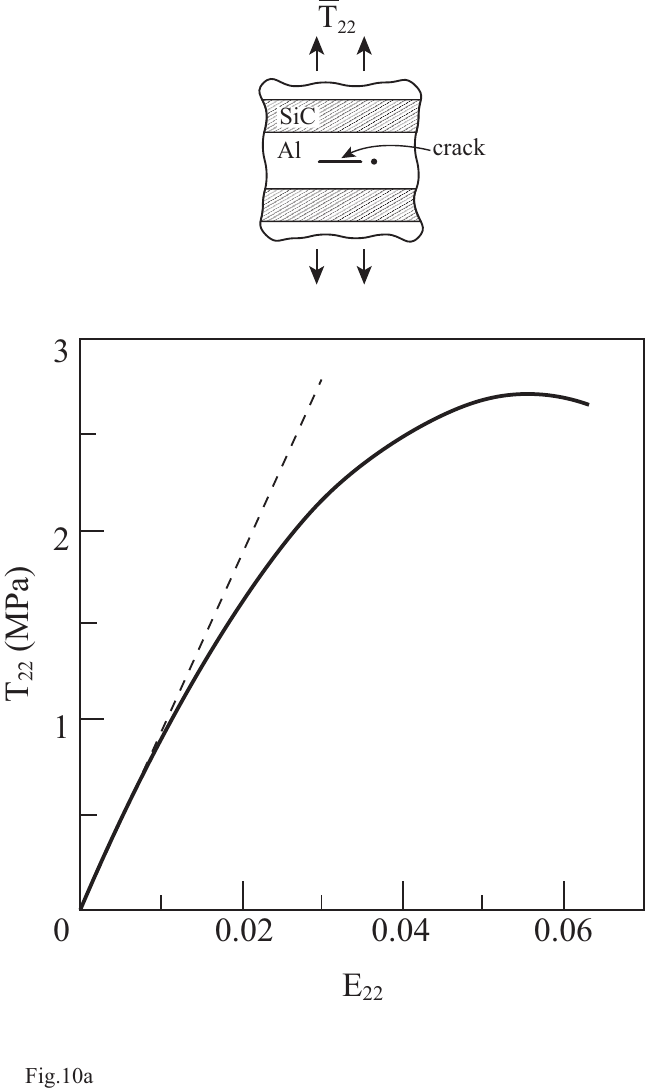} \\
\includegraphics[scale=0.5,trim={0.5cm 1cm 0cm 0.5cm},clip]{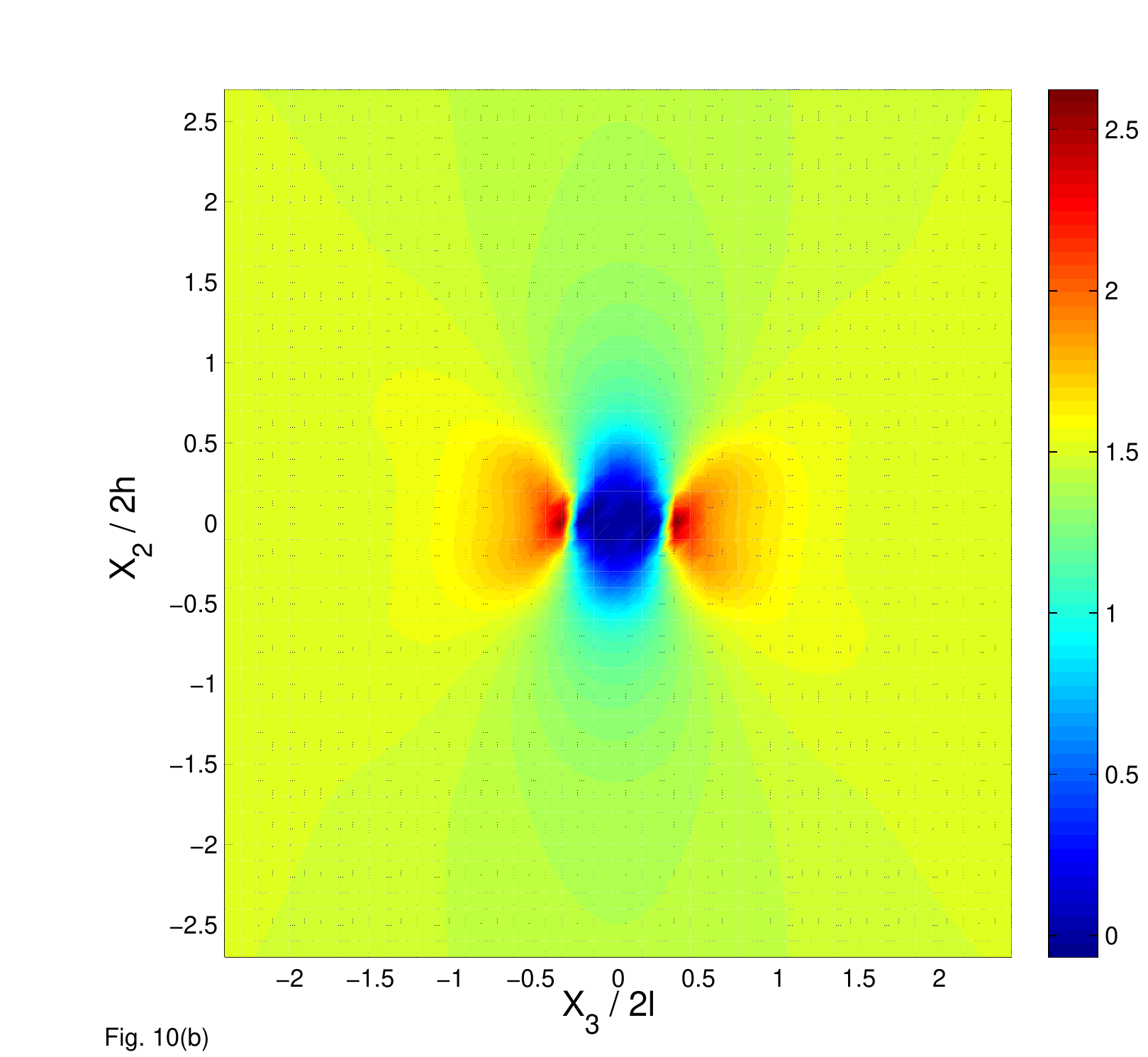} \\
\includegraphics[scale=0.5,trim={0.5cm 1cm 0cm 0cm},clip]{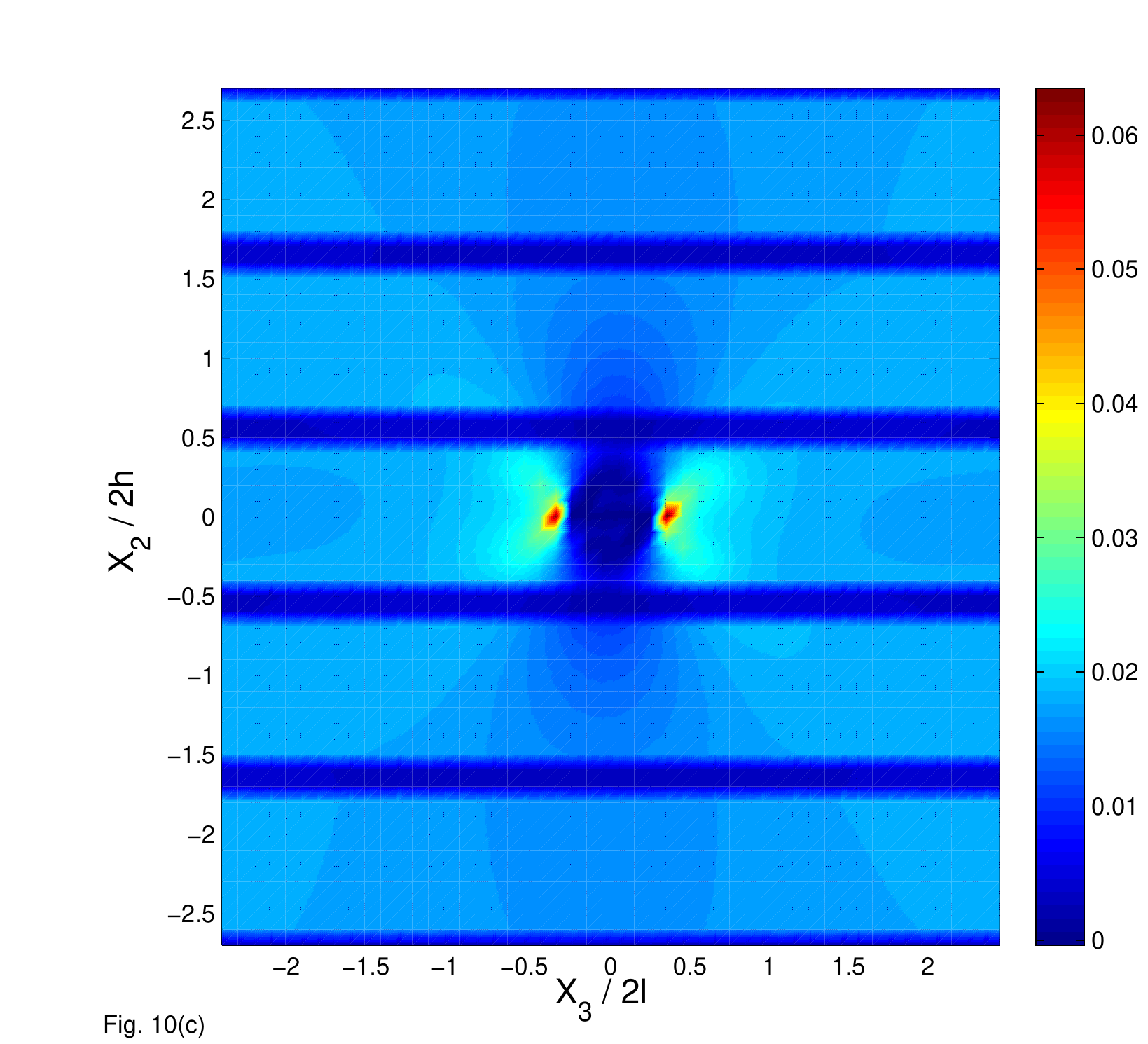}}

\end{center}
  \begin{picture}(0,0)
 \put(203,410){$T_{22}$ (MPa)}
  \put(212,207){$E_{22}$}
  \put(228,212){$X_3/2l$}
  \put(8,212){$(b)$}
  \put(8,420){$(a)$}
  \put(228,10){$X_3/2l$}
  \put(8,10){$(c)$}
  \end{picture}
\caption{ (a) A peridically bilayered `Silicon carbide'/`aluminum' composite with a ``crack'', subjected to remote stress $\bar T_{22}$.
                      Normal finite stress-strain response at the indicated point adjacent to the ``crack''.
                  (b) and (c) The distribution of the normal Piola-Kirchhoff stress $T_{22}$ and large strain $E_{22}$, respectively.   }
\label{Figure10}
  \end{figure}

\begin{figure}[tp]
\begin{center}
\includegraphics[scale=0.27,trim={0cm 1.1cm 0cm 0cm},clip]{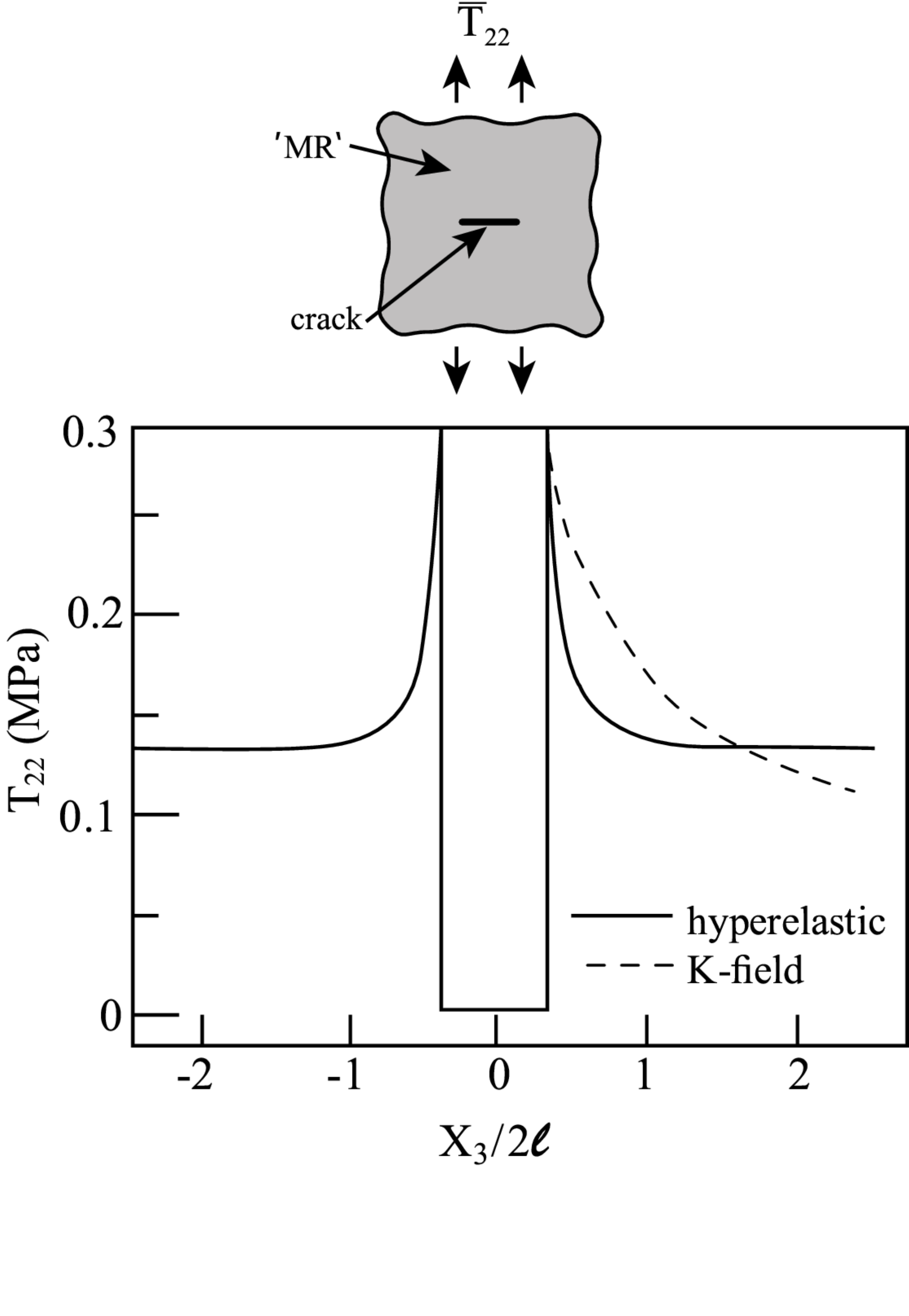}
\end{center}
\caption{The normal stress variation along the ``crack'' line. The dashed line shows the inverse square-root K-field in a linearly elastic material with a finite crack.   }
\label{Figure6d}
\end{figure}

\begin{figure}[tp]
\begin{center}
\includegraphics[scale=0.6,trim={0cm 0.75cm 0cm 0cm},clip]{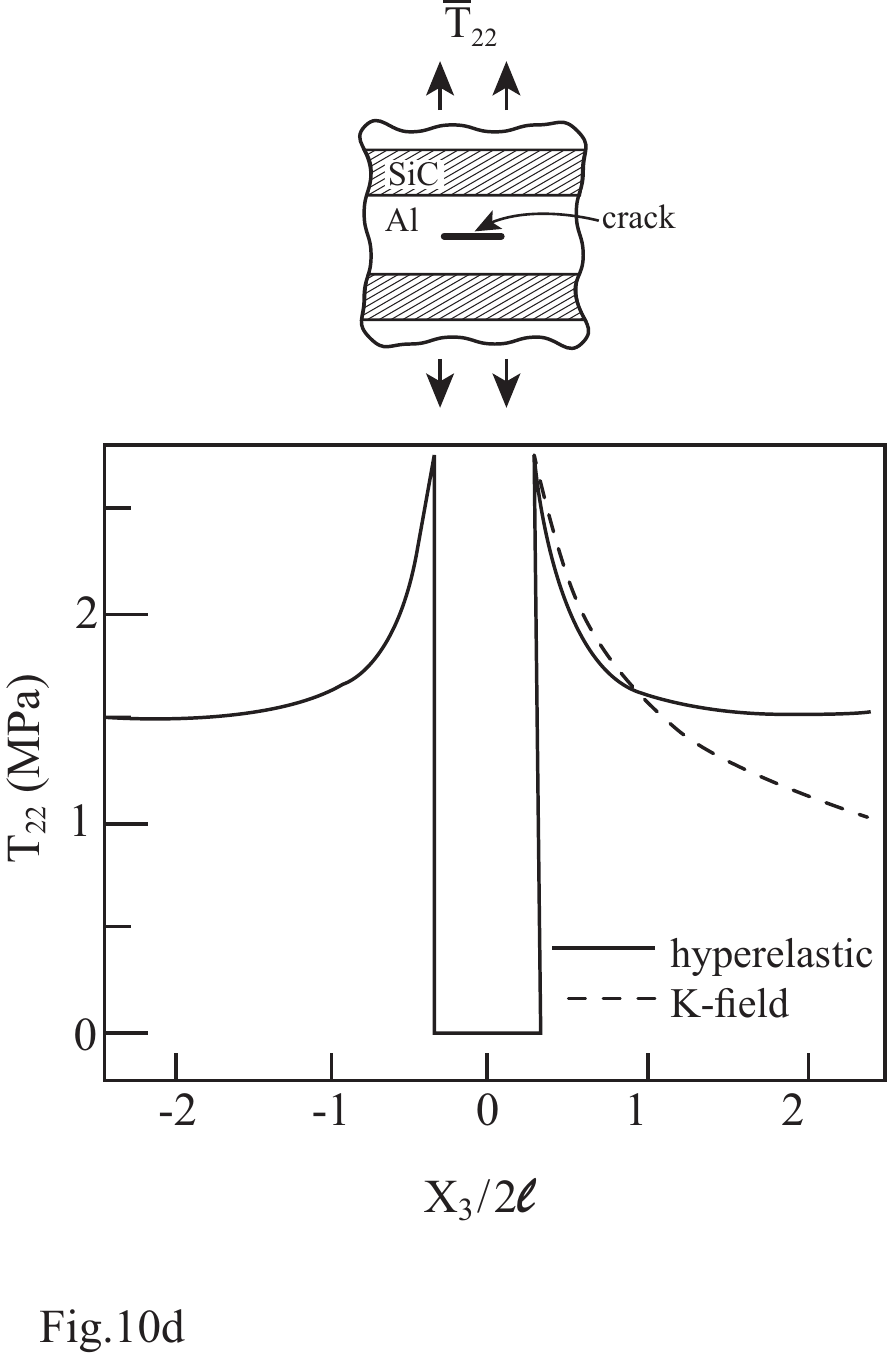}
\end{center}
\caption{The normal stress variation along the ``crack'' line. The dashed line shows the inverse square-root K-field in a linearly elastic material with a finite crack.  }
\label{Figure10d}
\end{figure}

Figure \ref{Figure10} presents the case of a laminate (rather than a fiber-reinforced matrix) of `aluminum' and `silicon carbide' described, as for the case of Figure \ref{Figure9}, by the Murnaghan constitutive equations. This time, however, instead of two ``cracks'' emergent below and above a fiber inclusion, a single ``crack'' (represented by a single horizontal line of 6 fully damaged subcells, positioned symmetrically in the central cell) is assumed in the middle of the aluminum layer occupying, symmetrically, 9 of the 11 rows of subcells in each cell. The ``crack'' is of finite length, oriented `horizontally', such that the loading opens it in the first mode. Of course, as before, the ``crack'' is prismatic, manifesting two narrow surfaces disentangling everywhere but at the edges, where they are attached tangentially. The stress-strain relation evolution near a ``crack'' tip is as for the previous case, qualitatively. The stress distribution map, as shown in Fig. \ref{Figure10}(b), is similar to the case of a finite crack in a uniform space stretched in the first mode -- the layering seems to have no effect on the stress, within the visible range. Of course, it is known that a major failure mechanism in laminates is delamination, which results from stress increase around the tips of existing micro-cracks at the laminate interfaces. However, within the idealization where aside from the explicitly-introduced ``crack'' there are no additional assumed cracks at the interfaces, it appears that due to stress equilibrium across interfaces, for sufficiently small strains, until nonlinearity kicks in, the lamination is not revealed in the stress map (clearly this is the result of the assumption of idealized interfaces between the layers). This is not surprising, since a layered composite loaded normally to the layering is nothing but a collection of generalized nonlinear springs arranged in a series, and in this case all the elements carry the same force. If the layers are homogeneous, this also means the same stress for all elements. For the middle layer, only the average stress is the same, with a non-uniform distribution, caused by the ``crack''. The stress concentration factor here has a maximum value of about 1.5, approximately the same as in the case of uniform material with a single crack. Figure  \ref{Figure10}(c) reveals the layered nature of the composite, showing a layered strain pattern, in addition to strain concentration at the ``crack'' tips. Again, the pattern is explained by analogy with an array of springs attached in a series, the elongation of which is inversely proportional to rigidity, as the more rigid `silicon carbide' exhibits less strain than the aluminum. Figure \ref{Figure10d} shows a stress profile on the ``crack'' axis, on top of the K-field -- the profile for linear material. As for the Mooney-Rivlin material, one observes here a higher level of stress localization compared to linear elastic fracture mechanics. However, the correspondence between the K-field and the Murnaghan case distribution is better than for the Mooney-Rivlin material -- one observes higher proximity between the solid and dashed curves close to the ``crack'' tip. This is unsurprising since Murnaghan materials are much more linear in terms of force-displacement relation than Mooney-Rivlin materials, as one expects metals to be, when compared to, say, rubber. The discrepancy in the `tails' of the curves in Fig. \ref{Figure10d} can, in turn, be attributed to the fact that a standard K-field does not contain the far-field stress term. The dashed curves of the K-field for Figs. \ref{Figure6d} and \ref{Figure10d} were obtained from the analytic formula for the stress around a finite-length crack tip, with the stress intensity factor determined by contour integration (J-integral) around the ``crack'' in the numerically-analyzed material. It is important to note that the K-field shown on the plots is regularized by the subcell size, since the inverse square root function starts not at zero but at an argument corresponding to distance of the order of the subcell length from the presumed ``crack'' tip. This regularization was chosen for better comparison with the hyperelastic material results. In the case of the hyperelastic calculation itself, the same regularization is present implicitly (as discussed above). The ``control'' points for collection of stress and strain histories were always located in the subcell positioned to the right of the damaged zone, vertically symmetrically with respect to the single or the upper damaged zone, one subcell to the right of the rightmost damaged subcell. Subcell-averaged values were used. The physical length dimension scales out in the elastostatic regime assumed here, but the presumed order of magnitude is 1mm for the cell length.

One last remark that should be made at this point concerns the issue of postprocessing, as manifested in the presented colormaps in Figs. \ref{Figure3}-\ref{Figure9}. The colormaps were produced in Matlab, using default plotting settings, one feature of which is smoothing interpolation, helpful for appreciation of asymptotic approach of the sought fields with resolution even before `infinite' resolution is reached. A shortcoming of such graphical smoothing interpolation is that it hides jumps of certain fields, such as normal strain, across interfaces between materials with different elastic coefficients. In the presented examples it was decided to favor smoothing for better appearance of fields distributions, on the account of exact representation of jumps across interfaces, having in mind that the distribution of distinct phases in the composite is rather regular (an opposite choice might have been made for random distribution of fibers. Here, the subcells belonging to the distinct phases are easy to spot and count. Thus it should be held in mind that if subcells adjacent to pore subcells, for instance, appear to have zero strain just as pore subcells themselves, it is only due to the graphical smoothing interpolation. Also, in regard with the pores and ``cracks'' being characterized by zero strain, it should be noted that the composite was solved as a simply-connected continuum, with points being characterized by damage-controlled zero stiffness, hence it was decided not to exclude the pores from the presentation, for self-consistence. The zero strain set in the pores helps to avoid presentation of large strain formally developed there due to vanishing stiffness and conditions of continuity. Such large strain would have skewed the range of the colormaps. Exclusion of the pore subcells at postprocessing could be a possible alternative, though it would conceal the idea of the method to model the material as simply-connected and undamaged with superimposed damage accounted for iteratively. In any case, the resulting ranges of the colormaps do not hide any special features weakly-distinct in color.

A final point regarding the colormaps is that in certain cases the fields appear to have slightly broken symmetry. This occurs due to the nonsmooth definitions entailed in the smoothing interpolation algorithms within the graphical software, activated by numerical round-off errors. The raw results do in fact respect the symmetry, and the aforementioned postprocessing feature is diminished with increased computational resolution. Thus, hopefully, the issue should bear limited significance on the comprehension of the results, as long as the said understanding is held in mind.

\section{Conclusions}
\label{sec:5}
This paper presents a computational method for multiscale analysis of hyperelastic composites with localized damage. A doubly-periodic composite with hyperelastic constituents is assumed, augmented by the presence of damage in a localized region, typically in a representative cell of the periodic medium. The analysis involves homogenization for derivation of `far' fields. A second step consists of decomposing the first Piola-Kirchhoff stress tensor into a uniform linear part and the remainder, comprised of damage-affected and purely-nonlinear terms. The values of the remainder terms are assumed to be zero and then corrected iteratively. A spatial double Fourier transform is employed and a linear mechanical problem with a predictor inhomogeneous part is then solved in Fourier space, separately for each Fourier harmonica. These emergent linear systems are of size approximately equal to the number of subcells discretizing a representative cell of the periodic composite. The independence between the different Fourier problems permits the use of parallel computing, if memory capabilities allow it. Alternatively, the same processor can be used, with the corresponding increase in CPU. Trade-off between memory and CPU can be made by use of programming optimization. The possibility for such optimization is a noteworthy feature of the method, based on its resorting to Fourier transform application, making it comparable in this sense to more standard approaches, such as the Finite Element method, for example, in which parallel computing can be employed for the stiffness matrix construction. A key step in the solution is the convergence of the iterative sequence of solutions of linear sub-problems. This convergence is obtained naturally, by use of a contracting Banach mapping, resulting from the stress decomposition, in the case of composites with linear-elastic constituents, and is lost for hyperelastic ones. For the latter case, instead of Banach's mapping, a general numerical nonlinear (Quasi-Newton) solver is employed (to be specific, the Good Broyden step algorithm of the Quasi-Newton family). Moreover, since the iterating vector is rather large, due to resolving two spatial scales, a low-memory version of the Good Broyden method is employed, in order to avoid storing in memory the elements of a large square matrix. The most economic low-memory algorithms require storing one vector for each iteration. The present paper, among other things, suggests such an algorithm, the noteworthy property of which is that the vectors to be stored in the memory are the values of the vector functionthe root of which is sought. For the case where such function values are related to stress imbalances in subcells discretizing the domain of a damaged periodic composite, storing the function values directly gives higher convergence control, as one can require convergence specifically or more exclusively at points of stress concentration.

The method of analysis thoroughly described and discussed in the first part of the paper is then applied to specific examples in its second part. Multiple cases of damaged composites are examined, with several materials, periodicity patterns and damage types. Interesting insights are obtained in regard with stress concentration in hyperelastic media. Compared to other approaches, beyond the already discussed features, one can also acknowledge finer points, as raised and discussed for example in \cite{Ch14} (among other things, the issue of subcell size, in view of the interpolation within the subcell being of higher order than in methods resorting to spatial linearization between computational grid points).

To conclude, the suggested approach can be used for quasi-static analysis of loaded periodic composites with solid hyperelastic constituents and spatially-localized damage, when an alternative may be in place to standard methods, such as, for example, the Finite Element method, which requires the use of loading increments sufficiently small for local monotonicity to hold, whereas the proposed method is suitable for increments within which the equations can be non-monotonic. This means that larger increments can be taken, in principle, without creating a problem of convergence to local minima, due to the present method being a strong-form approach to integration of PDEs, and once a solution was obtained, no further validation of convergence with smaller increments is required.

In regard with geometric flexibility in describing composite constituents of complex or non-periodic geometry, it should be said that there is room for further development, which the authors intend to take on in the future. One possible extension can be the combination of the employed iterative approach with non-cartesian meshing, such as the one employed in the Parametric HFGMC algorithm \cite{Ch14}. In any case, the enhancement of the approach of the HFGMC with the Higher Order Theory in analysis of periodic solid composites with localized damage allowing to account for nonlinear elasticity of the constituent phases appears to be of value to the development of this take on computational mechanics, which has its historical account.

%


\bibliographystyle{spmpsci}      


\renewcommand{\theequation}{{A.}\arabic{equation}}
\setcounter{equation}{0}
\setcounter{section}{0}

\gdef\thesection{Appendix \Alph{section}. }
\section{Derivation of the Quasi-Newton step} 
\label{AppendixA}

The exact Newton step uses current spatial directions, whereas the Quasi-Newton method uses the history of vectors tangential to the solution trajectory. For high-dimensional spaces and good initial guesses, the Quasi-Newton method is much cheaper (and also non-singular). Of course, using a smaller number (than $d$) of linearly-independent yet not orthogonal vectors one can never obtain the exact Jacobian inverse, and this has an effect on the stability of the fixed-point iteration mapping. Taking the gradient of the fixed-point in Eq. (\ref{B3}) for the exact Newton step and employing Eq. (\ref{B4}), we get:
\begin{equation} 
\label{B7}
\begin{split}
\left\lbrace\nabla[\mathcal{G}(\textbf{x})]^{\top}\right\rbrace^{\top}=-\sum_{n=1}^{d}{\left\lbrace (\textbf{f}^{\top}\textbf{v}_n)(\nabla\textbf{u}_n^{\top})^{\top}+(\textbf{u}_n\textbf{f}^{\top})(\nabla\textbf{v}_n^{\top})^{\top} \right\rbrace}
\end{split}
\end{equation}

For $\Vert\textbf{f}\Vert$ sufficiently small already at the starting point, the mapping in Eq. (\ref{B3}) will stay contracting.
On the other hand, if instead of using local orthogonal spatial finite-difference vectors, one uses a set of non-orthogonal non-local steps and function differences tangential to the solution trajectory, then the associated Quasi-Newton matrix could be expressed as:
\begin{eqnarray} \label{B8}
\ \ \ \ \ \ \tilde{\textbf{B}}_k=\tilde{\textbf{B}}_{n_0}+\sum_{n=n_0}^{k-1} \tilde{\textbf{u}}_n\tilde{\textbf{v}}_n^{\top}
\end{eqnarray}

Here it is assumed that an approximation of the inverse Jacobian exists at a given state, and then is updated using accumulated information, which is essentially the update in the function vector the root of which is sought, and as one would expect the matrix is updated by a new information encompassed in a vector in the only possible way, namely the addition of the dyadic product, or an outer product between two vectors, one of which depends on $\textbf{f}$ and the other on previous information and is only needed for constructing a matrix. The current Quasi-Newton matrix then comprises a number of such dyadic products -- this number should not be too large, since the approximation of the Jacobian should still be at least somewhat local.

If at a given iteration this decomposition is used, then at the next iteration, naturally a single dyadic product is added. This is a basic proposition of the explicit Quasi-Newton approach, which is employed here. Consequently, the update formula for the matrix approximating the inverse Jacobian would be:
\begin{eqnarray} \label{B9}
\ \ \ \ \ \ \ \ \ \ \ \tilde{\textbf{B}}_{k+1}=\tilde{\textbf{B}}_{k}+ \tilde{\textbf{u}}_k\tilde{\textbf{v}}_k^{\top}
\end{eqnarray}  

Taking the gradient of the Quasi-Newton step then yields:

\begin{equation} 
\begin{split}
\label{B10}
\left\lbrace\nabla[\tilde{\mathcal{G}}(\textbf{x})]^{\top}\right\rbrace^{\top}= \textbf{I}-\tilde{\textbf{B}}_{k}(\nabla\textbf{f}^{\top})^{\top} -\sum_{n=n_0}^{n_0+M}{\left\lbrace (\textbf{f}^{\top}\tilde{\textbf{v}}_n)(\nabla\tilde{\textbf{u}}_n^{\top})^{\top}+(\tilde{\textbf{u}}_n\textbf{f}^{\top})(\nabla\tilde{\textbf{v}}_n^{\top})^{\top} \right\rbrace}
\end{split}
\end{equation}

Next we define the step and the function change (decrease) as
\begin{eqnarray} \label{B11}
\ \ \ \ \ \ \ \ \ \ \ \textbf{s}_k\triangleq \textbf{x}_k-\textbf{x}_{k-1}, \ \textbf{y}_k\triangleq \textbf{f}_k-\textbf{f}_{k-1}
\end{eqnarray}

Multiplying Eq. (\ref{B10}) by the step and taking a 2-norm of the equation, and recalling Minkowski's inequality for $p=2$ (or the triangle inequality for a norm, which results from the Cauchy-Schwartz inequality), we get the following estimate for the Jacobian of the mapping: 
\begin{eqnarray} \label{B12}
\begin{split}
 \ \ \left\Vert\left\lbrace\nabla[\tilde{\mathcal{G}}(\textbf{x})]^{\top}\right\rbrace^{\top}\textbf{s}\right\Vert_2 \le \left\Vert\textbf{s}-\tilde{\textbf{B}}(\nabla\textbf{f}^{\top})^{\top}\textbf{s}\right\Vert_2+
\left\Vert \sum_{n=n_0}^{n_0+M}{\left\lbrace (\textbf{f}^{\top}\tilde{\textbf{v}}_n)(\nabla\tilde{\textbf{u}}_n^{\top})^{\top}+(\tilde{\textbf{u}}_n\textbf{f}^{\top})(\nabla\tilde{\textbf{v}}_n^{\top})^{\top} \right\rbrace}\textbf{s} \right\Vert_2
\end{split}
\end{eqnarray}

Next, under the reasonable assumption that the nonlinear system to be solved is smooth, one can acknowledge that at the solution, denoted by a star subscript, one has
\begin{eqnarray} \label{B13}
(\nabla\textbf{f}^{\top})^{\top}_*\textbf{s}_*=\textbf{y}_* \ , \ \left\lbrace\nabla[\tilde{\mathcal{G}}(\textbf{x})]^{\top}\right\rbrace^{\top}_*\textbf{s}_*=\tilde{\mathcal{G}}(\textbf{x}_*)-\textbf{x}_*
\end{eqnarray}
which yields the relation
\begin{eqnarray} \label{B14}
\ \ \ \ \ \ \ \ \ \ \ \ \left\Vert \tilde{\mathcal{G}}(\textbf{x}_*)-\textbf{x}_*\right\Vert_2\le\left\Vert\textbf{s}_*-\tilde{\textbf{B}}_{*}\textbf{y}_*\right\Vert_2
\end{eqnarray}

A natural definition for a stable fixed-point iteration mapping is one stating that at the fixed-point, the deviation vector from the fixed-point that the application of the mapping produces would be zero. A good stability condition is thus 
\begin{eqnarray} \label{B15}
\ \ \ \ \ \ \tilde{\mathcal{G}}(\textbf{x}_*)-\textbf{x}_*=\textbf{s}_*-\tilde{\textbf{B}}_{*}\textbf{y}_*=0\Rightarrow  \tilde{\textbf{B}}_{*}\textbf{y}_*=\textbf{s}_*
\end{eqnarray}

The straightforward way to satisfy this stability condition using only current and previous information without having the exact inverse Jacobian at hand, is to require that the relation holds not only at the solution but rather always. This would be an unnecessary but a sufficient condition. Moreover, this condition would give a constraint on the Quasi-Newton matrix at every iteration. This condition has the following form:
\begin{eqnarray} \label{B16}
\ \ \ \ \ \ \ \ \ \ \ \ \ \ \tilde{\textbf{B}}_k\textbf{y}_k=\textbf{s}_k \, \ \forall \ \  k >n_0
\end{eqnarray}

Clearly, another way of satisfying Eq. (\ref{B15}) is by using the exact Newton matrix $\textbf{B}=(\nabla\textbf{f}^{\top})^{-\top}$. An interesting feature of the exact Newton step, as can be learned from Eq. (\ref{B7}), is that for linear systems, convergence is guaranteed for any starting point. For Quasi-Newton methods, this is not true.

The condition in Eq. (\ref{B16}) guarantees convergence for a sufficiently good starting point and constraints the Quasi-Newton matrix. Substituting Eq. (\ref{B9}) into Eq. (\ref{B16}), and recalling Eqs. (\ref{B3}) and (\ref{B11}), one obtains a class of update formulas for the Quasi-Newton matrix:
\begin{equation} 
\label{B17}
\begin{split}
\tilde{\textbf{B}}_{k}=\tilde{\textbf{B}}_{k-1}+\frac{\textbf{s}_k-\tilde{\textbf{B}}_{k-1}\textbf{y}_k}{\textbf{y}_{k}^{\top}{\tilde{\textbf{v}}_{k-1}}}\tilde{\textbf{v}}_{k-1}^{\top}= \tilde{\textbf{B}}_{k-1}\left(\textbf{I}-\frac{\textbf{f}_{k}{\tilde{\textbf{v}}_{k-1}^{\top}}}{\textbf{y}_{k}^{\top}{\tilde{\textbf{v}}_{k-1}}}\right),\ \forall \ \  k > n_0 
\end{split}
\end{equation}
where  $\textbf{I}$ is the identity matrix of dimensions $d\times d$.

This is just the Quasi-Newton result, based on the fact that the matrix is not the exact inverse Jacobian and that it is constructed only from the solution history without performing local differentiation in various linearly-independent directions. The Quasi-Newton idea is to construct the approximation for the inverse Jacobian using information from iterations rather than dimension directions. The number of iterations used for this purpose should be smaller than the dimensionality of the problem for memory efficiency, however large enough for a good approximation of the matrix and consequently good convergence rate.

\renewcommand{\theequation}{{B.}\arabic{equation}}
\setcounter{equation}{0}

\gdef\thesection{Appendix \Alph{section}. }
\section{The Bad and the Good Broyden Methods} 
\label{AppendixB}

In order to be able to collect information from many solution steps for updating the Quasi-Newton matrix and using it for constructing the steps themselves, it is essential that the matrix will not change too sharply during the iterations, or else it would not have `time' to act in its `good' form. Clearly a single step does not provide sufficient information for approximation of the inverse Jacobian, so one would not want the matrix to change too much during a single iteration, since then it would not have a plateau, where it already approximates the inverse Jacobian well enough and can act in this capacity for significant amount of time to provide high convergence rate. 

Therefore, it would be advisable to add an additional constraint for determining the update of the Quasi-Newton matrix. This constraint will make sure that the change in the matrix during one iteration is small enough. Since Eq. (\ref{B17}) has a free vector, it would be natural to define the aforementioned constraint as a minimization problem of (some measure of) the change of the Quasi-Newton matrix during one iteration with respect to the free vector $\tilde{\textbf{v}}_{k-1}$. 

Now, if we combine Eqs. (\ref{B3}), (\ref{B11}) and (\ref{B16}), we get the following relation:.
\begin{equation} 
\label{B18}
\begin{split}
\textbf{s}_k=\textbf{x}_{k}-\textbf{x}_{k-1}=-\tilde{\textbf{B}}_{k-1}\textbf{f}_{k-1}=
\tilde{\textbf{B}}_{k}\textbf{y}_k= \tilde{\textbf{B}}_{k}\textbf{f}_{k}-\tilde{\textbf{B}}_{k}\textbf{f}_{k-1}\Rightarrow\left\lvert\left(\tilde{\textbf{B}}_{k}
-\tilde{\textbf{B}}_{k-1}\right)\textbf{f}_{k-1} \right\rvert=|\textbf{s}_{k+1}|
\end{split}
\end{equation}

This can be further reduced to 
\begin{eqnarray} \label{B19}
\ \ \ \ \ \ \ \ \ \ \left\Vert\tilde{\textbf{B}}_{k}-\tilde{\textbf{B}}_{k-1}
\right\Vert_{F}\ge\frac{|\textbf{s}_{k+1}|}{|\textbf{f}_{k-1}|} \ , \ \ \forall \ k  >n_0
\end{eqnarray}
where $\Vert\cdot\Vert_{F}$ is the Frobenius norm. On the other hand, if one first multiplies the last relation in Eq. (\ref{B18}) by $\tilde{\textbf{B}}_{k}^{-1}$, and then rearranges, one obtains the following equation:
\begin{eqnarray} \label{B20}
\ \ \ \ \ \ \ \ \ \ \left\Vert\tilde{\textbf{B}}_{k}^{-1}-\tilde{\textbf{B}}_{k-1}^{-1}
\right\Vert_{F}\ge\frac{|\textbf{f}_{k}|}{|\textbf{s}_{k}|} \ , \ \ \forall \ k  >n_0
\end{eqnarray}

Clearly the Frobenius norms in Eqs. (\ref{B19}) and (\ref{B20}) are feasible measures of how small the change per iteration is in the Quasi-Newton matrix. Obviously we cannot make these norms arbitrarily small only by varying $\textbf{v}_{k-1}$. The next best thing we can do, however, is to \emph{minimize} one of the aforementioned norms. Now, since Eqs. (\ref{B17}), (\ref{B19}) and (\ref{B20}) should all hold for a convex set of $k$ values, the optimization problem we would have in hand is a discrete variational minimization problem. The crucial idea is that we do not just set $\textbf{v}_{k-1}$ to obtain the optimum value for the left-hand side of Eqs. (\ref{B19}), (\ref{B20}) for some value of $k$, but rather is that we need to find all values $\textbf{v}_{m}$ for $n_0\le m\le k-1$, such that either $\left\Vert\tilde{\textbf{B}}_{m+1}-\tilde{\textbf{B}}_{m}
\right\Vert_{F}$ or $\left\Vert\tilde{\textbf{B}}_{m+1}^{-1}-\tilde{\textbf{B}}_{m}^{-1}
\right\Vert_{F}$ would be minimal for $n_0\le m\le k-1$.

Now, only due to the fact that this minimization is a (discrete) variational one, and it would hold for the entire history, Eqs. (\ref{B19}), (\ref{B20}) are valid. Had we only optimized for the current $\textbf{v}_{k-1}$, we could not have expected the minimization of the left-hand side in the aforementioned equations to affect the right-hand side, which locally does not depend on $\textbf{v}_{k-1}$. However, the solution to this seeming contradiction is indeed the understanding that the minimization is a (discrete) variational one, affecting the entire discrete solution trajectory (sequence), and thus the aforementioned right-hand sides are affected by $\textbf{v}_{m<k-1}$, and no contradiction arises. In practice, this understanding would be manifested in explicit update formulas for $\textbf{v}_{m}$, active for all the range $n_0\le m\le k-1$. We will next perform the aforementioned (discrete) variational minimization of the two Frobenius norms in question, analytically.

Substituting Eq. (\ref{B17}) into the left-hand side of Eq. (\ref{B19}), squaring, recalling that $\Vert\textbf{M}\Vert_F^2=\text{tr}(\textbf{M}^{\top}\textbf{M})$, and taking the variation, we get:
\begin{equation} 
\label{B21}
\begin{split}
\mathcal{N}_1\triangleq\left\Vert\tilde{\textbf{B}}_{k}-\tilde{\textbf{B}}_{k-1}
\right\Vert_{F}^2=\left|\textbf{s}_k-\tilde{\textbf{B}}_{k-1}\textbf{y}_k\right|^2\frac{\tilde{\textbf{v}}_{k-1}^{\top}\tilde{\textbf{v}}_{k-1}}{(\tilde{\textbf{v}}_{k-1}^{\top}{\textbf{y}_{k}})^2}  \Rightarrow  \delta\mathcal{N}_1 =2\mathcal{N}_1\tilde{\textbf{v}}_{k-1}^{\top} \frac{\textbf{y}_{k}\tilde{\textbf{v}}_{k-1}^{\top}-\tilde{\textbf{v}}_{k-1}\textbf{y}_{k}^{\top}}{\tilde{\textbf{v}}_{k-1}^{\top}{\textbf{y}_{k}}\tilde{\textbf{v}}_{k-1}^{\top}{\tilde{\textbf{v}}_{k-1}}}
\delta\tilde{\textbf{v}}_{k-1}
\end{split}
\end{equation}

Clearly, the only general root of the variation in Eq. (\ref{B21}) is $\tilde{\textbf{v}}_{k-1}^*=\textbf{y}_{k}$, which corresponds to the variational extremum. The second variation at the extremum then becomes:
\begin{equation} 
\label{B22}
\begin{split}
\ \ \ \ \delta^2\mathcal{N}_1^* =\frac{2\left|\textbf{s}_k-\tilde{\textbf{B}}_{k-1}\textbf{y}_k\right|^2}{(\textbf{y}_{k}^{\top}\textbf{y}_{k})^3}  \left[ (\textbf{y}_{k}^{\top}\textbf{y}_{k})(\delta\tilde{\textbf{v}}_{k-1}^{\top}{\delta\tilde{\textbf{v}}_{k-1}}) -(\textbf{y}_{k}^{\top}\delta\tilde{\textbf{v}}_{k-1})^2 \right]
\end{split}
\end{equation}
which is positive (for an arbitrary small variation) due to the Cauchy-Schwartz inequality.
Consequently, $\tilde{\textbf{v}}_{k-1}^*=\textbf{y}_{k}$ corresponds to the minimum of the squared Frobenius norms of the single-iteration update of the Quasi-Newton matrix. A minimum of a positive function coincides with the minimum of its square. Therefore the right-hand side of Eq. (\ref{B19}) can be minimized (as its supremum is being minimized) by the choice $\tilde{\textbf{v}}_{k-1}^*=\textbf{y}_{k}$. Moreover, since only one general extremum was found and the bound for the variable vector trajectory in question is zero, which is infeasible (since Eq. (\ref{B17}) places a unity-norm constraint on $\textbf{v}$), the obtained local minimum is also the global one.

Now, one notes that the minimum of $|\textbf{s}_{k+1}|/|\textbf{f}_{k-1}|$, as the minimum of any quotient, generally corresponds to maximizing the denominator and minimizing the numerator. This means that the obtained choice of the Quasi-Newton matrix would maximize the norm of the nonlinear equations the root of which is sought and minimize successive steps of the solution. This is clearly a bad idea. Even for a decrease in the norm of the equations, the step would decrease more rapidly than the solution attainment. This means that the solution may converge to a point which is not a root and get stuck there. And this is even if there is a way to minimize the norm of the equations by minimizing the ratio, which is questionable. Therefore minimizing the norm in Eq. (\ref{B19}) seems inadvisable. Indeed this approach is known as the `bad' Broyden method. It is suggested in \cite{Broyden1965}, although without the argumentation revealed here.
Let us see what can be obtained by working with Eq. (\ref{B20}) instead.

First, Eq. (\ref{B17}) can be inverted using the Sherman-Morrison (or the Householder) formula. Substitution in Eq. (\ref{B21}), while using Eq. (\ref{B3}) and (\ref{B11}), then yields:
\begin{equation} 
\label{B23}
\begin{split}
\ \ \ \mathcal{N}_2\triangleq\left\Vert\tilde{\textbf{B}}_{k}^{-1}-\tilde{\textbf{B}}_{k-1}^{-1}
\right\Vert_{F}^2=\left|\textbf{f}_k\right|^2\frac{\hat{\textbf{v}}_{k-1}^{\top}\hat{\textbf{v}}_{k-1}}{(\hat{\textbf{v}}_{k-1}^{\top}{\textbf{s}_{k}})^2}  \Rightarrow  \delta\mathcal{N}_2 =2\mathcal{N}_2\hat{\textbf{v}}_{k-1}^{\top} \frac{\textbf{s}_{k}\hat{\textbf{v}}_{k-1}^{\top}-\hat{\textbf{v}}_{k-1}\textbf{s}_{k}^{\top}}{\hat{\textbf{v}}_{k-1}^{\top}{\textbf{s}_{k}}\hat{\textbf{v}}_{k-1}^{\top}{\hat{\textbf{v}}_{k-1}}}
\delta\hat{\textbf{v}}_{k-1}
\end{split}
\end{equation}
where the independent vector trajectory 
(sequence) to be optimized for was replaced using the following substitution:
\begin{eqnarray} \label{B23b}
\ \ \ \ \ \ \ \ \ \ \ \ \ \ \hat{\textbf{v}}_{k-1}\triangleq\tilde{\textbf{B}}_{k-1}^{-\top}\tilde{\textbf{v}}_{k-1}
\end{eqnarray}
(the matrix multiplier with the simultaneous inversion and transposition operator on the right-hand side is constant in terms of the optimization as thus has zero variance).

As before, clearly, the only general root of the variation in Eq. (\ref{B23}) is $\hat{\textbf{v}}_{k-1}^{**}=\textbf{s}_{k}$, which corresponds to the variational extremum. The second variation at the extremum then becomes:
\begin{equation} 
\label{B24}
\begin{split}
\ \ \ \ \ \ \ \delta^2\mathcal{N}_2^{**} =\frac{2\left|\textbf{f}_k\right|^2}{(\textbf{s}_{k}^{\top}\textbf{s}_{k})^3}\left[ (\textbf{s}_{k}^{\top}\textbf{s}_{k})(\delta\hat{\textbf{v}}_{k-1}^{\top}{\delta\hat{\textbf{v}}_{k-1}})  -(\textbf{s}_{k}^{\top}\delta\hat{\textbf{v}}_{k-1})^2 \right]
\end{split}
\end{equation}

Again, as before, this quantity is positive (for an arbitrary small variation) due to the Cauchy-Schwartz inequality, subjected to the assumption that the Quasi-Newton step is of non-zero length.

Consequently, the choice $\hat{\textbf{v}}_{k-1}^{**}=\textbf{s}_{k}$ corresponds to the minimum of the squared Frobenius norms of the single-iteration update of the Quasi-Newton matrix. A minimum of a positive function coincides with the minimum of its square. Therefore the right-hand side of Eq. (\ref{B20}) can be minimized (as its supremum is minimized) by the choice 
\begin{eqnarray} \label{B24b}
\ \ \ \ \ \ \ \ \ \   \tilde{\textbf{v}}_{k-1}^{**}=\tilde{\textbf{B}}_{k-1}^{\top}\textbf{s}_{k}=-\tilde{\textbf{B}}_{k-1}^{\top}\tilde{\textbf{B}}_{k-1}\textbf{f}_{k-1}
\end{eqnarray}
(one notes that, as expected, only previous-step quantities are employed).

Moreover, for the same reasoning as before, the obtained local minimum is also the global one.

We therefore have obtained the result that the choice $\tilde{\textbf{v}}_{k-1}^{**}=\tilde{\textbf{B}}_{k-1}^{\top}\textbf{s}_{k}\ \forall \ k>n_0$, minimizes the ratio $|\textbf{f}_{k}|/|\textbf{s}_{k}|$ in each step of the solution. This is exactly the condition of fastest local relative norm decrease. In other words, this is the best result one can ask for. Now the constructed Quasi-Newton matrix, which uses only solution history, guarantees both the stability of the fixed-point at the end of the solution process, and the fastest relative equations norm decrease at every iteration (after some initial iteration). One can argue that this is the best one can do for the solution of a large system of nonlinear equations when a good starting point exists, since the choice of the free vectors is done such that the norm of the equations is minimized in the next iteration and the (backward directing) step size of the next iteration is maximized -- for fastest convergence (strictly speaking, the maximal-norm decrease and step increase would correspond to the infimum of the minimal value of the ratio of the former to the latter, but the key word is \emph{simultaneous} decrease of the equations-norm and increase of the step, for fastest approach to the solution -- this is what one seeks and this is what the minimization of the ratio supplies, within the constraint of a single vector sequence to optimize for). Indeed, this result is called the `good' Broyden method suggested in \cite{Broyden1965} and analyzed in more recent literature (although there the optimality of the method is not argued for the way it is done here). Anyhow, using the obtained result (the `good' Broyden method), the update formula for the Quasi-Newton matrix (or the Broyden matrix) becomes: 
\begin{equation} 
\label{B25}
\begin{split}
\ \ \ \tilde{\textbf{B}}_{k}=\tilde{\textbf{B}}_{k-1}-\frac{\tilde{\textbf{B}}_{k-1}\textbf{f}_{k}{{\textbf{f}}_{k-1}^{\top}\tilde{\textbf{B}}_{k-1}^{\top}\tilde{\textbf{B}}_{k-1}}}{{\textbf{f}}_{k-1}^{\top}{\tilde{\textbf{B}}_{k-1}^{\top}\tilde{\textbf{B}}_{k-1}(\textbf{f}_{k}-\textbf{f}_{k-1}})}, \ \ \ \forall  \ \ \ k > n_0 
\end{split}
\end{equation}
(where one observes the addition of a vector dependent on current function evaluation juxtaposed with a vector expressed through previous-iteration quantities only). One notes that the update only requires the previous Broyden matrix and the current (and previous) function value.

Such iterative update of the Broyden matrix requires an initial estimate. One can calculate the actual Jacobian at an initial point and invert it. If, however, calculations of this sort are too expensive even for the initial point, one can use another common route, which is performing the first solution step using the original, non-convergent mapping. In terms of the Broyden matrix, this would be equivalent to setting 
\begin{eqnarray} \label{B26}
\ \ \ \ \ \ \ \ \tilde{\textbf{B}}_{1}=-\textbf{I}
\end{eqnarray}

\renewcommand{\theequation}{{C.}\arabic{equation}}
\setcounter{equation}{0}

\gdef\thesection{Appendix \Alph{section}. }
\section{Derivation details for the Directly Function-Value Storing Low-Memory Good Broyden Method} 
\label{AppendixC}

The updating scheme in Eq. (\ref{B25}) requires sufficient memory for storing $d^2$ double type numeric values for the matrix entries, which may be a lot of information for large-scale problems.

A different approach is suggested below, where only a single matrix with $d\times k_{max}$ entries needs to be stored in the memory, $k_{max}$ being the maximum number of iterations used in the solution. This formulation is beneficial for large problems when a good initial guess exists.

First, we acknowledge that essentially what is required for updating the guess for the root is the step-size, rather than the Broyden matrix itself. Multiplying Eq. (\ref{B25}) by $-\textbf{f}_k$, the following updating formula for the step is obtained, represented as a recursion relation:
\begin{eqnarray} \label{B27}
 \ \ \ \ \ \ \ \ \ \ \  \textbf{s}_{n+1}=\frac{\textbf{s}_n^{\top}\textbf{s}_n\textbf{c}_n}{\textbf{s}_n^{\top}(\textbf{s}_n-\textbf{c}_n)}
\end{eqnarray}
(where $\textbf{c}_n\triangleq-\tilde{\textbf{B}}_{n-1}\textbf{f}_n$, but this relation is not explicitly used). Furthermore, substituting $k=n-1$ in Eq. (\ref{B25}) and multiplying it by $-\textbf{f}_{n}$ produces the updating formula for $\textbf{c}_n$: 
\begin{eqnarray} \label{B28}
 \ \ \ \ \ \ \ \ \  \textbf{c}_{n}=\textbf{d}_n^{(n-2)}+\frac{\textbf{c}_{n-1}\textbf{s}_{n-1}^{\top}\textbf{d}_n^{(n-2)}}{\textbf{s}_{n-1}^{\top}(\textbf{s}_{n-1}-\textbf{c}_{n-1})}
\end{eqnarray}
(where $\textbf{d}_n^{(n-2)}\triangleq-\tilde{\textbf{B}}_{n-2}\textbf{f}_n$, but, again, this relation is not explicitly used).

Finally,  substituting $k=n-j$ in Eq. (\ref{B25}) and multiplying it by $-\textbf{f}_{n}$ produces the updating formula for $\textbf{d}_n^{(n-2)}$ in the form of a set of recursion relations, as follows: 
\begin{eqnarray} \label{B29}
 \ \ \ \  \ \  \textbf{d}_n^{(n-j)}=\textbf{d}_n^{(n-j-1)}+\frac{\textbf{c}_{n-j}\textbf{s}_{n-j}^{\top}\textbf{d}_n^{(n-j-1)}}{\textbf{s}_{n-j}^{\top}(\textbf{s}_{n-j}-\textbf{c}_{n-j})}
\end{eqnarray}
(where $\textbf{d}_n^{(n-j)}\triangleq-\tilde{\textbf{B}}_{n-j}\textbf{f}_n$, and still, this relation is not explicitly used). Here $j$ should start at $2$ and increase up to $j=k-1$ for $n=k$.

The set of recursion relations given in Eqs. (\ref{B27})-(\ref{B29}) should be solved recursively for each iteration solution $k$, for values of $n$ starting from $k$ and decreasing up to $1,2$ or $3$, corresponding to Eqs. (\ref{B27}), (\ref{B28}) and (\ref{B29}), respectively, using Eq. (\ref{B30a}) and the following initial condition:
\begin{eqnarray} \label{B30}
 \ \ \ \ \  \ \ \ \ \ \  \textbf{d}_m^{(1)}=\textbf{f}_m, \ \forall \ m=3,..,k
\end{eqnarray}

The (recursive) solution of the system in Eqs. (\ref{B27})-(\ref{B30}) allows the calculation of the step $\textbf{s}_{k+1}$ for each iteration $k$, and the information it requires for this purpose is only the history of the values of the vector function the root of which is sought. In a matrix form, this information, which has to be kept in the memory during the iterative solution, can be encompassed in a single matrix whose columns are the consecutive values of the vector function the root of which is sought, constructed as shown in Eq. (\ref{B31}).

Thus the memory requirement of the algorithm is approximately $d\times k_{max}$ double precision numbers stored (overwritten) for each run (loading increment) of the solver.

\subsubsection*{C.1. Further modification -- decreasing CPU by avoiding recursive scalar products}

The recursive algorithm presented in Eqs. (\ref{B27})-(\ref{B30}) is a low-memory method. However, the successive function calls it implements may be high in CPU, especially due to computation of the various scalar products appearing in the aforementioned equations, which require double loops of size $d$, the number of equations, an outer loop for the components of the vectors in Eqs. (\ref{B27})-(\ref{B30}), and internal loops of the same size, for the computation of the scalar products. Normally, this might have been tolerable, however the low-memory version uses recursion, which multiplies the number of function calls considerably, employing the internal loops in each call. A natural remedy would be to avoid scalar product computations by directly updating the results of the scalar products in Eqs. (\ref{B27})-(\ref{B30}), instead of only updating the vectors and calculating the products every time. To this end, we define the following new variables:  
\begin{eqnarray} \label{B32}
 \ \ \ \ \  \alpha_n\triangleq \textbf{s}_n^{\top}\textbf{c}_n, \ \beta_n\triangleq \textbf{s}_n^{\top}\textbf{s}_n, \ \gamma_n^{(m)}\triangleq \textbf{s}_m^{\top}\textbf{d}_n^{(m-1)}
\end{eqnarray}

Consequently, the system in Eqs. (\ref{B27})-(\ref{B30}) is rewritten as shown in Eq. (\ref{B33}) and:
\begin{eqnarray} \label{B34}
 \ \ \ \ \ \ \ \textbf{c}_{n}=\textbf{d}_n^{(n-2)}+\frac{\gamma_n^{(n-1)}}{\beta_{n-1}-\alpha_{n-1}}\textbf{c}_{n-1} , \ n\ge3
\end{eqnarray}
\begin{eqnarray} \label{B35}
\ \ \ \textbf{d}_{n}^{(m)}=\textbf{d}_n^{(m-1)}+\frac{\gamma_n^{(m)}}{\beta_{m}-\alpha_{m}}\textbf{c}_{m} , \ n\ge3, \ m\ge2
\end{eqnarray}

Equation (\ref{B33}) is decoupled from the other two equations, and in itself is not required for the recursive sequence. It is therefore only needed as the update step for the next-iteration solution used in the end of the recursive function call. Thus the equation is written in terms of the `current' iteration step $k$, and not the general index $n$. The system of equations (\ref{B34})-(\ref{B35}) is relatively simple, as it allows updating the `current' quantities requiring only the current values of the vectors and scalar themselves (first-order update). Therefore this system can be written in a component-wise form, in a single `outer' loop over the component index, with all the recursive function calls happening inside this large loop, separately for each component. This is significant economy of time (complexity). The initial conditions for the vector quantities in the system are still given by Eqs. (\ref{B30a}$b$) and (\ref{B30}). 

What we need in order to implement the recursive scheme in Eqs. (\ref{B34})-(\ref{B35}), is recursive updating relations for the three scalar quantities defined above, namely, $\alpha_n,\beta_n$ and $\gamma_n^{(m)}$.

One can obtain such updating equations by taking six scalar products, corresponding to the three symmetric permutations of scalar products of the vector equations (\ref{B33}), (\ref{B34}) and (\ref{B35}).
First, one acknowledges that in order to write the aforementioned six equation in purely scalar quantities, one has to introduce three additional definitions, complementing Eq. (\ref{B32}), as follows:
\begin{eqnarray} \label{B36}
 \ \ \ \ \ \ \ \ \ \ \ \ \ \ \ \ \lambda_n\triangleq \textbf{c}_n^{\top}\textbf{c}_n, \ \mu_n^{(m)}\triangleq \textbf{c}_m^{\top}\textbf{d}_n^{(m-1)}
\end{eqnarray}

In addition to these one and two-argument recursive functions, a single three-argument function should be defined:
\begin{eqnarray} \label{B37}
 \ \ \ \ \ \ \ \ \  \ \ \ \ \ \ \ \ \  \nu_{nl}^{(m)}\triangleq [\textbf{d}_n^{(m-1)}]^{\top}\textbf{d}_l^{(m-1)}
\end{eqnarray}

Now we have six specially-defined scalar quantities, and we can write six recursion equations for them -- something we could not have done for the three first quantities alone (Eq.(\ref{B33}) itself should not be used in the recursive function calls, but a scalar equation derived from it should be used, and it is obtained by writing Eq. (\ref{B33}) with the index $n$, and taking its scalar product with itself). The six resulting explicit updating equations and initial conditions are given by Eqs. (\ref{B38})-(\ref{B44}).

One notes that this formulation requires the storage of $p^2$ numbers in memory for the sake of having initial conditions for the recursive relations, $p$ being the maximum number of iterations used by the algorithm (switching to notation more compact than $k_{\text{max}}$). This means that the total required memory is for $p N+p^2$ entries ($N$ replacing $d$ for the sake of more standard notation). As for CPU -- in implementing the proposed algorithm, there should be an outer loop for the iteration number, in which $p^2$ scalar products of $N$-dimensional vectors should be calculated in each iteration and added to existing arrays, then an inner loop on $N$, in which the vector functions presented above are called-for component-wise, such that the recursion only applies to scalars, and, of course, the`low-cost' update calls for the six scalar recursive functions. Consequently, the overall number of computations may be approximately exponential (combinatorial) in $p$, but only linear in $N$, which may be efficient for large $N$ and small enough loading increments allowing to use smaller values of $p$ (say of the order of 10). 

\subsubsection*{C.2. Iterative formulation for $p$-polynomial complexity}

The equations presented in the previous subsection lead to combinatorial (that is approximately exponential) complexity in $p$. Although for the case of convergence within several iterations the algorithm described above may be acceptable, even for efficient loading-stepping of the mechanical problem, tens of iterations may be required, for which exponential complexity may already render the procedure intractable. The reason for the exponential complexity is in the nature of the recursive formulation. A logical remedy is to interpret the system given in Eqs. (\ref{B34})-(\ref{B35}) and (\ref{B38})-(\ref{B43}) as an iterative scheme, rather than a recursive one, and iterating the values only once, starting from the initial condition, instead of calling the functions multiple times.

To this end, the scalar equations (\ref{B38})-(\ref{B43}) pose no problem. However, if the quantities $\textbf{c}_n$ and $\textbf{d}_n^{(m)}$ are treated not as functions but rather as iterated values, then they need to be stored in memory as two $p\times N$ matrices -- and then the twice as much memory economy is needed. The first step to overcome this problem is to solve system (\ref{B34})-(\ref{B35}) analytically for $\textbf{d}_n^{(m)}$, using the third initial condition in Eq. (\ref{B30}), which is possible due to linearity. Once this is done, the two equations (\ref{B34})-(\ref{B35}) can be replaced by a single vector equation, as follows:
\begin{eqnarray} \label{B45}
\ \ \ \ \ \textbf{c}_{n}=\textbf{f}_n+\sum_{m=2}^{n-1}\frac{\gamma_n^{(m)}}{\beta_{m}-\alpha_{m}}\textbf{c}_{m} , \ n\ge3
\end{eqnarray}

This system is still iterative. Although now only one quantity is iterated, namely, $\textbf{c}_n$, it still requires the storage of a $p\times N$ matrix in the memory, in addition to the already stored $p\times N$ matrix $\textbf{F}^{(k)}$. Thus the memory requirement here is $\mathcal{O}(2p N)$. In order to avoid the storage problem, one should avoid iterations. If only the last iteration is stored in the memory, the total memory requirement becomes more economic. To this end one only needs to obtain the analytical solution of the recursive relation in Eq. (\ref{B45}) for $\textbf{c}_n$. The solution is outlined in Eqs. (\ref{B49}) and (\ref{B46})-(\ref{B48}).

Equation (\ref{B45}) cannot be iterated component-wise, since the outer loop of the solver is the stepping in $k$ and the intermediate loop is on the vector dimension, $i$ (and the inner loop is on the sub-iteration index $m<k$). Consequently, one has to store the values of $\textbf{c}_m$ for all sub-iteration index values, for all the vector components, which means storing a $k\times N$ matrix, as large as $p\times N$, in addition to $\textbf{F}^{(k)}$. In contrast to that, the explicit formula in Eq. (\ref{B49}) does not require the storage of an additional $p\times N$ matrix beside $\textbf{F}^{(k)}$, but only the result of the application of a scalar product operator with a vector of size $p$ on the already stored matrix $\textbf{F}^{(k)}$.


\newpage
\clearpage

\newpage
\clearpage

\newpage
\clearpage

\newpage
\clearpage

\end{document}